\documentclass[aps,floats,twocolumn,prb,showpacs]{revtex4-1}
\usepackage{graphicx}
\usepackage{color}
\usepackage{upgreek}
\usepackage{setspace}
\usepackage{amssymb}
\usepackage{amsmath}
\usepackage{dsfont}
\usepackage{fancyhdr}
\usepackage{array}
\usepackage{multirow}

\allowdisplaybreaks[4]

\begin{document}
\title{Impact of nonlocal correlations over different energy scales: \\A Dynamical Vertex Approximation study}
\author{G. Rohringer$^{1,2}$ and A. Toschi$^1$}
\affiliation{$^1$Institute of Solid State Physics, Vienna University of Technology, 1040
Vienna, Austria\\
$^2$Russian Quantum Center, Novaya street, 100, Skolkovo, Moscow region 143025, Russia}
\date{Version 1, \today}

\begin{abstract}
In this paper, we investigate how nonlocal correlations affect, selectively, the physics of correlated electrons over different energy scales, from the Fermi level to the band-edges.  This goal is achieved by applying a diagrammatic extension of dynamical mean field theory (DMFT), the dynamical vertex approximation (D$\Gamma$A), to study several spectral and thermodynamic properties of the unfrustrated Hubbard model in two and three dimensions. Specifically, we focus first on the low-energy regime by computing the electronic scattering rate and the quasiparticle mass renormalization for decreasing temperatures at a fixed interaction strength. This way, we obtain a precise characterization of the several steps, through which the Fermi-liquid physics is progressively destroyed by nonlocal correlations. Our study is then extended to a broader energy range, by analyzing the temperature behavior of the kinetic and potential energy, as well as of the corresponding energy distribution functions. Our findings allow us to identify a smooth, but definite evolution of the nature of nonlocal correlations by increasing interaction: They either increase or decrease the kinetic energy w.r.t. DMFT depending on the interaction strength being weak or strong, respectively. This reflects the corresponding evolution of the ground state from a nesting-driven (Slater) to a superexchange-driven (Heisenberg) antiferromagnet (AF), whose fingerprints are, thus, recognizable in the spatial correlations of the paramagnetic phase. Finally, a critical analysis of our numerical results of the potential energy at the largest interaction allows us to identify possible procedures to improve the ladder-based algorithms adopted in the dynamical vertex approximation.
\end{abstract}

\maketitle

\let\n=\nu \let\o =\omega \let\s=\sigma


\section{Introduction}
\label{Sec:Intro}
The theoretical treatment of electronic correlations poses one of the major challenges to contemporary condensed matter physics. In fact, whenever the application of weak- or strong-coupling perturbative expansions is not possible, there are only a few situations for which rigorous analytical and/or numerical approaches are available. These include the limiting cases of one ($1d$) and infinite dimensions, where Bethe-Ansatz solutions\cite{Bethe1931,Hewson1993}, density matrix renormalization group (DMRG)\cite{Schollwock2005,Schollwock2011}/matrix product states (MPS)\cite{Perez2007} approaches and the dynamical mean field theory (DMFT)\cite{Georges1996}, respectively, have been successfully applied in the last decades.

No comparably powerful scheme, however, is presently available for treating the physically relevant cases of correlated electrons in two and three dimensions. Here, the intrinsic geometrical property of having more space directions at disposal changes radically the physics w.r.t. the one-dimensional case, preventing  any straightforward generalization of the rigorous approaches of $1d$ to higher dimensions. At the same time, the limited (finite) connectivity of $2d$ and $3d$ lattices does not justify any longer the application of mean-field approximations in space (such as DMFT), and their complete neglect of nonlocal spatial correlations. 

The same physical ingredients making hard the pathway towards a full theoretical understanding of electronic correlations, however, are responsible for some of the most exciting phenomena in solid state physics, such as, e.g., high-temperature superconductivity in cuprates\cite{Dagotto1994,Timusk1999,Lee2006}, iron-pnictides and chalchogenides\cite{Dai2015} and quantum criticality in transition metal oxides\cite{Buttgen2010} and heavy fermion materials\cite{Gegenwart2008}. This explains the huge quest for the development of new methods in quantum many-body theory, and the significant effort made by several research groups for improving the description of correlated electrons in two and three dimensions. Cutting edge approaches often exploit or extend the schemes, which worked successfully in different limiting situations. This is, for instance, the case of projected entangled pair states (PEPS)\cite{Orus2014} methods, which aim at extending the rigorous MPS treatment of the $1d$ physics to (at least) two dimensional systems. 

Here, however, we are interested in the opposite route of exploiting the exact description of correlations in the infinite dimensional limit (DMFT) as a starting point for nonperturbative approximations to the (more realistic) physics of interacting electrons confined in $3d$ solids or $2d$ layers. This route is paved by the extensions of DMFT. Cluster extensions\cite{Kotliar2001,Lichtenstein2000,Maier2005} allow for a rigorous treatment of short-range spatial correlations within a finite cluster size. In situations where long-range correlations prevail, however, diagrammatic extensions\cite{Toschi2007a,Kusunose2006,Rubtsov2008,Slezak2009,Valli2010,Toschi2011,Rohringer2013,Taranto2014,Wentzell2015,Ayral2015,Ayral2016a,Ayral2016} of DMFT are more suited as they treat fluctuations on {\sl all} length scales on equal footing. The latter exploit the purely local two-particle vertex functions and the (nonlocal) Green's functions computed in DMFT as a building blocks for different Feynman-diagrammatic expansions. The specific expansions chosen (typically finite order, ladder and/or parquet resummations of diagrams) thus allow for the inclusion of an important portion of long-range correlations neglected by DMFT and its cluster extensions in two and three dimensions. 

In this paper, by adopting one of the diagrammatic extensions of DMFT, the dynamical vertex approximation (D$\Gamma$A)\cite{Toschi2007a,Held2008},  we present a thorough  analysis of the mechanisms, through which the purely local physics of DMFT gets corrected by the nonlocal correlations of finite dimensions: We will go beyond the results of previous applications of D$\Gamma$A and other diagrammatic approaches, by investigating how spatial correlations in $2d$ and $3d$ selectively operate over different energy scales, depending on the parameter (interaction strength, temperature, dimension) region considered. In particular, as for the low-energy window, we will study how the Fermi-liquid properties of correlated electrons (at weak coupling) in $2d$ and $3d$ are progressively disfigured by low-temperature antiferromagnetic fluctuations of increasing strength and spatial extension. Further, we will expand our study to the whole energy domain, by analyzing how such fluctuations contribute to the internal energy of the electronic system and, specifically, to its kinetic and potential counterparts. The latter results will be also resolved in terms of the noninteracting dispersion of the system ($\varepsilon_{\mathbf{k}}$), by studying the evolution of the corresponding energy distribution functions [$n(\varepsilon)$]. In fact, while the low-energy window analysis helps to understand how the Fermi-liquid properties get altered by spatial correlations, the latter study provides also insights on how the Mott-Hubbard insulating physics is actually realized in finite dimensions. Our numerical results of D$\Gamma$A  will be also supplemented by analytical expressions derived by extracting the most relevant contributions of the corresponding D$\Gamma$A equations in the limit of a large correlation length. This way, we will be not only able to determine precisely the low-temperature behavior displayed by our numerical data, but also to compare it with that of complementary, semianalytic techniques such as the two-particle self-consistent\cite{Vilk1994,Vilk1997,Tremblay2011} (TPSC) approach or the composite operator method\cite{Matsumoto1992,Mancini2003,Mancini2004,Avella2007,Avella2014} (COM). 

Finally, we should also emphasize that the systematic study of the two- and three-dimensional physics presented in this paper is not only relevant for improving the understanding of correlated electrons in $2d$ and $3d$: It is also crucial to test the reliability of the diagrammatic approach adopted (here: the D$\Gamma$A)  to capture the overall physics of nonlocal correlations. In fact, diagrammatic schemes have been already applied hitherto to important but specific problems, such as the determination of the critical exponents in the Hubbard\cite{Rohringer2011,Hirschmeier2015} and the Falicov-Kimball model\cite{Antipov2014}, the observation of a pseudogap in two dimensions\cite{Huscroft2001,Moukouri2001,Jarrell2001a,Hafermann2009,Taranto2014,Schafer2015}, or the onset of competing superconducting instabilities\cite{Otsuki2015}. However, while the results of these selected applications have been quite successful, the lack of almost any kind of exact solution for the correlated physics in $2d$ and $3d$, strongly calls for an extensive benchmark of the overall physical description emerging from these approaches. In particular, our systematic study provides a comprehensive set of new numerical data and analytical trends in different coupling regimes: These can be used for future comparisons against the results of alternative techniques (including other diagrammatic or cluster extensions of DMFT, extrapolated lattice quantum Monte Carlo\cite{Blankenbecler1981,Schafer2015,Pudleiner2016a}, functional renormalization group\cite{Taranto2014,Eberlein2015}, etc.) in a similar spirit of the extensive benchmark review, recently presented for the $2d$ Hubbard model in Ref.~\onlinecite{Leblanc2015}. Even more important, our results can be also directly tested in terms of the internal consistency of the overall picture they are providing. This should fulfill, on its whole, specific physical expectations such as, e.g., various sum rules. In fact, as we will discuss in the final part of the paper, the critical cross-checked analysis of our D$\Gamma$A results will be also suggestive of possible, further algorithmic improvements for this approach.

The plan of the paper is the following:  In Sec.~\ref{sec:formalismphasediag} we will present the D$\Gamma$A formalism (focusing on the most used algorithm based on ladder diagrams resummations) and revisit the D$\Gamma$A phase diagrams of the two- and three-dimensional Hubbard model. In Sec.~\ref{sec:nearfermi} the impact of nonlocal correlations on Fermi liquid properties of the system is analyzed in detail. Afterward, in Sec.~\ref{sec:allenergies}, the investigation of the role of nonlocal correlations is extended to all energy scales by analyzing kinetic and potential energies and energy distribution functions. The critical analysis of the results of Sec.~\ref{sec:allenergies} inspired us to propose in Sec.~\ref{sec:lambdaimproved} an improved scheme for future ladder D$\Gamma$A calculations, which might be also of interest for other diagrammatic approaches. Finally, in Sec.~\ref{sec:conclusions}, we summarize our results and put them in perspective of further progresses in the nonperturbative treatment of electronic correlations in finite dimensions.

\section{nonlocal correlations from ladder D$\Gamma$A}
\label{sec:formalismphasediag}

The starting point of our study is the two- and three-dimensional Hubbard model on a simple square ($2d$) and cubic ($3d$) lattice, respectively
\begin{equation}
 \label{equ:defmodel}
 \hat{\mathcal{H}}=-t\sum_{<ij>\sigma}\hat{c}^{\dagger}_{i\sigma}\hat{c}_{j\sigma}+U\sum_i\hat{n}_{i\uparrow}\hat{n}_{i\downarrow},
\end{equation}
where $t$ denotes the hopping amplitude between nearest neighbors, $U$ is the Coulomb interaction, and $\hat{c}_{i\sigma }^{\dagger }$($\hat{c}_{i\sigma }$) creates (annihilates) an electron with spin $\sigma $ on site $i$; $\hat{n}_{i\sigma}\!=\!\hat{c}_{i\sigma }^{\dagger }\hat{c}_{i\sigma }$. We restrict ourselves to the paramagnetic phase with $n=1$ electron/site (half filling) at a finite temperatures $T\!=\!1/\beta>T_N$, where $T_N$ is the transition temperature to the low-$T$ antiferromagnetically ordered phase. For the sake of clarity, and in accordance with previous publications, we will define hereafter our energies in terms of a typical energy scale $D=2 \sqrt{2d}\, t$, where $d$ denotes the dimension of the system. This choice fixes the standard deviation $\sigma$ of the noninteracting density of states (DOS) to $0.5$ in all dimensions and, hence, allows for a better comparability between results for different dimensions.

A brief remark is in order here about the conventions adopted in the paper. As we will discuss derivations and results for self-energies, spectral and vertex functions on the imaginary as well as on the real frequency axis we resort to the following notation: $i\nu$ and $i\nu'$ will denote fermionic and $i\Omega$ bosonic (imaginary) Matsubara frequencies. If used as an argument in parenthesis, they are written in combination with the imaginary unit $i$, e.g., $\Sigma(i\nu)$. For the sake of readability we omit such $i$ when these frequencies are written in index notation (mostly for the vertex functions), e.g., $\Gamma^{\nu\nu'\Omega}$. A lower case $\omega$ will always denote real frequencies. For a detailed definition and an extensive discussion of the symmetry properties of all relevant one- and two-particle Green's and vertex functions considered in this paper we refer the reader to Ref.~\onlinecite{Rohringer2012} and the supplementary material of Ref.~\onlinecite{Gunnarsson2015}. 

All quantities which do not explicitly contain a momentum argument are purely local and can be obtained from the auxiliary Anderson impurity model (AIM) related to the DMFT solution of model (\ref{equ:defmodel}). For this task we have exploited an exact diagonalization (ED) solver using $N_s\!=\!4\!+\!1$ bath and impurity site(s), respectively. By comparing the AIM's Green's function with the local part of the corresponding lattice Green's function we have verified that the fitting procedure w.r.t. the ED discrete bath works accurately in the considered (intermediate-$T$) parameter regimes. Moreover, we have already tested several times in previous works\cite{Rohringer2012,Rohringer2013,Gunnarsson2015,Gunnarsson2016} that, for the DMFT one- and two-particle Green's functions, the deviations observed between our ED calculations and quantum Monte Carlo (QMC) results [both Hirsch-Fye and continuous-time\cite{Gull2011a} (CT) QMC] are only marginal in the low energy (frequency) regime. As for the high-frequency asymptotic region of the Green's and vertex functions, the ED solver has the intrinsic advantage of avoiding any statistical error compared to corresponding QMC methods. This is of particular importance both for the numerical solution of the D$\Gamma$A equations as well as for the analytic continuation of the Matsubara self-energies by means of Pad\'e fits.

\subsection{D$\Gamma$A formalism}
\label{subsec:formalism}

The D$\Gamma$A approach to the model (\ref{equ:defmodel}) has been derived in Refs.~\onlinecite{Toschi2007a,Katanin2009}. We briefly review here the principal equations, focusing on the algorithmic aspects which are most relevant for our work. The basic idea of D$\Gamma$A is to introduce nonlocal correlations beyond the local ones of DMFT in the self-energy $\Sigma(i\nu,\mathbf{k})$ of the system. This is achieved by means of the (Schwinger-Dyson) equation of motion (EOM)
\begin{align}
 \label{equ:EOM}
 \Sigma(i\nu,\mathbf{k})=&\frac{Un}{2}-\frac{U}{\beta^2}\underset{\mathbf{k'}\mathbf{q}}{\sum_{\nu'\Omega}}F_{\uparrow\downarrow,\mathbf{k}\mathbf{k'}\mathbf{q}}^{\nu\nu'\Omega}G(i\nu',\mathbf{k'})\nonumber\\&\times G(i\nu'+i\Omega,\mathbf{k'}+\mathbf{q})G(i\nu+i\Omega,\mathbf{k}+\mathbf{q})
\end{align}
In the full parquet-based version of D$\Gamma$A the vertex $F_{\uparrow\downarrow,\mathbf{k}\mathbf{k'}\mathbf{q}}^{\nu\nu'\Omega}$ and the Green's functions $G(i\nu,\mathbf{k})$ are constructed from the local fully irreducible vertex $\Lambda_{\sigma\sigma'}^{\nu\nu'\Omega}$ of DMFT \cite{Toschi2007a}. This requires the self-consistent solution of the numerically very demanding parquet- and Bethe-Salpeter-equations in {\sl all} scattering channels\cite{Chen1992a,Yang2009,Tam2013,Valli2015,Li2015a,Wentzell2016}. The convergence of the full D$\Gamma$A algorithm is made numerically very challenging also by possible divergences\cite{Schafer2013,Gunnarsson2016,Schafer2016} of the two-particle irreducible vertex function $\Lambda_{\sigma\sigma'}^{\nu\nu'\Omega}$ in the intermediate-to-strong coupling regime.

However, if we know {\sl a priori} which type of fluctuations prevails in the system, we can restrict ourselves to the corresponding Bethe-Salpeter equation in this channel, which corresponds to the {\sl ladder} version of D$\Gamma$A. In this case the Green's function $G(i\nu,\mathbf{k})$ in Eq.~(\ref{equ:EOM}) is just the one obtained from DMFT, i.e., $G(i\nu,\mathbf{k})=\left[i\nu+\mu-\varepsilon_{\mathbf{k}}-\Sigma(i\nu)\right]^{-1}$. Here, $\mu$ is the chemical potential of the system, $\varepsilon_{\mathbf{k}}=-2t\sum_{i=1}^{d}\cos(k_i)$ denotes the bare dispersion for the square (cubic) lattice in $2(3)$ dimensions, and $\Sigma(i\nu)$ is the purely local self-energy calculated within DMFT from the auxiliary AIM. In the ladder approximation the generalized susceptibility $\chi_{r,\mathbf{q}}^{\nu\nu'\Omega}$ (and from this of course the corresponding vertex $F_{r,\mathbf{q}}^{\nu\nu'\Omega}$) is constructed by a ladder (i.e., a Bethe-Salpeter equation) in the given channel(s) $r$. In the case of half filling particle-particle fluctuations are suppressed and, hence, we restrict ourselves to the two particle-hole channels $r=(d)$ensity and $r=(m)$agnetic in which the generalized susceptibilities read
\begin{align}
 \label{equ:bethesalp}
 \chi_{r,\mathbf{q}}^{\nu\nu'\Omega}=&\beta\delta_{\nu\nu'}\chi_{0,\mathbf{q}}^{\nu\Omega}-\frac{1}{\beta}\chi_{0,\mathbf{q}}^{\nu\Omega}\sum_{\nu_1}\Gamma_{r}^{\nu\nu_1\Omega}\chi_{r,\mathbf{q}}^{\nu_1\nu'\Omega} \nonumber\\
 =&\beta\delta_{\nu\nu'}\chi_{0,\mathbf{q}}^{\nu\Omega}-\chi_{0,\mathbf{q}}^{\nu\Omega}F_{r,\mathbf{q}}^{\nu\nu'\Omega}\chi_{0,\mathbf{q}}^{\nu'\Omega},
\end{align}
where $\Gamma_{r}^{\nu\nu'\Omega}$ is the local irreducible vertex in the channel $r=d,m$ from DMFT and $\chi_{0,\mathbf{q}}^{\nu\Omega}=-\sum_{\mathbf{k}}G(i\nu,\mathbf{k})G(i\nu+i\Omega,\mathbf{k}+\mathbf{q})$ is the bare susceptibility of DMFT. Inserting the vertex $F_{r,\mathbf{q}}^{\nu\nu'\Omega}$ into the EOM [Eq.~(\ref{equ:EOM})] yields the ladder-D$\Gamma$A self-energy $\Sigma_{\text{lad}}(i\nu,\mathbf{k})$. 

It turns out that the self-energy calculated by means of plain ladder D$\Gamma$A diagrams exhibits a violation of the asymptotic behavior. This means the $1/i\nu$-part of $\Sigma_{\text{lad}}(i\nu,\mathbf{k})$ would not have the (correct) prefactor $U^2\frac{n}{2}\left(1-\frac{n}{2}\right)$, differently from the exact and the DMFT solution. A more detailed discussion of this issue is reported in Appendix \ref{app:lambdacorr} and Ref.~\onlinecite{Rohringer2013a}. Specifically, there we demonstrate that the wrong asymptotics of $\Sigma_{\text{lad}}(i\nu,\mathbf{k})$ is a direct consequence of a violation of the sum rules of the physical susceptibilities $\chi_{r,\mathbf{q}}^{\Omega}=\frac{1}{\beta^2}\sum_{\nu\nu'}\chi_{r,\mathbf{q}}^{\nu\nu'\Omega}$. 

In order to overcome this problem, following the Moriya theory of itinerant magnetism\cite{Moriya1985}, one should introduce a so-called $\lambda$ correction in the theory (see Refs.~\onlinecite{Katanin2009,Rohringer2011}). In practice, this is performed by inserting one (or two) scalar parameter(s) $\lambda_m$ (and $\lambda_d$) into our ladder D$\Gamma$A equations. For the practical application of such a procedure, we have to consider two main questions:

(i) {\sl What is the condition to fix the value(s) of $\lambda_r$?} One necessary condition is of course to restore the correct asymptotic behavior of the ladder D$\Gamma$A self-energy. As shown in Appendix \ref{app:lambdacorr}, this requirement is equivalent to the fulfillment of the sum rule
\begin{equation}
 \label{equ:sumruleoccup}
 \frac{1}{\beta}\sum_{\Omega\mathbf{q}}\chi_{\uparrow\uparrow,\mathbf{q}}^{\Omega}\equiv\frac{1}{\beta}\sum_{\Omega\mathbf{q}}\frac{1}{2}\left[\chi_{d,\mathbf{q}}^{\Omega}+\chi_{m,\mathbf{q}}^{\Omega}\right]\equiv\frac{n}{2}\left(1-\frac{n}{2}\right)
\end{equation}
for the $\uparrow\uparrow$ spin susceptibility. Let us stress that the above condition is always fulfilled for the purely local susceptibilities of DMFT, but will be in general violated by approximate theories such as the ladder D$\Gamma$A. In this work we enforce condition (\ref{equ:sumruleoccup}) by applying a $\lambda$ correction only to $\chi_{m,\mathbf{q}}^{\Omega}$, i.e., we replace $\chi_{m,\mathbf{q}}^{\Omega}$ by $\chi_{m,\mathbf{q}}^{\lambda,\Omega}$ as defined in Eqs.~(\ref{equ:deflambdacorr}) or (\ref{equ:cap4lambdacorr}). The value of $\lambda$ is then fixed by the constraint\footnote{For the practical determination of $\lambda$, aiming at a minimization of the error due to the finite frequency cut-off in Eq. (\ref{equ:lambdafixused}), we have identified the left hand side of this equation with the corresponding DMFT expression $\frac{1}{\beta}\sum_{\Omega}\frac{1}{2}\left[\chi_{m}^{\Omega}+\chi_{d}^{\Omega}\right]$ summed over the same number of Matsubara frequencies. In this way Eq. (\ref{equ:lambdafixused}) is solved for a right hand side whose error is of the same order in $1/i\Omega$ as the left hand side.}
\begin{align}
 \label{equ:lambdafixused}
 \frac{1}{\beta}\sum_{\Omega\mathbf{q}}\frac{1}{2}\left[\chi_{m,\mathbf{q}}^{\lambda,\Omega}+\chi_{d,\mathbf{q}}^{\Omega}\right]&=\frac{n}{2}\left(1-\frac{n}{2}\right).
\end{align}
The above procedure raises obviously the question whether a similar correction should be applied to the charge susceptibility of D$\Gamma$A as well. However, if we want to correct both, the charge- {\sl and} the spin-propagator by means of a $\lambda_d$ {\sl and} a $\lambda=\lambda_m$, we need a second condition to fix both parameters. In a previous work\cite{Schafer2015} the assumption that the sum rules for both, the charge- and the spin-propagator, coincide independently with the corresponding local sum rules of DMFT has been made, i.e.,
\begin{equation}
 \label{equ:lambdafixusedomega}
 \frac{1}{\beta}\sum_{\Omega\mathbf{q}}\chi_{r,\mathbf{q}}^{\lambda_r,\Omega}=\frac{1}{\beta}\sum_{\Omega}\chi_{r}^{\Omega}.
\end{equation}
Note that an analogous condition is exploited in the dual boson approach\cite{vanLoon2014,vanLoon2016}. As {\sl local} DMFT correlation functions fulfill all (local) sum rules the above condition indeed ensures the correct particle number at the two-particle level [cf. Eq. (\ref{equ:sumruleoccup}) and Appendix \ref{app:lambdacorr}]. A more formal justification of condition (\ref{equ:lambdafixusedomega}) has been recently given in the context of the QUADRILEX method\cite{Ayral2016} which, however, requires the solution of an AIM containing a (three-frequency dependent) dynamically screened interaction.

The state-of-the-art $\lambda$-correction schemes discussed here, however, still present an intrinsic problem: They lead to ambiguous results for the kinetic and potential energy of the system within the ladder D$\Gamma$A scheme. Specifically, one obtains different values for the potential and kinetic energy when calculating these quantities from one- and two-particle propagators of the ladder D$\Gamma$A, respectively. This issue, which affects also DMFT calculations for finite dimensional systems as recently noted in Ref. \onlinecite{vanLoon2016}, is discussed in more detail in Sec.~\ref{sec:lambdaimproved}.

(ii) {\sl How can we introduce the parameters $\lambda_r$ into the ladder D$\Gamma$A equation?} The most natural way of introducing the $\lambda$ corrections into the ladder D$\Gamma$A approach is by applying it directly to the physical susceptibilities $\chi_{r,\mathbf{q}}^{\Omega}$. In practice this is achieved by just correcting the {\sl masses} of these propagators with $\lambda_r$, which can be implemented by the following equation (see Appendix \ref{app:lambdacorr}):
\begin{equation}
 \label{equ:deflambdacorr}
 \chi_{r,\mathbf{q}}^{\lambda,\Omega}=\left[\left(\chi_{r,\mathbf{q}}^{\Omega}\right)^{-1}+\lambda_r\right]^{-1}.
\end{equation}
Hence, in order to include the $\lambda$ corrections in the calculation of the self-energy within the ladder-D$\Gamma$A approach, we have to rewrite Eq.~(\ref{equ:EOM}) in such a way that the physical susceptibilities $\chi_{r,\mathbf{q}}^{\Omega}$ appear explicitly. Following Ref.~\onlinecite{Katanin2009}, this can be achieved by separating the D$\Gamma$A ladders by a bare interaction vertex $U_r=+/-U$ for $r=d/m$. Specifically, one defines the ladder quantity $\Phi_{r,\mathbf{q}}^{\nu\nu'\Omega}$ by the Bethe-Salpeter-like equation
\begin{equation}
 \label{equ:defphi}
 \chi_{r,\mathbf{q}}^{\nu\nu'\Omega}=\Phi_{r,\mathbf{q}}^{\nu\nu'\Omega}-\frac{1}{\beta^2}\sum_{\nu_1\nu_2}\Phi_{r,\mathbf{q}}^{\nu\nu_1\Omega}U_r\chi_{r,\mathbf{q}}^{\nu_2\nu'\Omega}.
\end{equation}
From this one can derive the so-called three-legs vertex $\gamma_{r,\mathbf{q}}^{\nu\Omega}$
\begin{equation}
 \label{equ:defgamma}
 \gamma_{r,\mathbf{q}}^{\nu\Omega}=\left(\chi_{0,\mathbf{q}}^{\nu\Omega}\right)^{-1}\frac{1}{\beta}\sum_{\nu'}\Phi_{r,\mathbf{q}}^{\nu\nu'\Omega},
\end{equation}
which is the same as the three-legs irreducible vertex appearing in the recently introduced TRILEX approach\cite{Ayral2015,Ayral2016a} (see Appendix \ref{app:threeleg}). Expressing now $F_{r,\mathbf{q}}^{\nu\nu'\Omega}$ in terms of $\Phi_{r,\mathbf{q}}^{\nu\nu'\Omega}$ allows us to rewrite the EOM in the case of ladder-D$\Gamma$A in the following form \cite{Katanin2009,Rohringer2011}:
\begin{align}
 \label{equ:EOMrewrite}
 \Sigma_{\text{lad}}&(i\nu,\mathbf{k})=\frac{Un}{2}-\frac{U}{\beta}\sum_{\Omega\mathbf{q}}\left[1+\frac{1}{2}\gamma_{d,\mathbf{q}}^{\nu\Omega}\left(1-U\chi_{d,\mathbf{q}}^{\Omega}\right)-\right.\nonumber\\&\left.\frac{3}{2}\gamma_{m,\mathbf{q}}^{\nu\Omega}\left(1+U\chi_{m,\mathbf{q}}^{\Omega}\right)+\right.\nonumber\\&\left.\frac{1}{2\beta}\sum_{\nu'}\chi_{0,\mathbf{q}}^{\nu'\Omega}\left(F_{d}^{\nu\nu'\Omega}-F_{m}^{\nu\nu'\Omega}\right)\right]G(i\nu+i\Omega,\mathbf{k}+\mathbf{q}),
\end{align}
where the term in the last line of this equations avoids the double counting of local DMFT diagrams.

Eventually, the application of our $\lambda$-correction scheme to the ladder D$\Gamma$A self-energy consists of replacing $\chi_{r,\mathbf{q}}^{\Omega}$ by the corresponding $\lambda$-corrected quantity $\chi_{r,\mathbf{q}}^{\lambda,\Omega}$ [see Eq.~(\ref{equ:deflambdacorr})] in Eq.~(\ref{equ:EOMrewrite}). Note that within this scheme the three-leg vertices $\gamma_{r,\mathbf{q}}^{\nu\Omega}$ remain {\sl uncorrected}. The latter assumption is justified, since $\gamma_{r,\mathbf{q}}^{\nu\Omega}$ does not contain any physical susceptibilities and, hence, is very little affected by nonlocal correlations as it is discussed in more detail in Appendix \ref{app:threeleg}.

\subsection{Phase diagrams revisited}
\label{subsec:phasediagram}

\begin{figure*}[t!]
 \centering
 \begin{tabular}{cc}
  \includegraphics[width=0.5\textwidth]{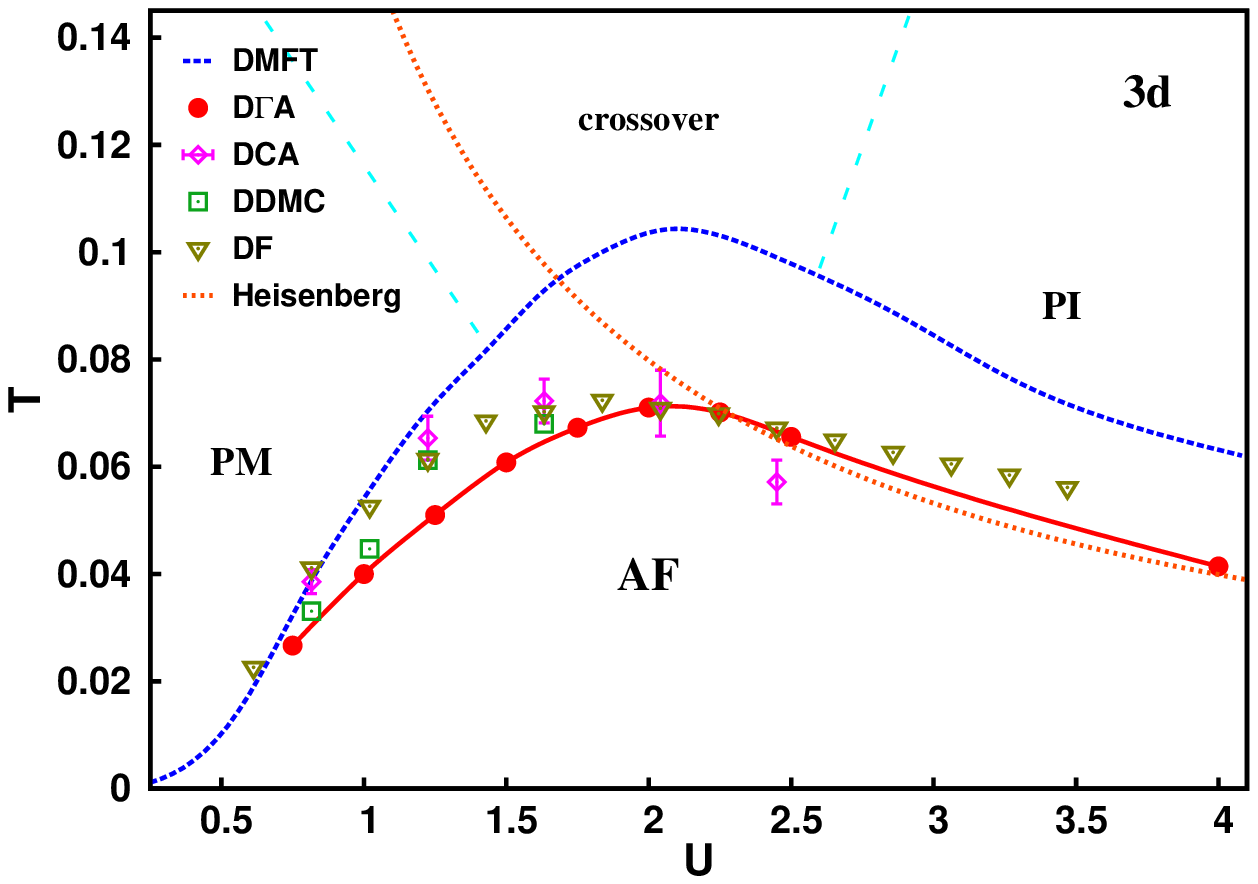} &
  \includegraphics[width=0.5\textwidth]{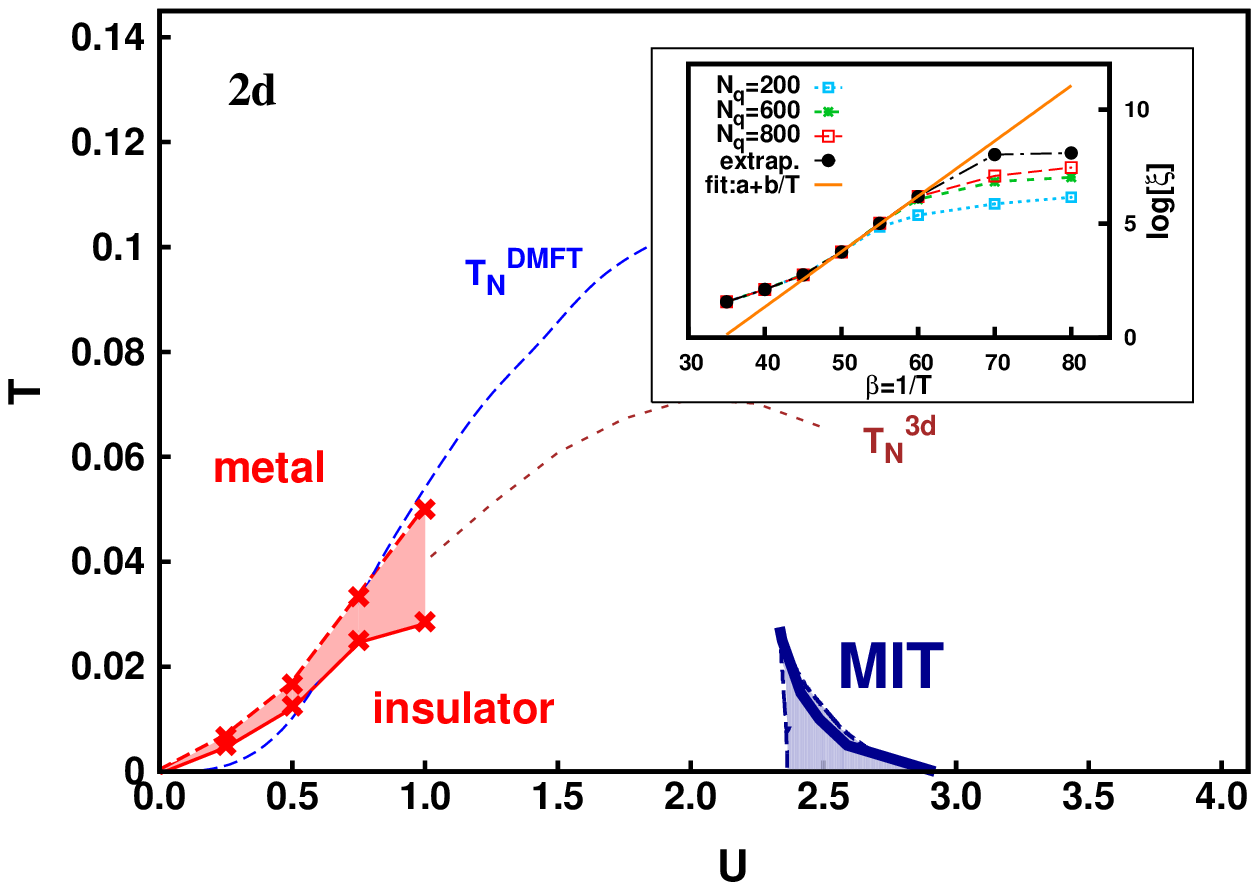}
	\end{tabular}
 \caption{Phase diagrams of the half-filled Hubbard model on a simple square and cubic lattice in 3d (left panel) and 2d (right panel), respectively. Data points of the main panels are reproduced from Refs.~\onlinecite{Rohringer2011,Hirschmeier2015} (3d) and \onlinecite{Schafer2015} (2d), respectively, except for the new D$\Gamma$A estimate of $T_N$ at $U=4.0$ in $3d$. Right panel: The metal-to-insulator (MIT) transition marked in blue is the one observed in paramagnetic DMFT calculations. Because of the strong nonlocal correlations captured by D$\Gamma$A calculations, in $2d$ the MIT is transformed into a sharp crossover (red shaded area) which extends down to $U\!=\!0$ at low $T$\cite{Schafer2015,Schafer2016a}. Inset: $\log\left[\xi\right]$, computed in D$\Gamma$A for $U=0.75$, as a function of the inverse temperature $\beta=1/T$ for different sizes of the momentum grid ($2N_{\mathbf{q}}$ is the number of $\mathbf{q}$ points in one direction) used for the determination of $\lambda$. The fit (orange line) has been performed in the low-$T$ regime according to the exponential function in Eq.~(\ref{equ:xi2d}).}
 \label{fig:phasediagrams}
\end{figure*}

To place the results of this work in the context of the previous D$\Gamma$A studies, let us briefly reconsider the D$\Gamma$A phase diagrams for the half-filled Hubbard model in two and three spatial dimensions as obtained in Refs.~\onlinecite{Rohringer2011,Schafer2015}.

In the {\sl three}-dimensional half-filled Hubbard model we observe a phase transition to an antiferromagnetically ordered phase at a finite temperature. In the left panel of Fig.~\ref{fig:phasediagrams} the $3d$ transition temperature of D$\Gamma$A (red dots) is shown as a function of the interaction parameter $U$ and compared to recent dynamical cluster approximation (DCA)\cite{Kent2005} (pink diamonds), determinantal diagrammatic Monte Carlo (DDMC)\cite{Kozik2013} (green squares), and dual fermion (DF)\cite{Hirschmeier2015} (brown triangles) results. All methods show a clear reduction of $T_N$ compared to its DMFT value (blue line). At weak coupling this reduction is slightly stronger in D$\Gamma$A than in the other methods though D$\Gamma$A results are rather close to the most recent DDMC estimates. Possible origins of this discrepancy have been discussed in detail in Ref.~\onlinecite{Rohringer2011}. At stronger coupling ($U=4.0$) D$\Gamma$A data agree well with the transition temperature of the corresponding Heisenberg model onto which the Hubbard model can be mapped at large couplings, while DF results find slightly higher transition temperatures. In this regime, where only spin fluctuations survive, the ladder D$\Gamma$A estimate of $T_N$ appears particularly accurate. We recall that in $3d$ nonlocal correlations play an important role for one-particle (spectral) properties in a relatively narrow temperature regime above $T_N$\cite{Rohringer2011}. At higher temperatures, thermal fluctuations become predominant, mitigating the D$\Gamma$A corrections to the local physics of DMFT. A more refined study of how this effect might occur differently for different physical observables will be, however, addressed in the next sections.  

In {\sl two} spatial dimensions the situation changes drastically as one can see from the corresponding phase diagram in the right panel of Fig.~\ref{fig:phasediagrams}: Due to the Mermin-Wagner theorem\cite{Mermin1966}, fulfilled in our D$\Gamma$A treatment, the antiferromagnetic phase is restricted to $T=0$. Previous D$\Gamma$A studies\cite{Katanin2009,Schafer2015} have shown that the long ranged antiferromagnetic fluctuations responsible for the suppression of the magnetic order induce at all values of $U$ a crossover (red shaded area in the right panel of Fig.~\ref{fig:phasediagrams}) to a low-$T$ insulating state. In fact, as discussed in Ref. \onlinecite{Schafer2015}, for small $U$ the crossover to an insulating state takes place approximately in the temperature regime where the rate of growth of the (AF) spin correlation length ($\xi$) becomes exponentially large. The low-$T$ behavior of $\xi(T)\sim e^{c/T}$ is explicitly shown in the inset of the right panel of Fig.~\ref{fig:phasediagrams}, which supports, refining them, the results of Ref.~\onlinecite{Schafer2015}: By plotting $\log\xi$ as a function of the inverse temperature ($\beta$), and studying the behavior as a function of increasingly denser momentum grids (for the precise determination of the $\lambda$ corrections), one sees how, at $U=0.75$, a direct proportionality sets in at an inverse temperature of $\beta\approx 45$, matching very well the onset of the crossover region in the $2d$ phase diagram.

In the following sections, we still start from these general D$\Gamma$A descriptions of the $2d$ and $3d$ physics, to analyze more profoundly the microscopic mechanisms at work and, in particular, how the underlying nonlocal correlations are operating selectively for different energy scales in the various cases considered.

\vskip 5mm
\section{D$\Gamma$A results near the Fermi surface}
\label{sec:nearfermi}

nonlocal correlations do not only affect two-particle response functions such as the magnetic susceptibility $\chi_{m,\mathbf{q}}^{\Omega}$ and, hence, phase transitions and the associated critical phenomena\cite{Rohringer2011,Antipov2014,Hirschmeier2015} described in the previous section. They also have a crucial impact on one-particle properties such as self-energies and spectral functions. At low energies (i.e., near the Fermi surface) nonlocal correlations change the Fermi liquid nature of the system described by DMFT, leading eventually to a breakdown of the Fermi liquid behavior. The way how this happens, however, is absolutely not trivial. In fact, the temperature at which the Fermi liquid breakdown can be observed in a specific physical observable is not necessarily the same for all observables. Specifically, we will show in the following that while the self-energies and spectral functions of the half-filled Hubbard model itself exhibit a Fermi-liquid-like structure down to moderately low temperatures above $T_N$, the temperature dependence of the self-energy and -more precisely- of the low-energy coefficients [for the definition see Eqs.~(\ref{subequ:defflparam})] extracted from it already feature a non-Fermi-liquid behavior at substantially higher temperatures. This scenario emerges clearly from our D$\Gamma$A data and will be discussed extensively in comparison with DMFT in the following two sections.  

\subsection{Self-energies and spectral functions}
\label{subsec:sigmaspectra}

\begin{figure*}
 \centering
  \includegraphics{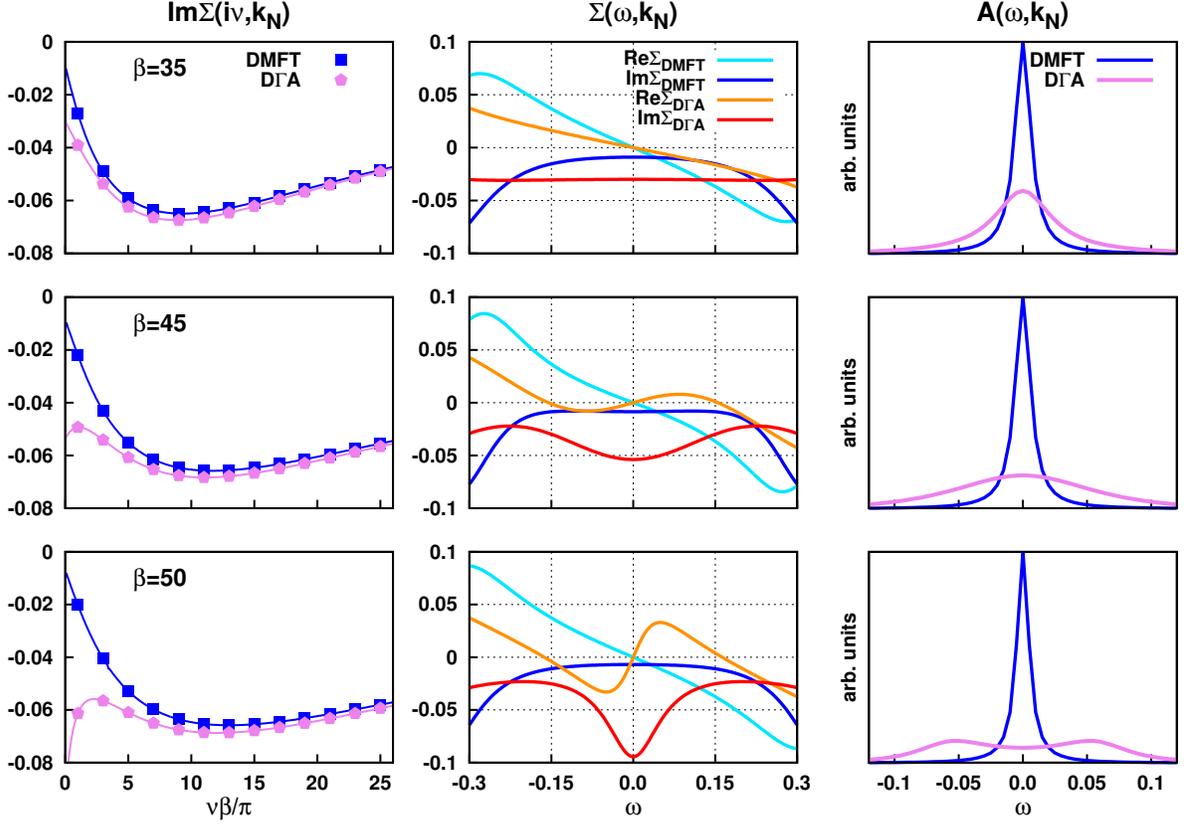}
 \caption{Self-energies and spectral functions of the $2d$ Hubbard model at the nodal point [$\mathbf{k}_N=(\pi/2,\pi/2)$] for three (decreasing) temperatures $\beta=35$ (first row), $\beta=45$ (second row) and $\beta=50$ (third row). In the {\sl first column} the Matsubara self-energy from DMFT (blue squares) and D$\Gamma$A (purple pentagons) and their corresponding Pad\'e fits (continuous lines) are shown as a function of the (fermionic) Matsubara index $2n+1=\nu\beta/\pi$. In the {\sl second column} the real and imaginary parts of the (retarded) self-energy are reported. The corresponding spectral functions of DMFT and D$\Gamma$A, respectively, are plotted in the {\sl third column} of the figure.}
 \label{fig:spectra_nodal}
\end{figure*}

We will start with an analysis of the retarded self-energy on the real frequency axis and the corresponding spectral functions in {\sl two dimensions} at $U=0.75$ obtained from DMFT and D$\Gamma$A. To this end we have performed Pad\'e fits (for details see Appendix \ref{app:pade}) for our self-energy data on the Matsubara axis for the two arguably most relevant $\mathbf{k}$ points on the Fermi surface, i.e., the so-called nodal point $\mathbf{k}_N=(\pi/2,\pi/2)$ and the antinodal point $\mathbf{k}_A=(\pi,0)$. For each of the two $\mathbf{k}$ points we have conducted calculations at three different temperatures. The results for the {\sl nodal} point are shown in Fig.~\ref{fig:spectra_nodal}. 
In the case of DMFT, one observes a Fermi-liquid-like structure of the self-energy for all temperatures, as it is expected for a small value of the interaction: At low energies, the slope of the real part of $\Sigma(\omega,\mathbf{k}_N)$ is always negative leading to the typical quasiparticle mass enhancement $m\rightarrow m^*=(1+\alpha_{\mathbf{k}})m$ [for the definition of $\alpha_{\mathbf{k}}$ see Eq. (\ref{equ:defalpha})]. Correspondingly, the imaginary part always exhibits a maximum (i.e., a minimum in absolute value) around $\omega=0$ indicating a minimal scattering rate for the quasiparticles at the Fermi energy. These observations are well reflected in the DMFT spectral functions where one observes the typical Fermi liquid quasiparticle peak at $\omega\!=\!0$ for all temperatures. 

nonlocal antiferromagnetic correlations taken into account by D$\Gamma$A change the situation drastically: At the {\sl highest temperature} $\beta=35$ these correlations are rather weak since the system is far away from the $T\!=\!0$ antiferromagnetic phase transition. In this case, no {\sl qualitative} difference to the DMFT data is found (in fact, though not visible on the scale of the plot, a very tiny maximum in the imaginary part of the self-energy at $\omega\!=\!0$ is found in the data). This Fermi liquid behavior can be also seen in the D$\Gamma$A spectral function, where a quasi particle peak -albeit strongly broadened compared to DMFT- is still present. Our results, hence, indicate that nonlocal fluctuations are still moderate at the considered temperature. This agrees well with the fact that at $\beta=35$ the system has not yet entered the critical regime in the 2d phase diagram (see right panel of Fig. \ref{fig:phasediagrams}) characterized by an exponential growth of the correlation length with decreasing temperature\cite{Schafer2015,Vilk1997} (see Sec. \ref{subsec:phasediagram}). The situation changes remarkably at the {\sl lower temperature} $\beta=45$, which is of the order of the crossover temperature in $2d$: While the Matsubara self-energy (first panel in the second line of Fig.~\ref{fig:spectra_nodal}) may still suggest a Fermi liquid behavior, the Pad\'e fit already displays a change in curvature when approaching zero frequency, i.e., it bends down slightly for $\nu\!\rightarrow\! 0$. This behavior is reflected in the real and the imaginary parts of the (retarded) self-energy on the real axis (second panel in the second line): $\mbox{Re}\Sigma(\omega,\mathbf{k}_N)$ shows already a positive -albeit very small- slope at $\omega\!=\!0$, while $\mbox{Im}\Sigma(\omega,\mathbf{k}_N)$ has a clear dip at the Fermi level. Both features are definite hallmarks of the breakdown of the Fermi liquid behavior of the system at the given temperature. Somewhat surprisingly, this breakdown is not clearly visible in the spectral function of the system, as it can be observed in the third panel of the second row in Fig.~\ref{fig:spectra_nodal}: At $\omega\!=\!0$ we still see a ``peak'' -albeit enormously broadened- suggesting the existence of quasiparticles compatible with a Fermi liquid description of the system. 

The unexpected dichotomy can be, however, understood by means of the following analytical considerations: The necessary and sufficient condition for the existence of a (non-Fermi-liquid/pseudogap) dip in the spectral function at $\omega=0$ is that its second derivative is larger than zero. Hence, expressing the spectral function in terms of the real and the imaginary part of the self-energy, one gets the following condition for the presence of a dip in the spectrum (for the explicit derivation see Appendix \ref{app:spectraldip}):
\begin{equation}
 \label{equ:dipcondition}
 \left[\frac{d^2}{d\omega^2}\text{Im} \Sigma(\omega,\mathbf{k})\right]_{\omega=0}>2\frac{1-\alpha_{\mathbf{k}}^2}{\gamma_{\mathbf{k}}},
\end{equation} 
where the coefficients $\alpha_{\mathbf{k}}$ and $\gamma_{\mathbf{k}}$ (mass renormalization and scattering rate in the Fermi liquid regime) are defined in the standard way as:
\begin{subequations}
\label{subequ:defflparam}
\begin{align}
 \label{equ:defalpha}
 &\alpha_{\mathbf{k}}=-\left[\frac{d}{d\omega}\text{Re} \Sigma(\omega,\mathbf{k})\right]_{\omega=0} \\
 &\gamma_{\mathbf{k}}=-\text{Im}\Sigma(\mathbf{k},\omega=0). \label{equ:defgammafl}
\end{align}
\end{subequations}
Equation (\ref{equ:dipcondition}), however, can be violated also for a non-Fermi-liquid self-energy (i.e., where the second derivative of the imaginary part of the self-energy at $\omega=0$ is already larger than $0$). This may happen if $\alpha_{\mathbf{k}}$ is sufficiently small and, hence, the right hand side of this inequality becomes considerably large. In our numerical data at $\beta\!=\!45$ we find indeed a very small slope of the real part of the D$\Gamma$A self-energy (see second panel in the second row of Fig.~\ref{fig:spectra_nodal}) consistent with the peculiar situation of the absence of a dip in the spectral function, as the right hand side of Eq. (\ref{equ:dipcondition}) becomes pretty large\footnote{Notably, the sign of the parameter $\alpha_{\mathbf{k}}$ is irrelevant for condition (\ref{equ:dipcondition}). Hence, in principle the reversed situation of a dip in the spectrum in a pure Fermi liquid regime, where $(d^2/d\omega^2)\text{Im} \Sigma(\omega,\mathbf{k})<0$, appears possible provided that $\alpha_{\mathbf{k}}$ is sufficiently larger than $1$ so that the right hand side of the inequality becomes negative. To determine whether such a situation is prohibited by other constraints of the theory requires, however, further investigations.}. 

On the other hand, in the standard case of a non-Fermi liquid with $(d^2/d\omega^2)\text{Im} \Sigma(\omega=0,\mathbf{k})>0$ {\sl and} $\alpha_{\mathbf{k}}<-1$ Eq.~(\ref{equ:dipcondition}) is always satisfied and, hence, the existence of a dip in the spectral function is guaranteed. This more conventional situation can be observed at the lowest considered temperature $\beta=50$ (third line of Fig. \ref{fig:spectra_nodal}): Here already the self-energy on the Matsubara axis clearly indicates the non-Fermi-liquid behavior of the system (first panel) by an abrupt change of curvature at the lowest Matsubara frequency. This matches the positive slope of the real part of the self-energy on the real axis (corresponding to a negative value of $\alpha_{\mathbf{k}}$) and the strong dip of the corresponding imaginary part at $\omega\!=\!0$ (second panel). Consistent with Eq. (\ref{equ:dipcondition}), thus, the non-Fermi-liquid dip in $A(\omega)$ becomes clearly visible in the numerical data. This demonstrates the full destruction of all Fermi liquid spectral properties by nonlocal correlations at $\beta=50$.   

An analogous analysis has been performed for the self-energies and spectral functions at the antinodal point. Because of the overall similarity, the data are reported in Appendix~\ref{app:spectraldip}. The only difference to be mentioned is that the appearance of the non-Fermi-liquid behavior is shifted to somewhat higher temperatures. The reason for this is that the physics at the antinodal point is strongly affected by the van Hove singularity of the $2d$ DOS and, hence, antiferromagnetic fluctuations are strongly effective to scatter electrons with this momentum vector. In Fig.~\ref{fig:spectra_antinodal} (Appendix \ref{app:spectraldip}) we indeed observe a breakdown of the Fermi liquid behavior already at $\beta=30$, while at $\mathbf{k}_N$ we found a Fermi liquid behavior for the self-energy down to $\beta=35$ as discussed before. This momentum differentiated feature is sometimes referred to as ``pseudogap'' behavior. It is intensely discussed in the literature, because it can be indeed experimentally observed, e.g., in the high-temperature superconducting cuprates\cite{Timusk1999,Huscroft2001,Gull2010,Scalapino2012,Gull2013,Gunnarsson2015}.

For the {\sl three-dimensional} half-filled Hubbard model the self-energies and spectral functions exhibit at weak-to-intermediate coupling always Fermi liquid behavior down to $T_N$. The corresponding spectral functions have already been reported for $U=1.0$ in Ref.~\onlinecite{Rohringer2011} and always show a peak at $\omega\!=\!0$ (see insets in Fig.~3 in this reference). In the same paper it has been suggested that a gap and, hence, non-Fermi-liquid spectral features, will always appear exponentially close to the phase transition. This statement is supported by the analytical (paramagnon-like) calculations presented, e.g., in Ref. \onlinecite{Vilk1997}. According to these results, the quasiparticle scattering rate $\gamma_{\mathbf{k}}$ should diverge at the phase transition. This effect is however {\sl not} visible in our numerical D$\Gamma$A results. This discrepancy motivated us to perform a detailed study of the temperature dependence of the scattering rate $\gamma_{\mathbf{k}}$ (and the low energy coefficients of $\Sigma$ in general), which will be presented in the following section.     

\subsection{Low energy coefficients}
\label{subsec:flparam}

\subsubsection*{Numerical results}

\begin{figure}[t!]
 \centering
  \includegraphics{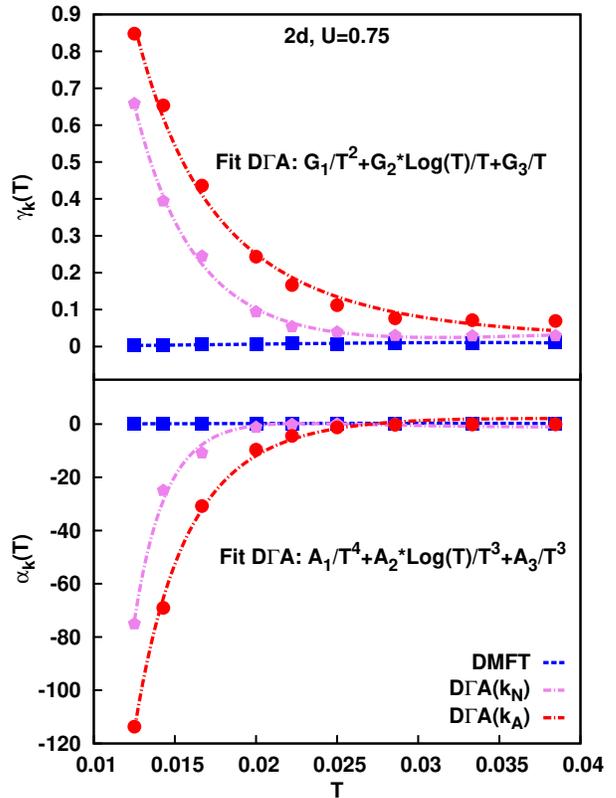}
 \caption{Temperature dependence of low-energy coefficients of the self-energy for $U=0.75$ in 2d. {\sl Upper panel:} Scattering factors $\gamma_{\mathbf{k}}(T)$ [for the definition see Eq.~(\ref{equ:defgammafl})] for D$\Gamma$A at the nodal [$\mathbf{k}_N=(\pi/2,\pi/2)$] and the antinodal [$\mathbf{k}_A=(\pi,0)$] point compared to the corresponding DMFT data; {\sl lower panel:} (Negative) slope of the real part of the self-energy at $\omega=0$, i.e., $\alpha_{\mathbf{k}}(T)$ [Eq. (\ref{equ:defalpha})] for D$\Gamma$A at $\mathbf{k}_N=(\pi/2,\pi/2)$ and $\mathbf{k}_A=(\pi,0)$ compared to the corresponding DMFT data. The continuous lines represent the fits of the data points with the analytical function given in the plot (for the analytical derivations see the main text).} 
 \label{fig:gamma_2d_U075}
\end{figure}
In this subsection, we present our numerical D$\Gamma$A results for the coefficients defining the low-frequency behavior of the self-energy ($\gamma_{\mathbf{k}}$ and $\alpha_{\mathbf{k}}$), defined in Eqs. (\ref{equ:defgammafl}) and (\ref{equ:defalpha}), as a function of $T$. The values for $\gamma_{\mathbf{k}}$ and $\alpha_{\mathbf{k}}$ have been extracted from the Pad\'e fits of our Matsubara data, where the stability of these fits (and, hence, of the final results) has been thoroughly checked by varying the set of Matsubara frequencies used for the fitting procedure (for details, see Appendix \ref{app:pade}). 

Let us start discussing the scattering factors $\gamma_{\mathbf{k}}(T)$ in $2d$, which are shown in the upper panel of Fig.~\ref{fig:gamma_2d_U075}. In a perfect Fermi liquid system, these scattering rates should decrease with decreasing temperature and eventually vanish at $T\!=\! 0$ as $\gamma_{\text{FL}}(T)\!~\!\sim T^2$. Within DMFT our model is a Fermi liquid at $U=0.75$ and, though hardly visible in the scale of the plot, we have checked that $\gamma(T)$ (blue points) indeed follows a $T^2$ behavior (at higher temperatures corrected by the next term in the Sommerfeld expansion, i.e., $T^4$). On the contrary, in D$\Gamma$A we observe a (at lower temperatures very strong) {\sl increase} of the scattering rate with decreasing temperature, which clearly signals the non-Fermi-liquid nature of the physics. These enhanced scattering rates at low temperatures can be attributed to an enhanced scattering of the electrons at nonlocal spin fluctuations, whose spatial extension grows exponentially with decreasing temperature. Remarkably, the non-Fermi-liquid behavior develops already at temperatures where the self-energies and spectral functions still exhibit a definite Fermi-liquid-like structure. In particular, for the nodal point we recognize an increase of $\gamma_{\mathbf{k}_N}(T)$ with decreasing temperature already at $\beta=35$ ($T=0.029$) while the corresponding self-energies and spectral functions in the first row of Fig.~\ref{fig:spectra_nodal} are clearly compatible with a Fermi liquid description. An analogous behavior can be observed for the antinodal point where, at the temperatures shown in the figure, $\gamma_{\mathbf{k}_A}(T)$ always increases with decreasing temperature. 

\begin{figure}
	\centering
		\includegraphics{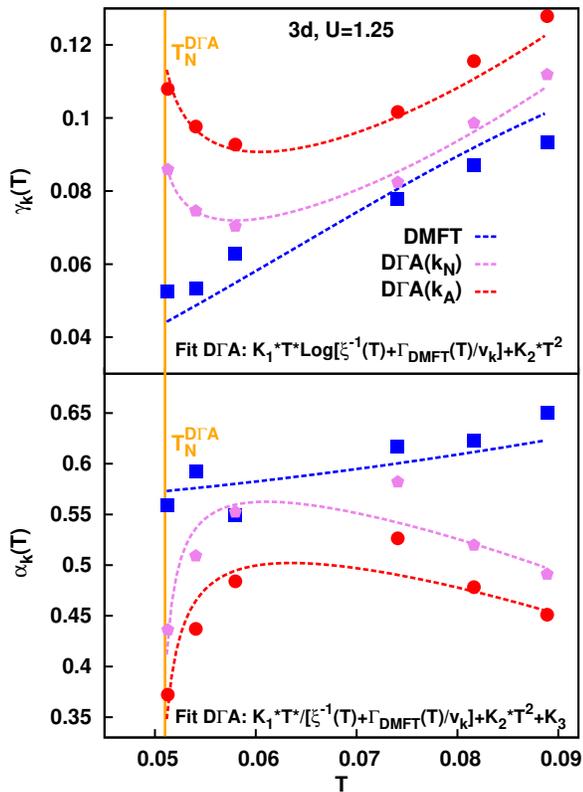}
	\caption{Same as in Fig. \ref{fig:gamma_2d_U075} but in $3d$ for $U=1.25$ with $\mathbf{k}_N=(\pi/2,\pi/2,\pi/2)$ and $\mathbf{k}_A=(\pi,\pi/2,0)$. $T_N$ denotes the finite transition temperature to the antiferromagnetic phase obtained in D$\Gamma$A (see Ref. \onlinecite{Rohringer2011}).}
	\label{fig:gamma_3d_U125}
\end{figure}

Let us now consider the case of $3d$ (upper panel in Fig. \ref{fig:gamma_3d_U125}). In DMFT the situation is rather conventional: $\gamma(T)$ is reduced with decreasing temperature as it is anticipated for a Fermi liquid. On the contrary, the D$\Gamma$A data clearly show a nonmonotonous behavior: While at high $T$ one observes a decrease of $\gamma_{\mathbf{k}}(T)$ with temperature, when approaching the phase transition we find an {\sl increase} of $\gamma_{\mathbf{k}}(T)$: The two regimes are connected by a well-defined {\sl minimum} of $\gamma_{\mathbf{k}}(T)$. As for $2d$, the enhanced scattering rate observed upon decreasing $T$ can be ascribed to an enhanced scattering of the electrons at nonlocal spin fluctuations. This effect is again more pronounced for the antinodal point but since in 3d no van Hove singularity is present in the noninteracting DOS, the difference between $\mathbf{k}$ points at the Fermi surface is less marked. As it is clear from the sign of $\alpha_{\mathbf{k}}(T)$ (lower panel of Fig.~\ref{fig:gamma_3d_U125}) -and already anticipated in the previous section- D$\Gamma$A always exhibits a Fermi-liquid-like self-energy and spectral function down to the transition temperature\cite{Rohringer2011}. Furthermore, our data suggest that the scattering rate will stay {\sl finite} even at the phase transition, in contrast to the arguments of Refs. \onlinecite{Rohringer2011,Vilk1997}. The temperature dependence of the very same observables, however, indicates a breakdown of Fermi liquid behavior at significantly higher temperatures above $T_N$. Some consequences of these observations will be discussed in more detail in Sec. \ref{subsec:contin}. 

\subsubsection*{Analytical results and interpretation}

In the following, we aim at an analytical understanding of the low-$T$ dependence of the D$\Gamma$A coefficients $\gamma_{\mathbf{k}}$ and $\alpha_{\mathbf{k}}$, whose numerical results have been discussed in the previous section. To this end, we single out the dominant contributions to the D$\Gamma$A self-energy $\Sigma_{\text{lad}}(i\nu,\mathbf{k})$ of Eq. (\ref{equ:EOMrewrite}) in the limit of a large (AF) correlation length ($\xi\rightarrow\infty$). This leads to the following simplifications, allowing us to derive (semi)analytical results for the D$\Gamma$A (cf. Appendix \ref{app:dgaanalyt} for all details):  (i) We restrict Eq.~(\ref{equ:EOMrewrite}) to the dominant classical spin fluctuations (represented by $\chi_{m,\mathbf{q}}^{\Omega=0}$) around the AF wave vector $\mathbf{q}=(\pi,\pi,(\pi))$ in $2(3)d$. $\chi_{m,\mathbf{q}}^{\Omega=0}$ can then be parameterized through the corresponding AF correlation length $\xi(T)$. (ii) We take into account the DMFT self-energy only close to the Fermi level where it can be described solely in terms of the low-frequency coefficients, and essentially by means of the DMFT quasiparticle scattering rate $\Gamma(T)\!=\!\gamma(T)\!/\![1\!+\!\alpha(T)]$. Considering, moreover, that the frequency and temperature dependence of the three-leg vertex is negligible (see Appendix \ref{app:dgaanalyt}) the above simplifications lead to the following approximate expressions for the D$\Gamma$A self-energy Eq. (\ref{equ:EOMrewrite}) in $2d$ and $3d$, whose derivation is provided in full detail in Appendix \ref{app:dgaanalyt}:
\begin{subequations}
\label{subequ:EOMapprox}
\begin{align}
 \Sigma_{\text{lad}}(\omega,\mathbf{k})&\cong C_{\mathbf{k}}^{\omega}T\int d^dq \frac{1}{\mathbf{q}^2+\xi^{-2}(T)}\nonumber\\&\times\frac{1}{\omega+v_{\mathbf{k}}q_x+i\Gamma(T)},\quad \text{if }v_{\mathbf{k}}\equiv\lvert\mathbf{v}_{\mathbf{k}}\rvert\neq 0 \label{equ:EOMapproxnodal}\\
\Sigma_{\text{lad}}(\omega,\mathbf{k})&\cong C_{\mathbf{k}}^{\omega}T\int d^dq \frac{1}{\mathbf{q}^2+\xi^{-2}(T)}\nonumber\\ \times&\frac{1}{\omega+(q_x^2-q_y^2)/m_{\mathbf{k}}+i\Gamma(T)},\quad \text{if }v_{\mathbf{k}}= 0, \label{equ:EOMapproxantinodal}
\end{align}
\end{subequations}
where $\mathbf{v}_{\mathbf{k}}$ denotes the Fermi velocity and $m_{\mathbf{k}}$ the effective mass of the noninteracting electrons renormalized by the quasiparticle weight of DMFT (for the explicit definitions see Appendix~\ref{app:dgaanalyt}). The frequency dependence of the prefactor $C_{\mathbf{k}}^{\omega}$ enters through the frequency dependence of the three-leg vertex [defined in Eq.~(\ref{equ:defgamma})]. In Appendix \ref{app:qintegrals} we give the definition of this prefactor explicitly and discuss the irrelevance of its frequency dependence for the temperature behavior of $\gamma_{\mathbf{k}}$ and $\alpha_{\mathbf{k}}$ in the critical regime of large $\xi(T)$.

We should stress that Eqs.~(\ref{subequ:EOMapprox}) are equivalent to the corresponding relations in the TPSC of Tremblay {\sl et al.} [Ref.~\onlinecite{Vilk1997}, see, e.g., Eq.~(55) therein] with the only significant difference being that the Green's function is not the bare one, but it contains the quasiparticle scattering rate of DMFT. As we will see in the following, this qualitatively alters the expressions for the temperature dependence of the $\gamma_{\mathbf{k}}(T)$ and $\alpha_{\mathbf{k}}(T)$ w.r.t. the results of the TPSC.

In order to determine the explicit temperature dependence of $\gamma_{\mathbf{k}}(T)$ and $\alpha_{\mathbf{k}}(T)$ from Eqs.~(\ref{subequ:EOMapprox}) we must also specify the temperature dependence of the correlation length -as computed by D$\Gamma$A- in the critical regime, which is of course different in two and three dimensions. In $3d$ we are dealing with a second order phase transition at finite temperatures and, thus, $\xi(T)$ diverges for $T\rightarrow T_N$ according to a power law\cite{Rohringer2011,Hirschmeier2015}, while in $2d$ the correlation length exhibits an exponential increase\cite{Vilk1997,Schafer2015} when approaching the phase transition at $T\!=\!0$. 

Hence, we can parametrize the correlations lengths in two and three dimensions in the following way:
\begin{subequations}
\label{subequ:xi}
\begin{align}
&\xi_{2d}(T)=c_1e^{c_2/T}\label{equ:xi2d}\\
&\xi_{3d}(T)=c_3(T-T_N)^{-\nu},\label{equ:xi3d}
\end{align}
\end{subequations}
where $c_1$, $c_2$ and $c_3$ are positive real constants, $T_N$ is the $3d$ transition temperature and $\nu$ the critical exponent. For the $3d$ system we will adopt in the following the values for $c_3$, $T_N$ and $\nu$ given in Ref.~\onlinecite{Rohringer2011} while for $2d$ we will leave $c_1$ and $c_2$ as free fit-parameters appearing in the final equations for the coefficients $\gamma_{\mathbf{k}}(T)$ and $\alpha_{\mathbf{k}}(T)$. 

The integrals in Eqs.~(\ref{subequ:EOMapprox}) can be now performed analytically. The detailed calculations are similar to those in the TPSC\cite{Vilk1997} and will be given in Appendix \ref{app:qintegrals}. Here, we will just discuss the final results for the parameters $\gamma_{\mathbf{k}}$ and $\alpha_{\mathbf{k}}$ as defined in Eqs. (\ref{subequ:defflparam}) in two and three dimensions for the two relevant $\mathbf{k}$ points $\mathbf{k}_N$ and $\mathbf{k}_A$ on the Fermi surface. In this way, we aim at improving our understanding how long-range antiferromagnetic fluctuations, parametrized by the correlation length $\xi(T)$, affect the one-particle spectral properties of the system.

\noindent
{\sl Two dimensions, nodal point $\mathbf{k}=\mathbf{k}_N$:} At the nodal point the Fermi velocity $v_{\mathbf{k}_N}$ is finite and, thus, we have to solve the integral in Eq.~(\ref{equ:EOMapproxnodal}) in $2d$ and extract from it the coefficients $\gamma_{\mathbf{k}_N}(T)$ and $\alpha_{\mathbf{k}_N}(T)$ according to Eqs. (\ref{subequ:defflparam}). The results read:
\begin{widetext}
\begin{subequations}
\label{subequ:fl2dnodal}
\begin{align}
 \gamma_{\mathbf{k}_N}&(T)=C_{\mathbf{k}_N}T\xi(T)\frac{2}{\sqrt{b_{\mathbf{k}_N}^2(T)-1}}\log\left(b_{\mathbf{k}_N}(T)+\sqrt{b_{\mathbf{k}_N}^2(T)-1}\right)+\mathcal{R}_{\gamma}(T), \label{equ:gamma2dnodal} \\
 \alpha_{\mathbf{k}_N}&(T)=-A_{\mathbf{k}_N}T\xi^2(T)\left[\frac{b_{\mathbf{k}_N}(T)\log\left(b_{\mathbf{k}_N}(T)+\sqrt{b_{\mathbf{k}_N}^2(T)-1}\right)}{\left[b_{\mathbf{k}_N}^2(T)-1\right]^{3/2}}-\frac{1}{b_{\mathbf{k}_N}^2(T)-1}\right]+\mathcal{R}_{\alpha}(T),
\end{align}
\end{subequations}
\end{widetext}
where $C_{\mathbf{k}_N}$ and $A_{\mathbf{k}_N}$ are temperature independent constants determined from $C_{\mathbf{k}}^{\omega}$ (see Appendix~\ref{app:qintegrals}) and $\mathcal{R}_{\gamma}(T)$ and $\mathcal{R}_{\alpha}(T)$ denote the regular terms which do not diverge (or even go to $0$) when $T\rightarrow 0$. The quantity $b_{\mathbf{k}_N}(T)$ is given by
\begin{equation}
 \label{equ:defak}
 b_{\mathbf{k}_N}(T)=\frac{\Gamma(T)\xi(T)}{v_{\mathbf{k}_N}}.
\end{equation}
Let us now analyze Eqs.~(\ref{subequ:fl2dnodal}) for {\sl two important limiting cases} of the parameter $b_{\mathbf{k}_N}(T)$: First, if $\Gamma(T)\equiv 0$ and, hence, $b_{\mathbf{k}_N}(T)\equiv 0$, we obtain $\gamma_{\mathbf{k}_N}(T)\sim\xi(T)$ and $\alpha_{\mathbf{k}_N}(T)\sim-\xi^2(T)$, i.e., an exponential growth of the coefficients  with decreasing temperature, since the correlation length $\xi(T)$ grows exponentially when lowering $T$  [see Eq.~(\ref{equ:xi2d})]. This corresponds to the results obtained in TPSC in Ref.~\onlinecite{Vilk1997} [see Eq.~(56) therein]. In this case, the strong antiferromagnetic fluctuations are completely ``transferred'' to the one-particle spectral properties without any ``damping''. This leads to a corresponding exponential decrease of spectral weight at the Fermi level upon lowering the temperature. In our D$\Gamma$A equations, on the other hand, we have to take into account a {\sl finite} DMFT quasiparticle scattering rate $\Gamma(T)$. Although this vanishes as $T^2$ when approaching $T\!=\!0$, the product $\xi(T)\Gamma(T)$ eventually diverges in this limit, because the correlation length grows much faster at low $T$. This means that for D$\Gamma$A, differently from TPSC, we have to analyze the low-temperature limit of $b_{\mathbf{k}_N}(T)\gg 1$. Applying this limit to Eqs.~(\ref{subequ:fl2dnodal}) one observes that the terms proportional to $\xi(T)$ and $\xi^2(T)$ in $\gamma_{\mathbf{k}_N}(T)$ and $\alpha_{\mathbf{k}_N}(T)$, respectively, cancel out. The resulting expressions then read:
\begin{subequations}
\label{subequ:fl2dnodalabig}
\begin{align}
\gamma_{\mathbf{k}_N}(T)&\cong \widetilde{C}_{\mathbf{k}_N}\frac{T\log\left(\frac{2\Gamma(T)\xi(T)}{v_{\mathbf{k}_N}}\right)}{\Gamma(T)}\nonumber\\&=\frac{G_1}{T^2}+G_2\frac{\log(T)}{T}+\frac{G_3}{T} \label{equ:fl2nodalabiggamma}\\
\alpha_{\mathbf{k}_N}(T)&\cong -\widetilde{A}_{\mathbf{k}_N}\frac{T\left[\log\left(\frac{2\Gamma(T)\xi(T)}{v_{\mathbf{k}_N}}\right)-1\right]}{\Gamma^2(T)}\nonumber\\&=\frac{A_1}{T^4}+A_2\frac{\log(T)}{T^3}+\frac{A_3}{T^3},\label{equ:fl2nodalbigalpha}
\end{align}
\end{subequations}
where the constants $\widetilde{C}_{\mathbf{k}_N}$ and $\widetilde{A}_{\mathbf{k}_N}$ are determined from $C_{\mathbf{k}_N}$ and $A_{\mathbf{k}_N}$, respectively, when taking the limit $b_{\mathbf{k}_N}(T)\gg 1$ in Eqs.~(\ref{subequ:fl2dnodal}). We neglect here the regular parts, as they are not relevant at low temperatures. In the second lines of Eqs.~(\ref{equ:fl2nodalabiggamma}) and (\ref{equ:fl2nodalbigalpha}) the explicit low-temperature dependence of $\gamma_{\mathbf{k}_N}(T)$ and $\alpha_{\mathbf{k}_N}(T)$ is obtained by inserting $\xi(T)$ [Eq.~(\ref{equ:xi2d})] and $\Gamma(T)$ [Eq.~(\ref{equ:tempdepgammadmft})] into the corresponding first lines of these equations. By hands of these analytical expressions, we could fit our D$\Gamma$A data for $\gamma_{\mathbf{k}_N}(T)$ and $\alpha_{\mathbf{k}_N}(T)$ in Fig.~\ref{fig:gamma_2d_U075}: A very good agreement is found between our D$\Gamma$A points and the corresponding fits (dashed lines in the figures). This defines a clear-cut physical interpretation of our numerical results for the $2d$ Hubbard model at weak coupling: The strong antiferromagnetic fluctuations, which are displayed by an exponentially large correlation length at low temperatures, lead to a destruction of the Fermi liquid indicated by a strong enhancement of the corresponding coefficients $\gamma_{\mathbf{k}_N}(T)$ and the (negative) $\alpha_{\mathbf{k}_N}(T)$ upon decreasing temperature. The actual growth rate of these parameters is, however, {\sl qualitatively} reduced compared to the exponential growth rate of the correlation length itself, being cut off by local DMFT correlations via the scattering factor $\Gamma(T)$. Specifically, instead of an exponential behavior, we obtain a power-law divergence of $\gamma_{\mathbf{k}_N}(T)~\sim 1/T^2$ and $\alpha_{\mathbf{k}_N}(T)~\sim 1/T^4$. Let us stress that this change of temperature behavior happens for arbitrarily small values of the local scattering rates $\Gamma(T)$ and, hence, modifies the corresponding scenario derived within the TPSC\cite{Vilk1997} also in the weak-to-intermediate coupling regime.  

\noindent
{\sl Two dimensions, antinodal point $\mathbf{k}=\mathbf{k}_A$:} In the following we perform the same investigation, but for the antinodal point $\mathbf{k}_A=(\pi,0)$. As for this momentum vector the Fermi velocity vanishes, and we must solve the integral (\ref{equ:EOMapproxantinodal}) instead of (\ref{equ:EOMapproxnodal}) for $d\!=\!2$. This is more cumbersome, because in the denominator of the second part of the integrand $q_x^2-q_y^2$ rather than just $q_x$ enters the equation. Hence, in the following we will only present our calculations for $\gamma_{\mathbf{k}_A}(T)$ in a very concise way and refer the reader to Appendix \ref{app:qintegrals} for more details. In particular, we will focus here on highlighting the differences to the results for the nodal point. By calculating the integral in Eq.~(\ref{equ:EOMapproxnodal}) for $\gamma_{\mathbf{k}_A}(T)$, one obtains the following expression (we neglect again any regular parts):
\begin{align}
 \label{equ:fl2antinodalgamma}
 \gamma_{\mathbf{k}_A}(T)\cong &C_{\mathbf{k}_A}T\xi^2(T)\left[\frac{\log\left[b_{\mathbf{k}_A}(T)\right]}{\sqrt{b_{\mathbf{k}_A}^2(T)+1}}\right.\nonumber\\&\left.-\frac{b_{\mathbf{k}_A}(T)}{\pi}\int_0^{\pi}d\phi\frac{\log\lvert\cos\varphi\rvert}{b_{\mathbf{k}_A}^2(T)+\cos^2(\varphi)}\right],
\end{align}
where $b_{\mathbf{k}_A}(T)$ is given as
\begin{equation}
\label{equ:defbk}
 b_{\mathbf{k}_A}(T)=\Gamma(T)\xi^2(T)m_{\mathbf{k}_A}.
\end{equation}
Note that in contrast to the nodal point in Eq.~(\ref{equ:defak}) $\xi(T)$ appears squared in the corresponding definition for $b_{\mathbf{k}_A}$. Let us again analyze the two different limiting cases for $b_{\mathbf{k}_A}(T)$: First, if $\Gamma(T)\equiv 0$ and, hence, $b_{\mathbf{k}_A}(T)=0$, we obtain an exponential growth of $\gamma_{\mathbf{k}_A}(T)$ proportional to $\xi^2(T)$ except for a diverging logarithm $\log(0)$ [which will in general be cut off by $\Gamma(T)$]. This coincides exactly with the results found in TPSC\cite{Vilk1997}. Hence, when neglecting the DMFT scattering factor, $\gamma_{\mathbf{k}_A}(T)\sim \xi^2(T)$ diverges much faster than $\gamma_{\mathbf{k}_N}(T)\sim\xi(T)$ with decreasing temperature. On the contrary, for $\Gamma(T)>0$ which implies $b_{\mathbf{k}_A}(T)\gg 1$ at low temperatures, we observe a cancellation of the contribution proportional $\xi^2(T)$ in Eq.~(\ref{equ:fl2antinodalgamma}). The explicit evaluation of Eq. (\ref{equ:fl2antinodalgamma}) for $b_{\mathbf{k}_A}\gg 1$ indeed shows that the result for temperature dependence of $\gamma_{\mathbf{k}_A}(T)$ is completely equivalent to the corresponding expression for the nodal point in the second line of Eq.~(\ref{equ:fl2nodalabiggamma}), except for a different prefactor. Thus, the very same fitting expression can be exploited for $\gamma_{\mathbf{k}_N}(T)$ and $\gamma_{\mathbf{k}_A}(T)$. The upper panel in Fig.~\ref{fig:gamma_2d_U075} shows that this fitting function describes well our numerical data for $\gamma_{\mathbf{k}_A}(T)$. A corresponding analysis for the much more complicated $\alpha_{\mathbf{k}_A}(T)$ leads to the analogous results, that its $T$-dependence has the same form as for the nodal point (for more details see Appendix \ref{app:qintegrals}). Hence, we draw the following conclusions: While for $\Gamma(T)=0$ there is a qualitative difference in the temperature behavior of the low energy coefficients $\gamma_{\mathbf{k}}(T)$ and $\alpha_{\mathbf{k}}(T)$ for different $\mathbf{k}$ points\cite{Vilk1997} we find that the introduction of a finite $\Gamma(T)$ significantly reduces this difference. This finding, which may be of interest for future studies of the separability\cite{Schafer2015a} of the temporal and spatial correlations in low dimensions, is consistent with the fact that, to some extent, local correlations reduce the nonlocal features of the physics as it also happens, e.g., in the Mott insulating phase observed at large values of the coupling $U$ in the Hubbard model.

\noindent
{\sl Three dimensions:} Let us finally discuss the results for the coefficients $\gamma_{\mathbf{k}}(T)$ and $\alpha_{\mathbf{k}}(T)$ in terms of the analytical expressions obtained from Eqs.~(\ref{subequ:EOMapprox}) for the $3d$ Hubbard model. We first note that, in this case, $v_{\mathbf{k}}\ne 0$ for all $\mathbf{k}$-point on the Fermi surface. Hence, we can stick to Eq.~(\ref{equ:EOMapproxnodal}) for both of the two chosen points at the Fermi surface ($\mathbf{k}_N$ and $\mathbf{k}_A$, for the definition see the caption of Fig.~\ref{fig:gamma_3d_U125}). In fact, the only difference between them is expressed in a slightly changed Fermi velocity, i.e., $v_{\mathbf{k}_A}=v_{\mathbf{k}_N}/\sqrt{3}$. The explicit evaluation of the integral in Eq.~(\ref{equ:EOMapproxnodal}) for $3d$ yields (see Appendix \ref{app:qintegrals})
\begin{subequations}
\label{subequ:fl3d}
\begin{align}
 \gamma_{\mathbf{k}}&(T)=C_{\mathbf{k}}T\log\left[\frac{b_{\mathbf{k}}(T)+\sqrt{1+\left[\pi\xi(T)\right]^2}}{b_{\mathbf{k}}(T)+1}\right]\nonumber\\&\hspace{5.4cm}+\mathcal{R}_{\gamma}(T), \label{equ:gamma3d} \\
 \alpha_{\mathbf{k}}&(T)=A_{\mathbf{k}}T\xi(T)\left[\frac{1}{b_{\mathbf{k}}(T)+\sqrt{1+\left[\pi\xi(T)\right]^2}}\right.\nonumber\\ &\hspace{3.2cm}\left.\vphantom{\frac{1}{b_{\mathbf{k}}(T)+\sqrt{1+\left[\pi\xi(T)\right]^2}}}-\frac{1}{b_{\mathbf{k}}(T)+1}\right]+\mathcal{R}_{\alpha}(T), \label{equ:alpha3d}
\end{align}
\end{subequations}
where $b_{\mathbf{k}}(T)$ is defined analogously to the $2d$ case in Eq.~(\ref{equ:defak}), i.e., $b_{\mathbf{k}}(T)=\frac{\Gamma(T)\xi(T)}{v_{\mathbf{k}}}$, and $\mathcal{R}_{\gamma}(T)$ and $\mathcal{R}_{\alpha}(T)$ denote the regular parts of $\gamma_{\mathbf{k}}(T)$ and $\alpha_{\mathbf{k}}(T)$, respectively.

As in $2d$, we will now analyze Eqs.~(\ref{subequ:fl3d}) for two different limiting cases: First, if we neglect the DMFT scattering factor $\Gamma(T)$, we must evaluate these equations for $b_{\mathbf{k}}(T)\equiv 0$. Analogous to the situation in $2d$, we observe also in $3d$ that the critical fluctuations are directly ``transferred'' from the two-particle susceptibility to the one-particle spectral properties. Indeed, the corresponding Fermi liquid parameters grow strongly when lowering the temperature as $\gamma_{\mathbf{k}}(T)\sim\log[\xi(T)]$ and $\alpha_{\mathbf{k}}(T)\sim\xi(T)$, eventually {\sl diverging} at the transition point. Such a behavior has been also found in the TPSC\cite{Vilk1997}. On the contrary, in our D$\Gamma$A calculations, we have a finite value for $\Gamma(T)$. Hence, approaching the phase transition at $T\!=\!T_N$ the quantity $\xi(T)\Gamma(T)~\sim\xi(T)T_N^2$ -and accordingly $b_{\mathbf{k}}(T)$- gets strongly enhanced and eventually diverges at $T\!=\!T_N$. In fact, considering the limit $b_{\mathbf{k}}(T)\gg 1$ in Eqs. (\ref{subequ:fl3d}) we find for the leading contributions to $\gamma_{\mathbf{k}}(T)$ and $\alpha_{\mathbf{k}}(T)$:
\begin{subequations}
\label{subequ:fl3dabig}
\begin{align}
\gamma_{\mathbf{k}}(T)&\cong \widetilde{C}_{\mathbf{k}}T\left[\log(\pi)-\log\left(\frac{\Gamma(T)}{v_{\mathbf{k}}}+\xi(T)^{-1}\right)\right] \label{equ:fl3dabiggamma}\\
\alpha_{\mathbf{k}}(T)&\cong \widetilde{A}_{\mathbf{k}}T\left[\frac{1}{\pi}-\frac{1}{\frac{\Gamma(T)}{v_{\mathbf{k}}}+\xi(T)^{-1}}\right],\label{equ:fl2dabigalpha}
\end{align}
\end{subequations}
where we neglected the corresponding regular contributions. Note that in order to derive Eqs.~(\ref{subequ:fl3dabig}) we have taken into account\footnote{This can be easily understood by the fact that in a weakly correlated situation (as described by DMFT) the electron propagates during its actual lifetime $\tau\sim 1/\Gamma(T)$ over length scales much larger then the lattice constant (given by $1$), i.e., $\tau v_{\mathbf{k}}\gg 1$.} that (within DMFT) $\Gamma(T)\ll v_{\mathbf{k}}$ in Eqs. (\ref{subequ:fl3d}).

We have then fitted our numerical data points for $\gamma_{\mathbf{k}}(T)$ and $\alpha_{\mathbf{k}}(T)$ in Fig.~\ref{fig:gamma_3d_U125} to the functions given in Eqs.~(\ref{subequ:fl3dabig}) (neglecting the term containing just the momentum cutoff $\pi$) and added regular parts to the corresponding expressions, which have to be taken into account for the moderately high temperatures above $T_N$. Specifically, for $\gamma_{\mathbf{k}}(T)$ we consider a standard Fermi liquid contribution $\mathcal{R}_{\gamma}(T)\sim K_2T^2$ which is certainly originated from the regular (DMFT-like) contributions in Eq.~(\ref{equ:EOMrewrite}). On the other hand, for $\alpha_{\mathbf{k}}$ the corresponding regular contribution is not known in general. However, since $\alpha_{\mathbf{k}}(T)$ does not vanish at $T=0$ [in contrast to $\gamma_{\mathbf{k}}(T)$] we have simply assumed a Sommerfeld-like functional form $\mathcal{R}_{\alpha}(T)\sim K_2T^2+K_3$. The resulting fits are shown by dashed lines in Fig.~\ref{fig:gamma_3d_U125} where we have used the corresponding fit functions for $\Gamma(T)$ and $\xi(T)$ given in Ref.~\onlinecite{Rohringer2011}. Hence, also in the $3d$ case, we find a satisfactory agreement between the numerical data and the analytic estimate for $\alpha_{\mathbf{k}}(T)$ and $\gamma_{\mathbf{k}}(T)$, allowing for a transparent interpretation of our D$\Gamma$A results: At high $T$, the temperature behavior of $\gamma_{\mathbf{k}}(T)$ is determined by its regular Fermi liquid part, as in DMFT, because the argument of the logarithm in Eq.~(\ref{equ:fl3dabiggamma}) is slightly smaller than $1$, i.e., $\Gamma(T)/v_{\mathbf{k}}+\xi^{-1}(T)\leq 1$. Upon decreasing temperature the regular part of $\gamma_{\mathbf{k}}(T)$ decreases as $T^2$. Since $\xi^{-1}(T)\rightarrow 0$ for $T\rightarrow T_N$ also the argument of the logarithm decreases and, hence, becomes much lower than $1$. Consequently, the logarithmic contribution $-\log\left[\Gamma(T)/v_{\mathbf{k}}+\xi^{-1}(T)\right]$ itself {\sl increases} upon lowering the temperature. This competition between the regular part of $\gamma_{\mathbf{k}}(T)$ (decreasing when lowering $T$), and the (increasing) logarithmic part of Eq.~(\ref{equ:fl3dabiggamma}) leads to the emergence of a minimum of the scattering rate at a given temperature $T^*$ close to $T_N$. This minimum is indeed well visible for both $\mathbf{k}$ points in our numerical data in Fig.~\ref{fig:gamma_3d_U125}. As such a nonmonotonous behavior of $\gamma_{\mathbf{k}}(T)$ is clearly {\sl not} compatible with a Fermi liquid description of the system we argue that antiferromagnetic fluctuations in $3d$ are strong enough to destroy the Fermi liquid properties when approaching the phase transition, even if the (metallic) quasiparticle peak remains visible in the spectra. As in $2d$, the effect of these nonlocal fluctuations is, however, damped by the local scattering rate $\Gamma(T)$ of DMFT. While in $2d$ such a reduced effect of fluctuations leads nevertheless to a non-Fermi-liquid shape of the self-energies and the spectral function themselves, we observe for $3d$ a Fermi liquid behavior for the frequency dependence of the one-particle correlation functions for all temperatures $T>T_N$. This is consistent with the fact that $\alpha_{\mathbf{k}}$ is always positive in $3d$, where the contribution in Eq.~(\ref{equ:fl2dabigalpha}) represents just a correction of the regular part leading to a nonmonotonous behavior also for $\alpha_{\mathbf{k}}(T)$.   

From a physical perspective one could summarize our findings as follows: In $3d$ the effect of the nonlocal fluctuations caused by the (diverging) correlation length $\xi(T)$ on spectral properties is cut off by local correlations leading to Fermi-liquid-like self-energies and spectra for all $T>T_N$. The competition between high-temperature thermal fluctuations (which decrease with decreasing temperature) and low-temperature antiferromagnetic fluctuations (which increase with decreasing temperature), on the other hand, leads to nonmonotonous behavior of the temperature dependence of $\gamma_{\mathbf{k}}(T)$ and $\alpha_{\mathbf{k}}(T)$ clearly indicating a break-down of the Fermi liquid at $T=T^*>T_N$ (cf. also Ref. \onlinecite{Fuchs2011a}). 

Let us finally remark, that our findings of different effects of correlations in different observables (such as here, the $T$-dependence of the scattering rate $\gamma$ and the correlation length $\xi$) are of importance also {\sl beyond} the specific calculations of this work. In fact, somewhat related differentiation effects have been observed in the most recent transport experiments in cuprates\cite{Barisic2015}, where a universal (Fermi-liquid-like) scattering rate is found at all (hole) dopings in spite of quite diverse $T$ behaviors of different response functions.

\subsection{Continuity of spectral properties at the phase transition}
\label{subsec:contin}

The above discussions of the D$\Gamma$A results raise the interesting question whether the one-particle spectral properties, i.e., self-energies and spectral functions, are continuous when crossing the (here: AF) phase transition.

 In the somehow ``extreme'' case of {\sl two dimensions} the magnetic phase transition occurs exactly at $T=0$. At this temperature the system exhibits, hence, a finite order parameter (i.e., a finite staggered magnetization) which gives rise to a (full) finite gap in the spectral function. For $T>0$, in D$\Gamma$A we have found a  divergent $1/T^2$ behavior for the electronic scattering rate $\gamma_{\mathbf{k}}(T)$. This leads to an opening of a perfect gap (i.e., with zero spectral weight) for $T\rightarrow 0$ indicating that the spectral functions of the system are indeed {\sl continuous} at the phase transition. One can, thus, state that the antiferromagnetic fluctuations in the paramagnetic phase above $T_N=0$ prevent a discontinuity in the {\sl normal} part of the one-particle Green's functions at the transition. This obviously does not hold for the {\sl anomalous} part of the one-particle Green's function whose $(t,\mathbf{r})\!=\!0$ contribution corresponds to the (AF) order parameter. The latter, because of the Mermin-Wagner theorem, jumps abruptly from $0$ in the paramagnetic phase for $T>0$ to a finite value at $T=0$. 

The situation is somewhat different in {\sl three dimensions}, where we are dealing with a second-order phase transition at finite $T$. Here, one-particle quantities such as the order parameter itself are --per definition--- continuous at the transition temperature, as they are obtained from first derivatives of the corresponding thermodynamic potential. Consequently, we would reasonably expect that also the one-particle spectral properties exhibit no discontinuity upon crossing $T_N$. 
We recall, however, that a static mean-field treatment of the ordered phase (which corresponds to the infinite dimensional limit for a classical model) leads to an abrupt opening of the (full) gap right at the transition temperature, with a point-like jump of the spectral function at zero frequency (i.e., exactly at the Fermi level).
In fact, such a jump well matches the result of a logarithmically divergent scattering rate $\gamma_{\mathbf{k}}(T\rightarrow T_N)$ as obtained, e.g., in the TPSC\cite{Vilk1997}. This agreement is consistent with the fact that both approaches adopt a {\sl static} two-particle irreducible vertex and a noninteracting Green's function for calculating the magnetic susceptibilities. On the contrary, the D$\Gamma$A is based on the {\sl dynamical}, i.e., frequency-dependent, irreducible self-energy and vertex functions of DMFT. In particular, the former difference w.r.t.~the static mean-field-like approaches modifies the situation qualitatively, as it was demonstrated in the previous subsections: The divergences of $\gamma_{\mathbf{k}}(T)$ [and of $\alpha_{\mathbf{k}}(T)$] at $T=T_N$ are cut off by the finite scattering rate of DMFT, leading to a {\sl finite} weight for the spectral function at the Fermi level across the magnetic transition. This physically more plausible behavior is actually also consistent with DMFT calculations in the symmetry-broken phase\cite{Peters2015} where a finite spectral weight is found at $T=T_N$, while a full gap is predicted only for $T=0$. These findings suggest, partly different from the situation in $2d$, a continuous behavior of {\sl both} the normal and the anomalous part of the Green's function across the phase transition in $3d$. As the $3d$ physics should lie in between the $2d$ and the $d\!=\!\infty$ one of DMFT, the observed continuity of the one-particle spectral properties at $T=T_N$, as suggested by D$\Gamma$A, looks to be quite a convincing result. These findings highlight the importance of including  dynamical local correlations for an accurate description of second order phase transitions, also at finite temperature.

\vskip 5mm
\section{D$\Gamma$A at all energy scales}
\label{sec:allenergies}

In this section we extend the analysis of effects originating from nonlocal fluctuations in the two- and three-dimensional Hubbard model to all energy scales. Specifically, we will consider thermodynamic observables, such as the kinetic ($E_{\text{kin}}$) and the potential ($E_{\text{pot}}$) energies, whose values incorporate contributions from all energy scales in the system. Let us remark that the calculation of these energies within a given approximation scheme might become -to a certain extent- ambiguous, as $E_{\text{kin}}$ and $E_{\text{pot}}$ can be expressed either in terms of one- or two-particle quantities. While in exact as well as in approximate, but two-particle self-consistent, theories both results coincide, this is not the case for the approximate schemes applied in this paper (DMFT and ladder D$\Gamma$A for {\sl finite} $d$) due to a lack of self-consistency at the two-particle level. Hence, in principle, one is left with the problem of selecting which expression for the kinetic and potential energy, respectively, is more reliable to capture the physics of the system. 

Let us here discuss how such ambiguities occur already in DMFT when this is applied to finite $d$ systems. To this end, we first recall that for {\sl $d\!=\!\infty$} (where DMFT corresponds to the exact solution of the system) the DMFT self-consistency condition between the momentum-summed lattice Green's function and the impurity one [$\sum_{\mathbf{k}} G(i\nu,\mathbf{k})\!=\!G(i\nu)$] implies automatically the fulfillment of analogous relations at the two-particle level, e.g., for the susceptibilities $\chi_{r,\mathbf{q}}^{\Omega}$, i.e., $\sum_{\mathbf{q}}\chi_{r,\mathbf{q}}^{\Omega}\!=\!\chi_r^{\Omega}$ (for a detailed discussion see Ref.~\onlinecite{Georges1996}, Sec.~IV A). In this situation, provided that the numerical solution of the auxiliary AIM is accurate, no ambiguity can be encountered as it is expected for an exact theory. On the contrary, in {\sl finite} spatial dimensions, where DMFT represents an approximation, the (one-particle) self-consistency condition of DMFT does not guarantee any longer that the corresponding relation at the two-particle level holds. In fact, the $\mathbf{k}$-summed lattice susceptibilities do in general {\sl not} coincide anymore with the corresponding local AIM susceptibilities. Such inconsistency between the one- and the two-particle level leads to the occurrence of ambiguities in the calculations of $E_{\text{kin}}$ and $E_{\text{pot}}$ already at the DMFT level.

This important issue for the calculation of thermodynamic observables within an approximate (not two-particle self-consistent) theory, as well as its implications, will be discussed in detail for both $E_{\text{kin}}$ and $E_{\text{pot}}$ at the beginning of the corresponding subsections.

\subsection{Kinetic Energy}
\label{sec:ekin}

\begin{figure*}
	\centering
		\includegraphics{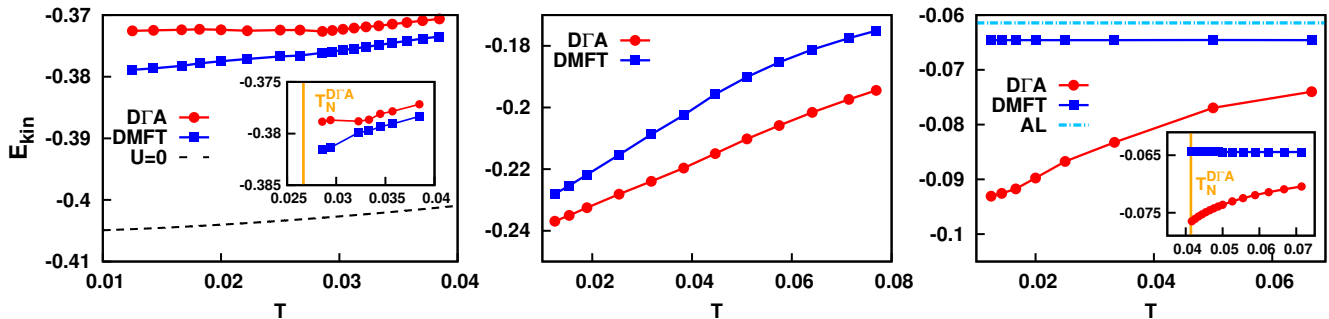}
	\caption{Kinetic energies for the noninteracting case ($U=0$), DMFT, D$\Gamma$A and the atomic limit (AL) as a function of temperature in two (main panels) and three (insets) dimensions at half filling for different values of the interaction parameter $U=0.75$ (left panel), $U=2.0$ (middle panel) and $U=4.0$ (right panel).}
	\label{fig:Ekin2dAllU}
\end{figure*}

The kinetic energy is given by the expectation value of the noninteracting part of the (here: Hubbard) Hamiltonian, i.e., the first term on the right hand side of Eq.~(\ref{equ:defmodel}). Hence, in terms of the one-particle Green's function, it reads
\begin{equation}
 \label{equ:ekin}
 E_{\text{kin}}=\frac{2}{\beta}\sum_{\nu\mathbf{k}}\varepsilon_{\mathbf{k}}G(i\nu,\mathbf{k}),
\end{equation}
where the factor $2$ in front of this equation is due to the two different spin components of the system. Note that for the D$\Gamma$A case the Green's function obviously contains the corresponding D$\Gamma$A self-energy as given in Eq.~(\ref{equ:EOMrewrite}). At the same time, $E_{\text{kin}}$ can be also obtained through a two-particle expression, exploiting the so-called $f$-sum rule [see Eq.~(A25) in Ref.~\onlinecite{Vilk1997} and Eq. (\ref{equ:fsumrule}) in Sec. \ref{sec:lambdaimproved}]. For an approximate theory it is in general not guaranteed that both expressions of $E_{\text{kin}}$ yield the same result (i.e., that the $f$-sum rule holds), unless they are conserving in the Baym-Kadanoff sense\cite{Baym1961,Baym1962,Hettler2000}.

While a systematic numerical study of these discrepancies is not the focus of this paper, their occurrence can inspire, nonetheless, further improvements of the ladder D$\Gamma$A algorithm, which will be presented in Sec. \ref{sec:lambdaimproved}. Here we argue that, since the kinetic energy is originally defined as the expectation value of a {\sl one-particle} operator, its evaluation in terms of one-particle Green's functions should be the most natural choice. Hence, as for the data presented in this section, our calculations have been performed by means of Eq.~(\ref{equ:ekin}), under careful treatment of the high-frequency tails of the Green's functions in the corresponding frequency sum (see Appendix~\ref{app:energies} for details). We note that this is consistent with the usual choice made in DMFT calculations for both model and realistic systems (e.g., DFT+DMFT\cite{Kotliar2006,Held2007}).

In Fig.~\ref{fig:Ekin2dAllU} we present our kinetic energy results for three (two) different values of $U$ ranging from weak- to strong-coupling in two (three) dimensions. In the leftmost panel of Fig.~\ref{fig:Ekin2dAllU} the temperature dependence of the kinetic energies is shown for the weak-coupling $U=0.75$. Starting with the $2d$ case (main panel) we observe for DMFT an increase of $\lvert E_{\text{kin}}\rvert(T)$ with decreasing temperature. This is indeed compatible with a typical Fermi liquid temperature behavior of the kinetic energy, where the (maximal) $T=0$ electronic mobility is reduced by Sommerfeld-like $T^2$ corrections.  In this light, it is not surprising that the DMFT data (blue squares) for $E_{\text{kin}} (T)$ display the same trends as the noninteracting case, only visibly renormalized by the local correlations. D$\Gamma$A (red circles) leads to a further reduction of (the absolute value of) the kinetic energy w.r.t. DMFT. In this case, however, the corresponding $T-$dependence is also qualitatively different from DMFT: While at higher $T$ the D$\Gamma$A curve exhibits still an increase of $\lvert E_{\text{kin}} \rvert(T)$ upon lowering $T$,  it saturates at $T\!\sim\!0.025$ which corresponds roughly to the temperature where the system enters the $2d$ ``critical'' regime with exponentially long-ranged antiferromagnetic fluctuations (see previous sections and Ref.~\onlinecite{Vilk1997}). Thus, the deviation from the corresponding DMFT results grows upon decreasing temperature. An analogous situation is observed in $3d$ (see inset in the leftmost panel of Fig.~\ref{fig:Ekin2dAllU}) where $\lvert E_{\text{kin}}\rvert(T)$ is also systematically smaller in D$\Gamma$A than in DMFT for all $T$ and their difference grows with decreasing temperature (becoming particularly large when approaching the finite-temperature phase transition of the $3d$ system). Hence, we conclude that, at weak-coupling, antiferromagnetic correlations as described by D$\Gamma$A reduce (the absolute value of) the kinetic energy compared to DMFT. This directly reflects the properties of the antiferromagnetic ground state of the system in this parameter regime, and is also (qualitatively) consistent with the thermodynamical properties of the underlying antiferromagnetic phase described by the DMFT\cite{Gull2008}: In fact, at small $U$, the antiferromagnetic phase of DMFT is stabilized by a reduction of the potential energy in the ordered phase, with a simultaneous {\sl increase} of the kinetic energy w.r.t. the normal paramagnetic phase. This defines the {\sl Slater} mechanism\cite{Toschi2005,Toschi2005b,Taranto2012,Tagliavini2016} for the antiferromagnetism of the single band Hubbard model. Our D$\Gamma$A data demonstrate that the Slater-like energetics becomes already well visible in the corresponding antiferromagnetic fluctuations of the paramagnetic phase, i.e., for temperatures {\sl above} the actual transition temperature, as it was just briefly mentioned in Ref.~\onlinecite{Schafer2015} (cf. also the DCA results of Ref.~\onlinecite{Gull2008}). Obviously, this effect is visible for $2d$ in a much broader temperature regime compared than in $3d$, because in $2d$  the ordered phase is restricted to $T\!=\!0$, but strong antiferromagnetic correlations affect a large temperature range of the phase diagram. In $3d$, instead, the major effects of antiferromagnetic fluctuations are usually confined to a narrower temperature region\cite{Rohringer2011,Vilk1997} above the (finite) $T_N$.

At $U=2.0$ (middle panel of Fig. \ref{fig:Ekin2dAllU})  we observe a very different situation. Here, the magnitude of $E_{\rm kin}$ in DMFT and in D$\Gamma$A is reversed compared to the weak-coupling regime: $\lvert E_{\text{kin}}\rvert(T)$ is {\sl larger} in D$\Gamma$A than in DMFT for all $T$ considered, whereas both D$\Gamma$A and DMFT exhibit a significant enhancement of the (absolute value of the) kinetic energy upon decreasing temperature. Again an improved understanding of this behavior is obtained relating these results to corresponding DMFT data for the antiferromagnetically ordered phase\cite{Toschi2005,Toschi2005b,Gull2008,Taranto2012,Tagliavini2016}: At intermediate values of $U$, $\lvert E_{\text{kin}}\rvert(T)$ gets enhanced by the onset of a symmetry-broken phase in DMFT. Hence, for such values of $U$, the kinetic energy starts helping to stabilize the AF ordered phase, differently than at weak coupling: This represents a first precursor of the Heisenberg mechanism for stabilizing the AF phase, where the order is set among already preformed local magnetic moments. Significantly, the change in nature of the underlying AF state, is well reflected by our D$\Gamma$A data, where antiferromagnetic fluctuations in the paramagnetic phase enhance $\lvert E_{\text{kin}}\rvert$ compared to DMFT.

Let us point the attention to another very interesting feature of the kinetic energies at $U=2.0$: While for $U=0.75$ and $U=4.0$ the difference between the DMFT and the D$\Gamma$A results increases with decreasing temperature, the situation is reversed for the $U=2.0$, where such difference gets smaller. This reversed trend actually reflects the intermediate coupling nature of $U=2.0$, where the DMFT low-T metallic increase of $\lvert E_{\text{kin}}\rvert(T)$ coexists (and competes) with an already Heisenberg-like increase driven by the nonlocal AF fluctuations in D$\Gamma$A. The twofold aspect of the intermediate coupling fluctuations will be discussed in more detail in Sec. \ref{subsec:occupations}. 

Finally, for $U=4.0$ (right panel of Fig. \ref{fig:Ekin2dAllU}) we observe a typical strong coupling scenario: In $2d$ the kinetic energy of DMFT is very small and almost temperature independent: In the local moment regime almost no hopping of electrons occurs at temperatures scales smaller than the Hubbard interaction. This is nicely exemplified by comparing the DMFT curve with the one for the atomic limit $U\rightarrow\infty$ (AL in Fig.~\ref{fig:Ekin2dAllU}) for which in Eq.~(\ref{equ:ekin}) the DMFT self-energy has been replaced by the corresponding expression for the atomic limit (see next section and Appendix \ref{appsub:al} for more details): In the AL, we observe a very similar behavior as in DMFT except for even stronger suppression of the electron mobility. The similarity between the DMFT and the AL results, hence, indicates the almost atomiclike nature of the system within DMFT. The D$\Gamma$A data, instead, exhibit significantly {\sl larger} values for $\lvert E_{\text{kin}}\rvert(T)$ than in DMFT at all considered temperatures. Moreover, the corresponding difference is enhanced upon decreasing temperature as antiferromagnetic fluctuations grow stronger. This scenario is a hallmark of a pure Heisenberg-like nature of the related antiferromagnetic phase at $T\!=\!0$, which becomes, once again, visible already in the corresponding fluctuations in the paramagnetic phase. Similar conclusions can be drawn in $3d$ (inset in the right panel of Fig. \ref{fig:Ekin2dAllU}) although the effect is -as expected- less pronounced compared to $2d$, due to the reduction of the impact of spatial fluctuations in higher dimensions.

\subsection{Energy distribution function}
\label{subsec:occupations}

An improved understanding of our results for the kinetic energy can be gained by analyzing the energy distribution function $n(\varepsilon)$. This is defined as:
\begin{equation}
 \label{equ:defoccup}
n(\varepsilon)=\sum_{\mathbf{k}}n_{\mathbf{k}}\delta(\varepsilon-\varepsilon_{\mathbf{k}})=\frac{2}{\beta}\sum_{\nu\mathbf{k}}\delta(\varepsilon-\varepsilon_{\mathbf{k}})G(\nu,\mathbf{k}),
\end{equation}
where $n_{\mathbf{k}}$ is the occupation of the single-particle momentum eigenstate with the energy $\varepsilon_{\mathbf{k}}$ of the corresponding noninteracting system and the factor $2$ accounts for the two spin-projections. $n(\varepsilon)$ fulfills of course the sum rule $\int_{-\infty}^{+\infty}d\varepsilon\;n(\varepsilon)\!=\!n$. Note that for the noninteracting case $n(\varepsilon)$ coincides with the $\mathbf{k}$-integrated (and spin-summed) spectral function $n(\varepsilon)=2f(\varepsilon)D(\varepsilon)$ where $f(\varepsilon)=(1+e^{\beta \varepsilon})^{-1}$ is the Fermi function and $D(\varepsilon)=\sum_{\mathbf{k}}\delta(\varepsilon-\varepsilon_\mathbf{k})$ is the DOS of the noninteracting system. We recall that this equivalence is no longer true for interacting electrons: Here the spectral function $A(\omega)$ describes the redistribution of the original single particle excitation energies of the noninteracting system due to electronic scattering. At strong coupling this leads, among other features, to Hubbard bands at much higher energies than the upper edge of the noninteracting DOS. $n(\varepsilon)$, on the other hand, describes the correlated state of the system for $U>0$ solely in terms of a redistribution of the {\sl occupation} of the {\sl original} single-electron eigenstates $\varepsilon_{\mathbf{k}}$. Hence, while $A(\omega)$ can be directly measured in direct/inverse photoemission experiments $n(\varepsilon)$ can be only extracted from the latter by means of Eq.~(\ref{equ:defoccup}). Nevertheless, $n(\varepsilon)$ helps considerably in gaining a more profound understanding of our kinetic energy results, since the latter can be also written as 
\begin{equation}
 \label{equ:ekinoccup}
 E_{\text{kin}}=\int\limits_{-\infty}^{+\infty}d\varepsilon\; \varepsilon\; n(\varepsilon).
\end{equation}
In this way $n(\varepsilon)$ allows for an identification of the energy scale(s) from which the differences between the kinetic energies of D$\Gamma$A  and DMFT originate and, hence, for a transparent physical description of the energy-scale selective effects of spatial correlations on the energetics of the correlated electron system.

\begin{figure}[t!]
  $d=2,U=0.75$, $\beta=80$
 \includegraphics[width=0.45\textwidth]{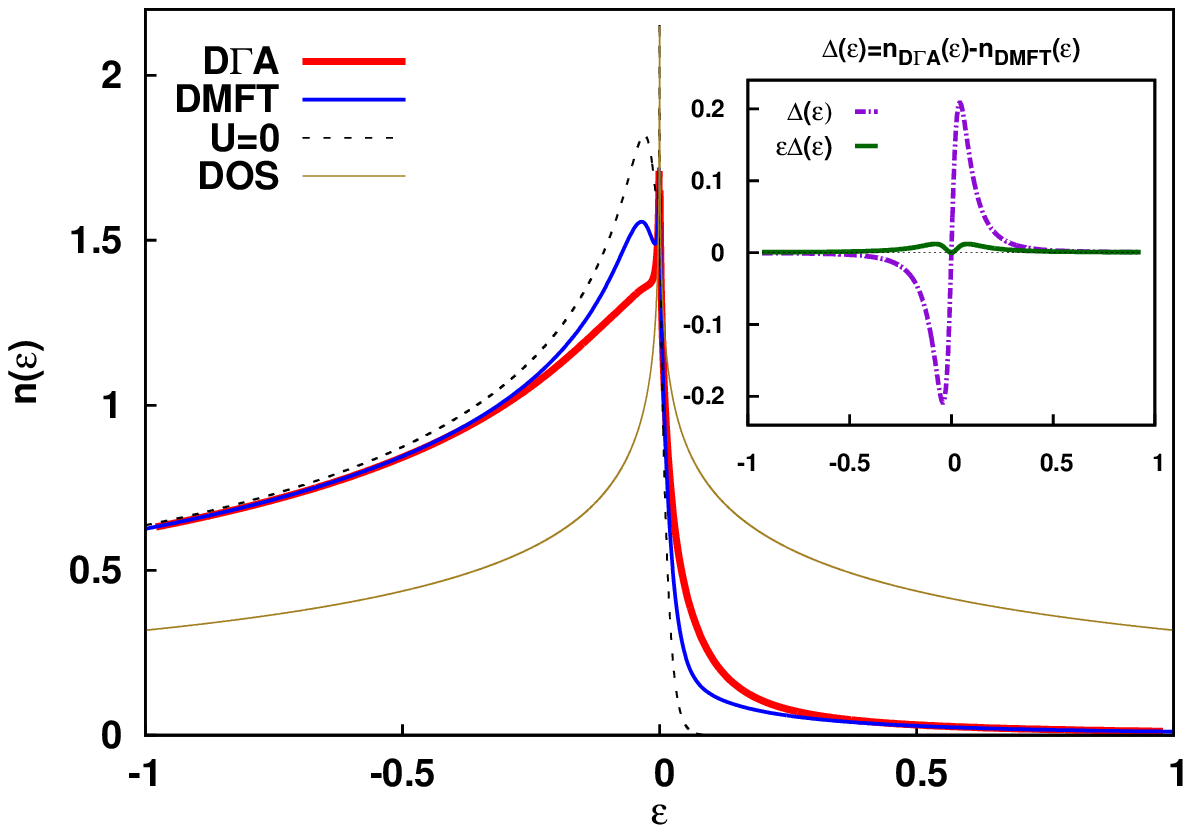}
  $d=2,U=2.0$, $\beta=80$
 \includegraphics[width=0.45\textwidth]{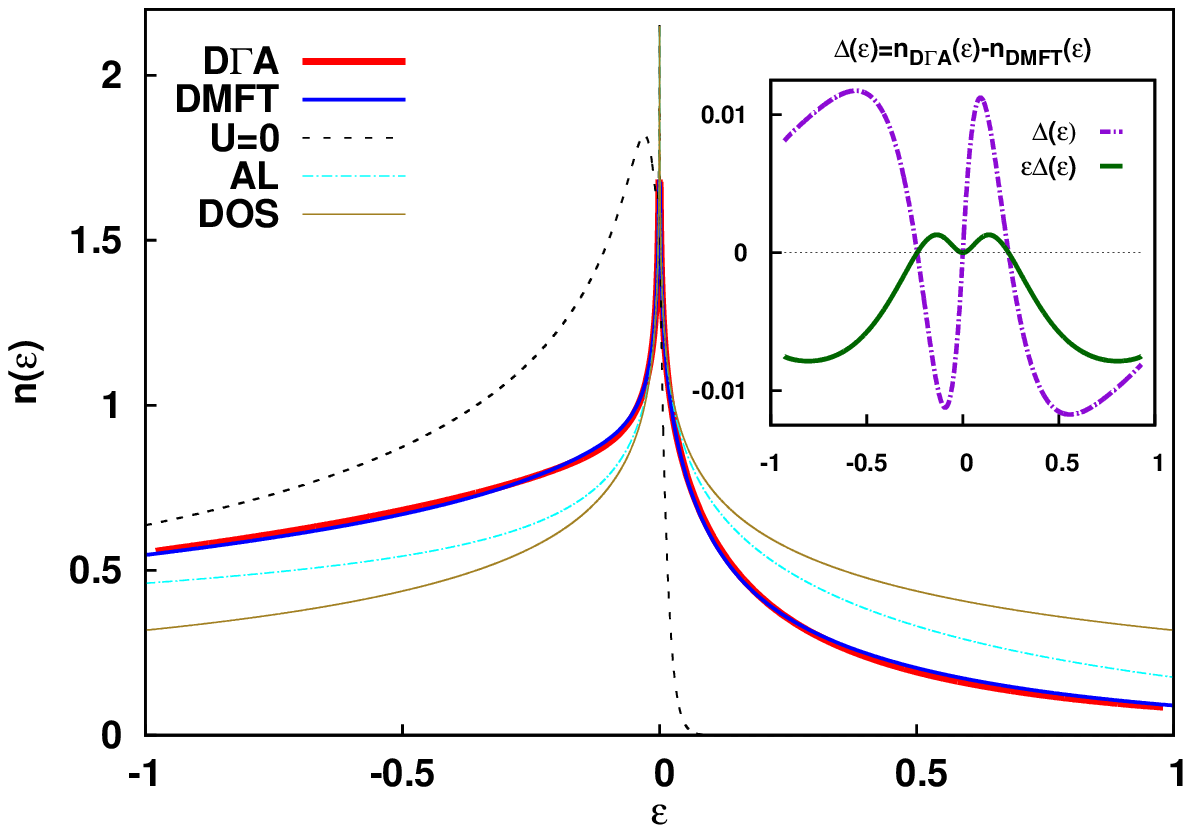}
  $d=2,U=4.0$, $\beta=80$
 \includegraphics[width=0.45\textwidth]{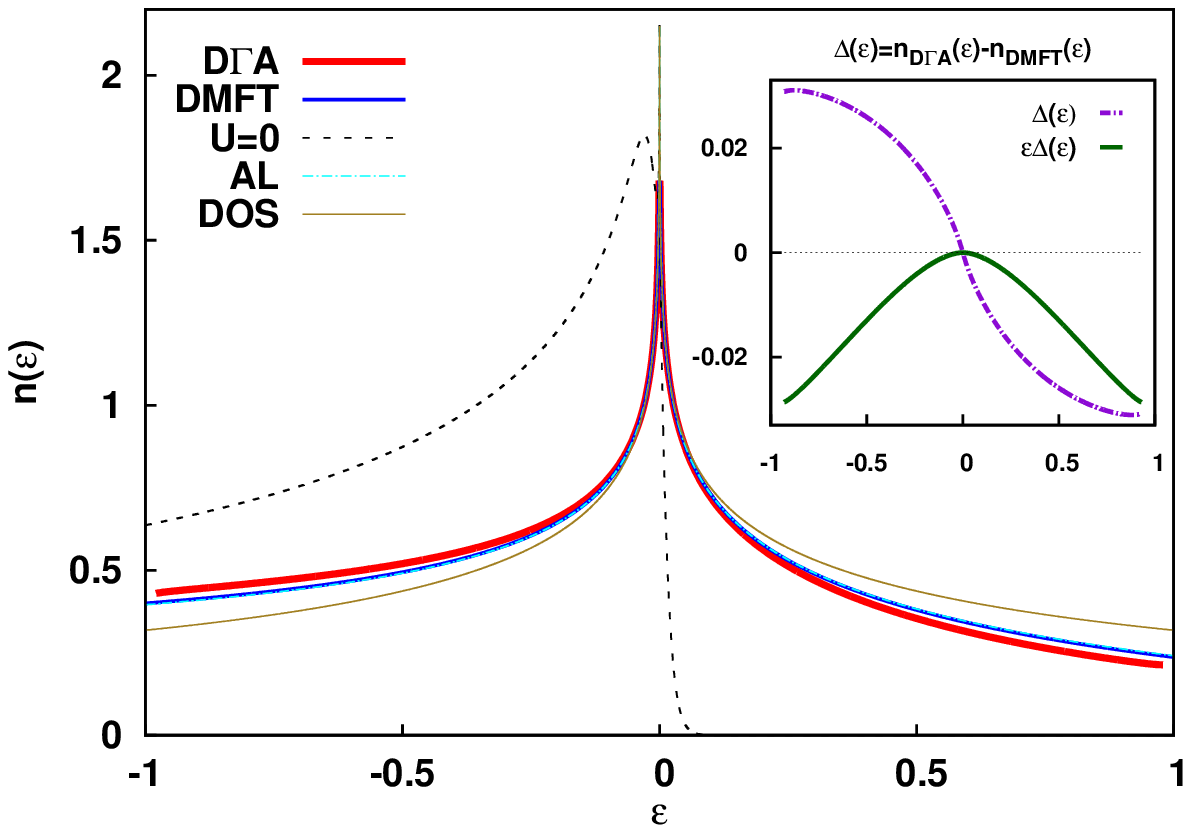}
\caption{Energy distributions $n(\varepsilon)$ for the $2d$ Hubbard model at three different values of the interaction parameter $U$ for the lowest considered temperature ($\beta=80$). In the insets the difference $\Delta(\varepsilon)$ between the energy distributions of DMFT and D$\Gamma$A as well as the contribution $\varepsilon\Delta(\varepsilon)$ to the corresponding difference of the kinetic energies are shown.}
\label{fig:occup2dbeta80}
\end{figure}

In the upper panel of Fig.~\ref{fig:occup2dbeta80} the energy distribution $n(\varepsilon)$ for $2d$, $U\!=\!0.75$ and $\beta\!=\!80$ obtained by D$\Gamma$A is compared to the DMFT results, the noninteracting case $U\!=\!0$ and the corresponding DOS. First, we observe a spectral weight shift from negative to positive energies around the Fermi level between DMFT and the noninteracting case. The distribution function $n(\varepsilon)$ of DMFT, however, still exhibits a very pronounced decrease (which will eventually turn into a discontinuity, i.e., a jump, at $T\!=\!0$) at the Fermi level, consistent with the presence of a well-defined Fermi surface\cite{Tocchio2012} in the DMFT data. The D$\Gamma$A results show an additional shift of weight in $n(\varepsilon)$ from negative to positive $\varepsilon$  w.r.t. DMFT. At the small value of $U$ considered here, this shift affects only a small energy region around the Fermi energy $\varepsilon=0$ (see inset). Hence, the reduction of the electronic mobility (i.e., $\lvert E_{\text{kin}} \rvert$) due to spatial correlations (difference between DMFT and D$\Gamma$A) observed at weak-coupling is realized entirely through a shift of weight in $n_{\text{D}\Gamma\text{A}}(\varepsilon)$ in a narrow energy window {\sl close}  to the Fermi level (i.e., for $\lvert\varepsilon\rvert\ll 1$) w.r.t $n_{\text{DMFT}}(\varepsilon)$. 

\begin{figure}
  $d=2, U=2.0$, $\beta=31.4$
 \includegraphics[width=0.45\textwidth]{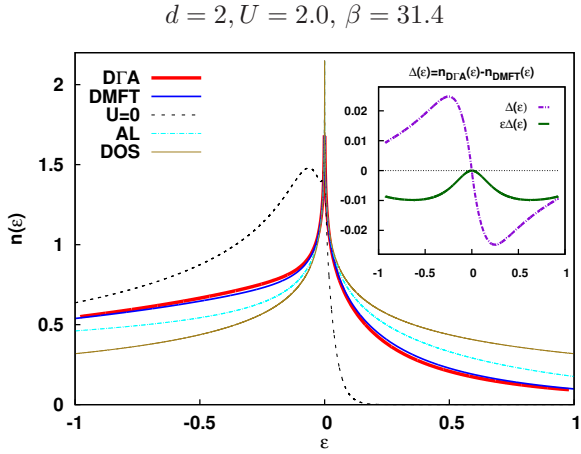}
 \caption{Same as in the middle panel of Fig. \ref{fig:occup2dbeta80}, but for $\beta=31.4$.}
 \label{fig:occup2dbeta314}
\end{figure}

In the middle panel of Fig.~\ref{fig:occup2dbeta80} we perform a similar comparison in the intermediate coupling regime, i.e., $U=2.0$. As expected for this value of the interaction,  both the D$\Gamma$A and DMFT curves deviate significantly from the noninteracting one. As for the difference between D$\Gamma$A and DMFT (inset),  we observe an interesting coexistence of {\sl opposite} trends, depending on the energy scale considered: (i)  At large energies, for $\lvert\varepsilon\rvert\gg 0$, a shift of weight in $n_{\text{D}\Gamma\text{A}}(\varepsilon)$ w.r.t.\ $n_{\text{DMFT}}(\varepsilon)$ from {\sl positive to negative} $\varepsilon$ takes place. At the same time, (ii) at the Fermi level, i.e., for low values of $\lvert\varepsilon\rvert\sim 0$, an opposite shift of weight from {\sl negative to positive} energies occurs. The latter trend is similar to the (weak-coupling) one found at $U=0.75$. This means that, at intermediate coupling, the effect of nonlocal correlations has a {\sl twofold} nature:  Spatial correlations simply reduce the mobility of the electrons at the Fermi level, while -at the same time- they induce an enhanced occupation of large-energy one-particle states well {\sl inside} the Fermi surface. The latter, broader energy scales affected by the fluctuations might be associated to the formation of coherent spin-polaron excitations\cite{Sangiovanni2006,Taranto2012} typical of Heisenberg antiferromagnets, which involve electronic states at all $\varepsilon$. In D$\Gamma$A, their presence can be detected, indirectly, {\sl also} in the paramagnetic phase. Hence, from the opposite trends in the  weight shifts in $n(\varepsilon)$, a compensation between overall gains and losses in the kinetic energy is induced\footnote{This allows a clear-cut interpretation of the temperature evolution of the difference in $\lvert E_{\text{kin}}\rvert$ between DMFT and D$\Gamma$A for $U=2.0$: At high $T$ (corresponding to $\beta=31.4$) the shift of weight in $n_{\text{D}\Gamma\text{A}}(\varepsilon)$ from {\sl positive to negative} energies w.r.t. to $n_{\text{DMFT}}(\varepsilon)$ for $\lvert\varepsilon\rvert\gg 1$ leads to a corresponding enhancement of $\lvert E_{\text{kin}} \rvert$ for D$\Gamma$A compared to DMFT (see middle panel in Fig. \ref{fig:Ekin2dAllU}). At low $T$ (corresponding to $\beta=80$) on the other hand, we observe  an additional {\sl inverse} shift at lower energies around the Fermi level, which partially compensates the gain of $\lvert E_{\text{kin}}\rvert$ in D$\Gamma$A w.r.t. DMFT which is well reflected in the smaller difference between these two values at lower $T$ as it can be observed in the middle panel of Fig. \ref{fig:Ekin2dAllU}.}, when the energetics of spatial correlations in D$\Gamma$A and/or the underlying long-range (antiferromagnetically) ordered ground state are considered.   We should also note that for the highest temperature data presented in Fig.~\ref{fig:occup2dbeta314} the change of $n(\varepsilon)$ around $\varepsilon\!=\!0$ disappears. This can be understood by the fact that within DMFT at higher $T$ the system enters in the so-called crossover regime between the metallic and the insulating phase, where the spectral weight at the Fermi level is already strongly suppressed by purely local correlations. Hence no particular kinetic energy loss of mobility can affect the (already incoherent) Fermi-surface electrons.

In the strong coupling limit ($U=4.0$, lowest panel in Fig. \ref{fig:occup2dbeta80}) we observe the same situation as for $U=2.0$, $\beta=31.4$, albeit the gain of $n(\varepsilon)$ for $-1\leq\varepsilon< 0$ in D$\Gamma$A w.r.t. DMFT gets much more pronounced. This is the origin of the enhanced $\lvert E_{\text{kin}}\rvert$ of D$\Gamma$A w.r.t. DMFT, demonstrating the clear-cut Heisenberg-type nature of the nonlocal antiferromagnetic fluctuations in this regime.

The above discussion also demonstrates how $n(\varepsilon)$ marks a fundamental difference between the low-$T$ antiferromagnetic gap at small $U$ and the large-$U$ Mott gap of the $2d$ system.

In the former situation the spectral weight is suppressed by nonlocal fluctuations only in a tiny region around the Fermi energy. Indeed, at weak coupling, our D$\Gamma$A results for the spectral functions exhibit sharp peaks at $\omega=\varepsilon_{\mathbf{k}}$ for $\mathbf{k}$ vectors away from the Fermi surface, while the $\omega=\varepsilon_{\mathbf{k}_F}=0$ spectral weight is clearly reduced by antiferromagnetic fluctuations w.r.t.\ DMFT (see, e.g., third rows of Figs.~\ref{fig:spectra_nodal} and \ref{fig:spectra_antinodal}). This behavior is reflected in $n(\varepsilon)$ which deviates between DMFT and D$\Gamma$A only in a small region around $\varepsilon=0$. Upon lowering the temperature, we have verified by our D$\Gamma$A data that just the depth of the dip in the spectral function, but not its extension in the $\mathbf{k}$ space, increases. Consistently, only the difference between $n_{\text{D}\Gamma\text{A}}(\varepsilon)$ and $n_{\text{DMFT}}(\varepsilon)$ [or $n_{U=0}(\varepsilon)$] gets enhanced for $\lvert \varepsilon \rvert\ll 1$ but {\sl not} the $\varepsilon$-region, where such a difference is observed. This behavior can be further interpreted in terms of a spin-fermion model\cite{Katanin2009}, where the effective interaction between the electrons is governed by the spin fluctuations, i.e., $U_{\text{eff},\mathbf {q}}^{\omega}\sim U^2\gamma_{m,\mathbf{q}}^{\nu\omega}\chi_{m,\mathbf{q}}^{\omega}$. As the effective interaction between the particles is strongly $\mathbf{q}$ (and $\omega$) dependent, with a maximum at $\mathbf{q}$ equal to the nesting vector of the noninteracting Fermi surface  and $\omega\!=\!0$, the effect of $U_{\text{eff}}$ will be particularly strong only at the Fermi level.

On the contrary, the progressive suppression of coherent spectral weight at strong coupling can be attributed to a strong enhancement of the effective interaction $U_{\text{eff}}^{\omega}\sim U^2\gamma_{m}^{\nu\omega}\chi_{m}^{\omega}$ which is (within DMFT) mainly governed by the {\sl purely local} spin susceptibility $\chi_{m}^{\omega}$. In fact, in this case, mainly $\mathbf{k}$-independent local fluctuations prevail, which are typically well captured by the local self-energy of DMFT. Indeed, when $U$ becomes of the order of the bandwidth of the system, i.e., upon approaching the Mott transition of DMFT where $\chi_m^{\omega=0}$ diverges (for $T=0$), a spectral gap at $\omega=\varepsilon_{\mathbf{k}}$ opens at {\sl all} $\mathbf{k}$ points in the Brillouin zone. This situation is well reflected by the difference of $n(\varepsilon)$ in DMFT and D$\Gamma$A w.r.t. to the noninteracting case which appears at all values of $\varepsilon$ in the strong coupling regime (see lowest panel of Fig. \ref{fig:occup2dbeta80}).  

Let us finally turn our attention to another pertinent observation: For $U=4.0$, $n(\varepsilon)$ of both D$\Gamma$A and DMFT is almost identical to the corresponding distribution function obtained from the self-energy of the atomic limit (AL) which reads (at half filling)
\begin{equation}
\label{equ:alsigma}
 \Sigma(i\nu)=\frac{U^2}{4i\nu}+\frac{Un}{2}. 
\end{equation}
Moreover, $n_{\text{AL}}(\varepsilon)$ has a very similar shape as the noninteracting DOS (lower panel of Fig. \ref{fig:occup2dbeta80}). A corresponding calculation can be performed analytically (see Appendix \ref{appsub:al}) and yields (for $T=0$):
\begin{equation}
 \label{equ:nal}
 n_{\text{AL}}(\varepsilon)=D(\varepsilon)\left[1-\frac{\varepsilon}{\sqrt{U^2+\varepsilon^2}}\right],
\end{equation}
where $D(\varepsilon)$ denotes the noninteracting DOS of the system. Hence, for $U\rightarrow\infty$ the energy distribution of the electrons indeed coincides exactly with the DOS. This observation allows for a ``complementary'' interpretation of the insulating Mott phase of the strong-coupling regime, in terms of the occupation of the noninteracting single-particle energy-levels. For $U=0$, we are dealing with a filled Fermi sea (for $T=0$): $n(\varepsilon)=2D(\varepsilon)\theta(-\varepsilon)$, i.e., all single-particle energy states with $\varepsilon\le 0$ are doubly occupied (by one $\uparrow$ and one $\downarrow$ electron). Upon increasing the interaction $U$, electrons are gradually shifted from these negative energy-states to corresponding eigenstates with positive energies until at $U=\infty$ {\sl all} possible original one-particle states are occupied, consistent with $n(\varepsilon)=D(\varepsilon)$.  This is reflected in a corresponding change of the degeneracy (and, hence, the entropy) of the system: At $U=0$ the ground state  is nondegenerate as all negative energy eigenstates are doubly occupied by one $\uparrow$ {\sl and} one $\downarrow$ electron. Upon shifting electron to positive energy-states one has the freedom to shift an $\uparrow$ or $\downarrow$ electron. Hence, the number of available states for $\varepsilon >0$ increases until $n(\varepsilon)=D(\varepsilon)$. This corresponds to the well-known (DMFT) entropy/site of $\log(2)$ in the Mott-insulating phase of the system. Note that, while conventionally this $\log(2)$ refers to the possible spin projections at a given lattice site in real space, here we observe this degeneracy for the different $\mathbf{k}$ states of the system. This degeneracy in $\mathbf{k}$, however, corresponds to the possibility of forming linear combinations of (momentum) states that are localized in real space. Indeed, for $U\gg 0$ the system forms linear combinations that avoid any double occupations (e.g., resonating valence bond states\cite{Baskaran1987}) while for $U\ll 0$ (attractive Hubbard model) the system would build linear combinations of $\mathbf{k}$ states which comprise only doubly or not occupied sites. These considerations provide a somewhat alternative view on the Mott insulating phase, from a $\mathbf{k}$ space rather than a real space perspective, and are perfectly reflected in our numerical (and analytical) results for $n(\varepsilon)$ at strong coupling. In such a perspective, this view could be exploited as a basis to interpret the corrections of nonlocal correlations to the purely local physics of DMFT at strong coupling, as the latter provides an accurate, but not exact, description of the Mott-Hubbard insulating state in terms of high entropy ground states. 

Let us finally mention that a similar analysis of the energy distribution function has been also performed for three dimensions. However, since the results are qualitatively analogous to the corresponding two-dimensional ones we refer the interested reader to Appendix~\ref{appsub:furtherres} for the corresponding numerical data.  

\subsection{Potential Energy}
\label{sec:epot}

\begin{figure*}
	\centering
		\includegraphics{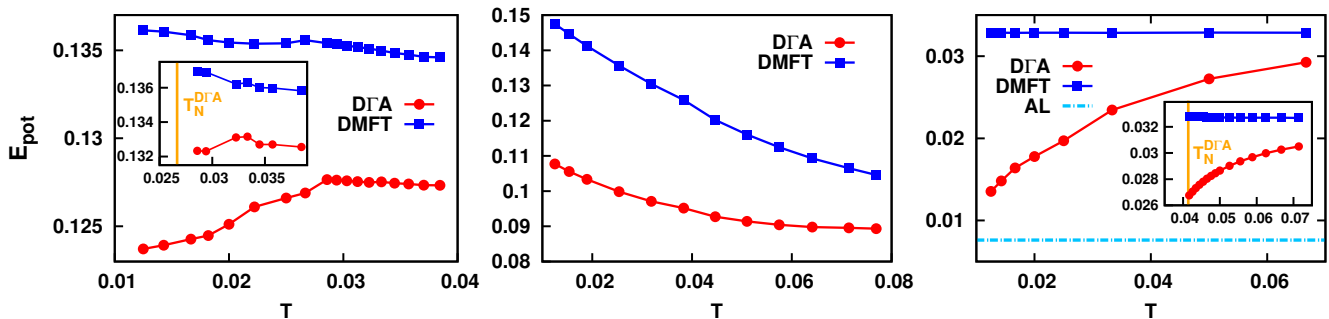}
	\caption{Potential energy of the half-filled Hubbard model computed in DMFT, D$\Gamma$A, and the atomic limit (AL) as a function of temperature in two (main panels) and three (insets) dimensions for different values of the interaction parameter $U=0.75$ (left panel), $U=2.0$ (middle panel), and $U=4.0$ (right panel).}
	\label{fig:Epot2dAllU}
\end{figure*}

The potential energy of the system is given by the expectation value of the interacting part of the Hubbard Hamiltonian [second term on the right hand side of Eq. (\ref{equ:defmodel})]. Hence, in terms of {\sl two-particle} Green's functions -or, more precisely, physical susceptibilities- it reads:
\begin{equation}
 \label{equ:epot1}
E_{\text{pot}}=\frac{U}{\beta}\sum_{\Omega\mathbf{q}}\chi_{\uparrow\downarrow,\mathbf{q}}^{\Omega}+U\left(\frac{n}{2}\right)^2,
\end{equation}
where $n$ denotes the number of particles and $\chi_{\uparrow\downarrow,\mathbf{q}}^{\Omega}$ is the physical susceptibility of the system in the $\uparrow\downarrow$ spin sector. At the same time, according to the EOM, $E_{\text{pot}}$ can be expressed only by means of one-particle quantities as
\begin{equation}
 \label{equ:epot2}
 E_{\text{pot}}=\frac{1}{\beta}\sum_{\nu\mathbf{k}}G(\nu,\mathbf{k})\Sigma(\nu,\mathbf{k}),
\end{equation}
which gives rise to an analogous ambiguity as for the calculation of the kinetic energy (cf. Sec. \ref{sec:ekin}). 

For the kinetic energy we argued that as $E_{\text{kin}}$ corresponds to a one-particle operator, it is natural to calculate it from one- rather than two-particle Green's functions. Consistent with this argument, one should evaluate the potential energy, which corresponds to the expectation value of a {\sl two-particle} operator, from two-particle Green's functions\cite{vanLoon2016}, i.e., from Eq.~(\ref{equ:epot1}). For DMFT, however, this leads to the following problem: The $\uparrow\downarrow$ susceptibility can be expressed through the charge and spin susceptibilities, respectively, $\chi_{\uparrow\downarrow,\mathbf{q}}^{\Omega}=1/2(\chi_{d,\mathbf{q}}^{\Omega}-\chi_{m,\mathbf{q}}^{\Omega})$ where for DMFT $\chi_{d/m,\mathbf{q}}^{\Omega}$ is obtained from the charge/spin ladder (Bethe-Salpeter equation) in Eq.~(\ref{equ:bethesalp}). The DMFT spin susceptibility $\chi_{m,\mathbf{q}}^{\Omega}$, however, diverges at the (rather high) DMFT transition temperature to the antiferromagnetically ordered phase and eventually becomes negative -and, hence, unphysical- below that temperature. This renders a calculation of $E_{\text{pot}}$ via Eq.~(\ref{equ:epot1}) highly questionable in large parameter regions. Maybe also for this reason, the typical evaluation of the potential energy in DMFT, and DFT+DMFT\cite{Kotliar2006,Liu2012}, is performed by means of the one-particle expression of Eq.~(\ref{equ:epot2}). The latter is algorithmically quite straightforward, because this expression, within the DMFT self-consistency, formally coincides with the double occupancy of the impurity site (multiplied by $U$) in the auxiliary AIM. As it was recently observed\cite{vanLoon2016}, on the other hand, in situations (i.e., for $T> T_N$), where both procedures for evaluating $E_{\rm pot}$ in DMFT are applicable, the difference between the corresponding results are physically significant. 

In principle, for $T< T_N$ of DMFT, one could perform antiferromagnetic DMFT calculations for determining the potential energy by means of Eq.~(\ref{equ:epot1}). This, however, makes conceptually problematic a comparison with the corresponding paramagnetic results of D$\Gamma$A, needed to identify the effect of nonlocal antiferromagnetic fluctuations on the potential energy in the {\sl paramagnetic phase}. Hence, in the following, we will exploit the standard ``one-particle'' expression, Eq.~(\ref{equ:epot2}), rather than Eq.~(\ref{equ:epot1}), for the determination of the potential energy. This allows for a paramagnetic calculation of $E_{\text{pot}}$ in DMFT down to (the lower) $T_N$ of D$\Gamma$A. Obviously, we should keep in mind the occurrence of possible inconsistencies\cite{vanLoon2016}, whose importance and consequences we will address in more detail at the end of this and in the next section.

In the leftmost panel of Fig.~\ref{fig:Epot2dAllU} we compare the potential energies of DMFT and D$\Gamma$A for $U=0.75$ for $2d$ (main panel) and $3d$ (inset), respectively. First of all, one observes that the potential energy of D$\Gamma$A is clearly reduced w.r.t. the DMFT result. Moreover, for DMFT we observe the typical Fermi-liquid like increase of the potential energy with decreasing temperature as the weakly coupled system gets more coherent at lower temperatures. In D$\Gamma$A, on the other hand, we see the opposite behavior, i.e., $E_{\text{pot}}$ decreases with decreasing $T$ enhancing, hence, the difference between $E_{\text{pot}}$ of DMFT and D$\Gamma$A upon lowering $T$. This is indeed consistent with the fact that the (underlying) antiferromagnetic phase at weak coupling is stabilized by a decrease of potential energy w.r.t. to the paramagnetic phase\cite{Toschi2005,Toschi2005b,Taranto2012}, while the kinetic energy is enhanced in the symmetry broken regime (see Sec.~\ref{sec:ekin}). Let us stress that this (Slater-like) mechanism is well reflected by our D$\Gamma$A data for $E_{\text{pot}}$ already above the transition temperature, i.e., in the {\sl paramagnetic} phase, similarly as our results for $E_{\text{kin}}$ in the same parameter regime.

At an intermediate coupling of $U=2.0$ (middle panel of Fig. \ref{fig:Epot2dAllU}) $E_{\text{pot}}$ of D$\Gamma$A is still lower than the corresponding DMFT value for all temperatures. Taking into account the enhanced $\lvert E_{\text{kin}}\rvert$ in D$\Gamma$A w.r.t. DMFT (see Sec. \ref{sec:ekin}) this is physically consistent with the fact that at this value of $U$ the ordered phase is stabilized by an decrease of {\sl both}, the kinetic {\sl and} the potential energy\cite{Toschi2005,Toschi2005b,Taranto2012,Tagliavini2016}. We emphasize, once again, that the energy balance between paramagnetic and antiferromagnetically ordered phase appears already encoded in the corresponding nonlocal fluctuations {\sl above} the transition temperature as it is indicated by our D$\Gamma$A results.  At the same time, we also observe for D$\Gamma$A an increase of $E_{\text{pot}}$ with decreasing temperature in contrast to the situation at weak coupling ($U=0.75$). This indicates that the corresponding antiferromagnetic fluctuations are not any more purely Slater-like (as also suggested by the increase of $\lvert E_{\text{kin}}\rvert$ of D$\Gamma$A w.r.t. to DMFT) but already display the first strong-coupling (Heisenberg like) features at $U=2.0$. Our D$\Gamma$A results, hence, perfectly reflect the {\sl intermediate}-coupling nature of the Hubbard model at $U=2.0$ in the paramagnetic phase.

According to the discussion above we would expect at the higher value of $U=4.0$ a typical strong coupling situation where $E_{\text{pot}}$ of D$\Gamma$A is enhanced w.r.t. DMFT and increases upon decreasing temperature. However, while for $E_{\text{kin}}$ we indeed observed the typical strong coupling behavior in our D$\Gamma$A data (see Sec. \ref{sec:ekin}), D$\Gamma$A clearly shows the {\sl opposite} trend  for $E_{\text{pot}}$ (see rightmost panel of Fig. \ref{fig:Epot2dAllU}): The potential energy of D$\Gamma$A is {\sl reduced} w.r.t. to DMFT and decreases upon decreasing temperature. While the second unexpected behavior at low $T$ might be still due to numerical inaccuracies related to the very small values of $E_{\text{pot}}$ at $U=4.0$, we attribute the first one to the ambiguities in the calculation of the potential energy discussed at the beginning of this section.

As demonstrated in Ref.~\onlinecite{vanLoon2016}, the potential energy obtained from Eq.~(\ref{equ:epot2}) in DMFT becomes {\sl higher} than the corresponding one obtained with methods including nonlocal correlations beyond DMFT in the strong-coupling regime. Hence, the authors of  Ref.~\onlinecite{vanLoon2016} logically suggest that the comparison should be better made with the corresponding DMFT double occupancy obtained from Eq.~(\ref{equ:epot1}), which yields the more physically plausible results of an enhancement\cite{Fuchs2011a} of double occupancies (in D$\Gamma$A) due to Heisenberg-like nonlocal AF fluctuations.

While such considerations need certainly to be carefully taken into account in all future DMFT and realistic/{\sl ab initio} DFT+DMFT\cite{Kotliar2006,Held2007} studies where the total energy is computed (e.g., for structure optimization, determination of competing phases, etc.) in our case, evidently, the option of using Eq.~(\ref{equ:epot1}) is {\sl not} viable, because we are below the transition temperature of DMFT, which renders -as discussed above- the corresponding DMFT susceptibility unstable. 

However, we want to point out here, that, even if we restricted ourselves to temperatures above $T_N$ of DMFT, where Eq.~(\ref{equ:epot1}) is still applicable, we would get unphysical trends, although in a different parameter regime. In fact, since above the $T_N$ of DMFT $\chi_{m}$ of D$\Gamma$A is {\sl reduced} w.r.t. the corresponding DMFT value by the $\lambda$ corrections, we would observe,  at all values of $U$, an enhanced potential energy in D$\Gamma$A compared to DMFT. While this restores the physically correct trend at strong coupling,  unfortunately, the very same trend would be now observed {\sl also} at weak coupling, where it is not physically consistent. In other words, the trends for $E_{\text{pot}}$ shown above for the calculation with Eq.~(\ref{equ:epot2}) would be just reversed if we adopted Eq.~(\ref{equ:epot1}) for the determination of the potential energy in DMFT and D$\Gamma$A, moving the problematic comparison from strong- to weak-coupling.

These new observations might actually suggest novel possible strategies for future improvements of the algorithmic schemes used.  For the specific case of the ladder D$\Gamma$A with Moriya $\lambda$ corrections, the results for $E_{\text{pot}}$ may be improved, if we take into account, in a fully two-particle consistent way,  such corrections also for the charge fluctuations, since they are most likely underestimated by the DMFT ladder in Eq.~(\ref{equ:bethesalp}). This would lead, e.g., to a correction of the above mentioned spurious ``hierarchy'' of the potential energies between DMFT and D$\Gamma$A at weak coupling, when exploiting Eq.~(\ref{equ:epot1}) for the calculations. The algorithmic implementation of new schemes goes evidently beyond the scope of this paper. Nonetheless, in the next section we will sketch new ideas for improving the $\lambda$ corrections of the ladder D$\Gamma$A, whose general features might be of interest also for other diagrammatic extensions of DMFT, based on ladder resummations.  

\section{Towards an improved version of ladder D$\Gamma$A}
\label{sec:lambdaimproved}

In this section, inspired by the inconsistencies in the determination of the kinetic and the potential energy in DMFT and its extensions (see previous section), we propose an improvement of the state-of-the-art ladder D$\Gamma$A algorithm, overcoming these ambiguities. In particular, a thorough analysis of the intrinsic sources of inconsistencies in approximated many-body treatments will naturally suggest the conditions which need to be enforced. We stress here, that while the algorithmic modifications proposed in this section are thought to be specifically applicable to future calculations within the ladder D$\Gamma$A scheme, the underlying concepts might be inspiring also for further developments of other quantum many-body approaches based on DMFT and its extensions. As for our specific case, we recall that the $\lambda$ corrections of ladder D$\Gamma$A have been originally introduced\cite{Katanin2009} in order to correct the spurious $1/i\nu$-asymptotic behavior of the ladder D$\Gamma$A self-energy $\Sigma_{\text{lad}}(i\nu,\mathbf{k})$. As discussed in Sec.~\ref{sec:formalismphasediag} and Appendix~\ref{app:lambdacorr}, this is equivalent to the fulfillment of the sum rule
\begin{equation}
 \label{equ:sumruleupup}
 \frac{1}{\beta}\sum_{\Omega\mathbf{q}}\chi_{\uparrow\uparrow,\mathbf{q}}^{\lambda=\lambda_m,\Omega}=\frac{1}{\beta}\sum_{\Omega\mathbf{q}}\frac{1}{2}\left[\chi_{d,\mathbf{q}}^{\Omega}+\chi_{m,\mathbf{q}}^{\lambda_m,\Omega}\right]\equiv\frac{n}{2}\left(1-\frac{n}{2}\right),
\end{equation}
for the $\uparrow\uparrow$ susceptibility $\chi_{\uparrow\uparrow,\mathbf{q}}^{\Omega}$ [see Eq.~(\ref{equ:cap4sumchi})]. Let us stress that Eq.~(\ref{equ:sumruleupup}) is always fulfilled for the purely local DMFT susceptibilities (i.e., for those of the auxiliary AIM). 

A corresponding summation over the $\uparrow\downarrow$ susceptibility is related to the double occupancy of the system [c.f. Eq. (\ref{equ:epot1})], i.e.,
\begin{equation}
 \label{equ:double2p}
 \frac{1}{\beta}\sum_{\Omega\mathbf{q}}\chi_{\uparrow\downarrow,\mathbf{q}}^{\lambda,\Omega}=\frac{1}{\beta}\sum_{\Omega\mathbf{q}}\frac{1}{2}\left[\chi_{d,\mathbf{q}}^{\Omega}-\chi_{m,\mathbf{q}}^{\lambda_m,\Omega}\right]\equiv\langle\hat{n}_{i\uparrow}\hat{n}_{i\downarrow}\rangle_1\!-\!\frac{n^2}{4}.
\end{equation}
As discussed in the previous section [see Eq.~(\ref{equ:epot2})], the double occupancy of the system can be also calculated solely in terms of one-particle quantities:
\begin{equation}
 \label{equ:double1p}
 \frac{1}{U\beta}\sum_{\nu\mathbf{k}}G(i\nu,\mathbf{k})\Sigma(i\nu,\mathbf{k})=\langle\hat{n}_{i\uparrow}\hat{n}_{i\downarrow}\rangle_2,
\end{equation}
In an exact theory the double occupations obtained in the two different ways of course coincide, i.e., $\langle\hat{n}_{i\uparrow}\hat{n}_{i\downarrow}\rangle_1\equiv\langle\hat{n}_{i\uparrow}\hat{n}_{i\downarrow}\rangle_2$. While this relation also holds for the full (parquet-based) D$\Gamma$A, it is in general {\sl not} fulfilled in the ladder version of D$\Gamma$A. In fact, taking into account renormalization effects only for the spin susceptibility by adopting a single parameter $\lambda=\lambda_m$ allows only for fixing {\sl one} sum rule. This is, in the case of the state-of-the-art ladder-D$\Gamma$A, the one for $\chi_{\uparrow\uparrow}$ [see Eq.~(\ref{equ:sumruleupup})], which corresponds to fix the charge density at the two-particle level. Hence, in order to render the double occupancies (and, accordingly, the potential energies) calculated at the one- and the two-particle level equal, we have to introduce {\sl another} degree of freedom. Since in ladder D$\Gamma$A also the charge susceptibility is treated in a not (two-particle) self-consistent way, it is reasonable to apply a $\lambda$ correction also to this response function in order to take into account corresponding nonlocal renormalization effects beyond DMFT. A new Moriya-corrections parameter $\lambda_d$ can be thus introduced to the corresponding charge susceptibility in the same way as for the spin channel [Eq.~(\ref{equ:cap4lambdacorr})]. Taking into account, in addition, a chemical potential $\mu$ to fix the charge density of the system at the one-particle level, we eventually obtain the following system of three equations for determining, unambiguously, $\lambda_d$, $\lambda_m$, and $\mu$:  
\begin{subequations}
\label{subequ:newlambda}
\begin{align}
 &\frac{1}{\beta}\sum_{\Omega\mathbf{q}}\frac{1}{2}\left(\chi_{d,\mathbf{q}}^{\lambda_d,\Omega}+\chi_{m,\mathbf{q}}^{\lambda_m,\Omega}\right)=\frac{n}{2}\left(1-\frac{n}{2}\right)\label{equ:newlambda1}\\
 &\frac{1}{U\beta}\sum_{\nu\mathbf{k}}G_{\mu}^{\lambda}(i\nu,\mathbf{k})\Sigma^{\lambda}(i\nu,\mathbf{k})\nonumber\\&\hspace{2cm}=\frac{1}{\beta}\sum_{\Omega\mathbf{q}}\frac{1}{2}\left(\chi_{d,\mathbf{q}}^{\lambda_d,\Omega}-\chi_{m,\mathbf{q}}^{\lambda_m,\Omega}\right)+\frac{n^2}{4}\label{equ:newlambda2}\\
 &\frac{1}{\beta}\sum_{\nu\mathbf{k}}G_{\mu}^{\lambda}(i\nu,\mathbf{k})=n,\label{equ:newlambda3}
\end{align}
\end{subequations}
where $G_{\mu}^{\lambda}(\nu,\mathbf{k})=\left[i\nu+\mu-\varepsilon_{\mathbf{k}}-\Sigma^{\lambda}(\nu,\mathbf{k})\right]^{-1}$ and $\Sigma^{\lambda}$ denotes the ladder D$\Gamma$A self-energy obtained from Eq.~(\ref{equ:EOMrewrite}) where both $\chi_d$ and $\chi_m$ are replaced by their $\lambda$-corrected counterparts. Note that if the $\mu$ is strongly altered compared to its DMFT value, an ``outer'' self-consistency likely needs to be performed in our calculations, because the local AIM should be readapted to the local part of the ladder D$\Gamma$A Green's function.

It is obvious that the $\lambda$ corrections introduced by Eqs.~(\ref{subequ:newlambda}) avoid -per construction- the ambiguity in the calculation of $E_{\text{pot}}$, since the accordance of this observable calculated from one- and two-particle quantities is enforced by Eq.~(\ref{equ:newlambda2}).
 
It is worth noticing that the proposed scheme follows, to some extent, the ideas of the TPSC approach for the Hubbard model\cite{Vilk1997,Tremblay2012}. There, RPA ladders have been constructed with different values $U_d$ and $U_m$ in the charge and spin channels, respectively. The {\sl free} parameters $U_d$ and $U_m$ of the TPSC calculations are in fact determined by requiring consistency between the one- and the two-particle level. In this perspective, this theory might be seen as the limiting case of the ladder D$\Gamma$A approach with the newly proposed $\lambda$ corrections for irreducible vertices $\Gamma_r^{\nu\nu'\omega}\rightarrow U_r$ and the DMFT scattering factor $\Gamma\rightarrow 0$. Hence, one would expect that both theories lead to similar results in the weak-to-intermediate coupling regime, while the D$\Gamma$A is applicable also at strong coupling. However, despite the formal similarity, the newly proposed $\lambda$-correction scheme for D$\Gamma$A will modify the results for the one-particle spectral functions w.r.t. TPSC {\sl also} in the weak-coupling regime. This is due to the presence of a finite quasiparticle scattering rate, as we have discussed in Sec. \ref{subsec:flparam}.

Let us finally address the problem of the consistent treatment of the kinetic energy within D$\Gamma$A. As already mentioned at the beginning of Sec.~\ref{sec:ekin}, $E_{\text{kin}}$ can be obtained also from two-particle quantities by means of the so-called $f$-sum rule\cite{Vilk1997}:
\begin{align}
 \label{equ:fsumrule}
 \lim_{\eta\rightarrow 0}\frac{1}{\beta}&\sum_{\Omega}\left(e^{i\Omega\eta}-e^{-i\Omega\eta}\right)i\Omega\;\chi_{d/m,\mathbf{q}}^{\Omega}\nonumber\\&=\frac{2}{\beta}\sum_{\nu\mathbf{k}}(\varepsilon_{\mathbf{k}+\mathbf{q}}+\varepsilon_{\mathbf{k}-\mathbf{q}}-2\varepsilon_{\mathbf{k}})G(\nu,\mathbf{k}).
\end{align}
In fact, one can immediately see that summing Eq.~(\ref{equ:fsumrule}) over $\mathbf{q}$ yields twice the negative value of the kinetic energy on its r.h.s. as the terms proportional to $\sum_{\mathbf{q}}\varepsilon_{\mathbf{k}\pm\mathbf{q}}$ vanish. This, hence, allows us to calculate $E_{\text{kin}}$ purely from either the charge or the spin susceptibility by just summing the left hand side of this equation over $\mathbf{q}$. In general, this will yield a different result for the kinetic energy as that obtained from Eq.~(\ref{equ:ekin}) in Sec.~\ref{sec:ekin} for D$\Gamma$A. We may be actually able to sidestep this ambiguity by introducing $\mathbf{q}$-dependent $\lambda$ corrections $\lambda_{\mathbf{q}}$, determined by the fulfillment of the $f$-sum rule in Eq.~(\ref{equ:fsumrule}). Specifically, we can introduce parameters $\lambda_{d,\mathbf{q}}$ and $\lambda_{m,\mathbf{q}}$ into $\chi_{d,\mathbf{q}}^{\Omega}$ and $\chi_{m,\mathbf{q}}^{\Omega}$ [see Eq.~(\ref{equ:cap4lambdacorr})] and determine their values (for each single $\mathbf{q}$) by Eq.~(\ref{equ:fsumrule}). Let us note that this modification would lead to a strongly coupled set of nonlinear self-consistent equations for $\lambda_{d,\mathbf{q}}$ and $\lambda_{m,\mathbf{q}}$. In fact, the self-energy $\Sigma^{\lambda_{\mathbf{q}}}(\nu,\mathbf{k})$, which enters in the Green's function on the r.h.s. of Eq.~(\ref{equ:fsumrule}), would depend on the values of the $\lambda$'s at {\sl all} momenta $\mathbf{q}$, since it should be calculated from the D$\Gamma$A Eq.~(\ref{equ:EOMrewrite}) including, in turn, a $\mathbf{q}$ sum over the $\lambda_{\mathbf{q}}$-corrected susceptibilities $\chi_{d,\mathbf{q}}^{\lambda_{d,\mathbf{q}},\Omega}$ and $\chi_{m,\mathbf{q}}^{\lambda_{m,\mathbf{q}},\Omega}$. Let us finally recall that the $f$-sum rule is automatically fulfilled in any conserving \cite{Baym1961,Baym1962} approximation in the Baym-Kadanoff sense.

A crucial question would be now whether it is possible to enforce Eqs.~(\ref{subequ:newlambda}) and Eq.~(\ref{equ:fsumrule}) at the same time, or if -within the ladder D$\Gamma$A- one should, instead, stick to a self-consistent treatment of either the potential {\sl or} the kinetic energy. In this respect, we note that the contribution of $\chi_{d/m,\mathbf{q}}^{\Omega}$ to the frequency sum on the left hand side of Eq.~(\ref{equ:fsumrule}) vanishes for $\Omega\equiv 0$. Hence, one can modify the values of $\lambda_{d/m,\mathbf{q}}$ only for $\Omega=0$ by adding a corresponding $\mathbf{q}$-independent constant $\lambda_{d/m}$ as in Eqs.~(\ref{subequ:newlambda}) without violating the validity of the sum rule. This would yield eventually an $\Omega$- and $\mathbf{q}$-dependent Moriya $\lambda$ correction:
\begin{equation}
 \label{equ:newlambdaomq}
 \lambda_{r,\mathbf{q}}^{\Omega}=\lambda_{r,\mathbf{q}}+\delta_{\Omega 0}\lambda_{r},
\end{equation}
where $r=d,m$. The parameters $\lambda_{r,\mathbf{q}}^{\Omega}$ can now be determined by the Eqs.~(\ref{subequ:newlambda}) and Eq.~(\ref{equ:fsumrule}) which indeed render {\sl both} the kinetic {\sl and} the potential energy consistent at the one- and the two-particle level. From a physical perspective the ansatz for the new $\lambda$-correction scheme in Eq.~(\ref{equ:newlambdaomq}) is well justified for $T>0$ where classical ($\Omega\!=\!0$) fluctuations are dominating. Hence, the discontinuity of $\lambda_{r,\mathbf{q}}^{\Omega}$ reflects to some extent the corresponding sharp increase of the physical observables, i.e., in the physical susceptibilities at $\Omega\!=\!0$. To which extent, however, these improved schemes are applicable for describing quantum fluctuations at $T=0$ remains an open question. 

We should also note that the above introduced momentum-dependent $\lambda$-correction scheme is in some sense complementary to the dual boson method, where the local retarded interaction ($\Lambda_{\Omega}$) can be interpreted as a {\sl frequency-dependent} $\lambda$ correction of the corresponding bosonic propagator\cite{vanLoon2014}. Let us, however, stress that in the latter approach the quantity $\Lambda_{\Omega}$ is determined by relating the susceptibility to the corresponding one of an auxiliary AIM [see Eq.~(36) in Ref.~\onlinecite{vanLoon2014}], while in our case it could be fixed by using only consistency conditions between one- and two-particle observables of the physical system.  

We should point out that the above defined $\mathbf{q}$-dependent $\lambda$-correction scheme might also overcome the problem of the analyticity violation of the imaginary part of the ladder-D$\Gamma$A self-energy for $\mathbf{k}$ points far away from the Fermi surface which has been recently reported in Ref.~\onlinecite{Valli2015}. In fact, as it has been argued there, that the state-of-the-art ($\mathbf{q}$-independent) $\lambda$ corrections work more accurately for $\mathbf{k}$ points at the Fermi surface, corresponding (in our case) to the $\mathbf{q}=\mathbf{\Pi}$ contribution of the susceptibilities, but are too large for $\mathbf{k}$ vectors far away from the Fermi surface (related to $\mathbf{q}\sim \mathbf{\Pi}/2$).

Let us finally state that the above introduced $\lambda$ corrections allow also for a correction of the anomalous critical exponent $\eta$ within the ladder D$\Gamma$A scheme, which was previously frozen to its mean-field value ($\eta=0$). In fact, while the $\mathbf{q}$-independent part of $\lambda_{r,\mathbf{q}}^{\Omega}$ accounts for the modification of the mean field value of the critical exponent $\nu$ (or $\gamma$, see Appendix~\ref{app:lambdacorr}) the $\mathbf{q}$-dependent part can modify the functional form of the $\mathbf{q}$ dependence of Eq.~(\ref{equ:ornstein}) and, thus, the critical exponent $\eta$.

\section{Conclusions and Outlook}
\label{sec:conclusions}

In this paper we have analyzed by means of the dynamical vertex approximation (and its comparison with DMFT) how nonlocal correlations selectively affect the physics of the two- and three-dimensional (unfrustrated) Hubbard model on different energy scales, over the whole phase diagram. Specifically, we found that, at low energies close to the Fermi level, antiferromagnetic fluctuations give rise to a strong suppression of spectral weight, reflected in an enhanced D$\Gamma$A quasiparticle scattering rate w.r.t. DMFT. This effect disfigures the typical Fermi-liquid temperature-dependencies of the physical quantities in a much broader $T$ regime than the one, where coherent quasiparticle excitations are eventually destroyed by the nonlocal correlations.

At the same time, the low-temperature enhancement of the D$\Gamma$A scattering rates, though significant, is weaker than that predicted by theories based on bare Green's functions and frequency-independent (static) vertices, such as TPSC, where, e.g., in $2d$ the scattering rate grows directly proportional to the correlation length of the system: In the D$\Gamma$A all temperature dependencies, are -to some extent- mitigated by local correlation effects, i.e., by the finite quasiparticle scattering rate of DMFT. Such corrections yield more physically plausible results, which look consistent with an overall continuity of the temperature evolution of the (normal part of the) spectral functions across the AF phase transition, both in $3d$ and $2d$.

Our findings are also of potential interest for the interpretation of the most recent transport experiments\cite{Barisic2015} on the superconducting cuprates, in that we have shown how the very same physical mechanism (here: AF correlations with extended correlation length) can manifest itself quite differently in different physical observables and at different energy scales.

In the second part of the paper, we have extended our analysis to all energy scales by calculating the energetics of the system, resolved in its kinetic and potential counterparts. In D$\Gamma$A, we found a reduction/enhancement of the electronic mobility w.r.t.\ DMFT due to nonlocal correlations at weak/strong coupling, consistent with the corresponding stabilization mechanism of the low-temperature magnetic phase (Slater vs. Heisenberg). A detailed study of the energy-distribution functions  has allowed us to identify the origin of this difference as a loss of kinetic energy in D$\Gamma$A w.r.t.\ DMFT around the Fermi level at weak coupling, and a corresponding gain at large energies, away from the Fermi level, at strong coupling. These trends have been, thus, interpreted in terms of a progressive destruction of low-energy coherent excitations induced by extended AF fluctuations at weak coupling, and of the emergence of coherent high-energy magnetic excitations from the Mott insulating phase at strong coupling, respectively.

However, by performing the corresponding analysis for the potential energy (which should display a perfectly specular behavior within this scenario) deviations from the expected trends have been found at strong-coupling. After a critical analysis of our results and of the energy expressions commonly used in quantum many-body algorithms, we attributed this (likely) spurious behavior of $E_{\rm pot}$  to the intrinsic ambiguity of the corresponding energy expressions.  In fact, the latter --both for the kinetic and the potential terms-- can be obtained, complementarily,  from one- {\sl and} two-particle Green’s functions.  

The emergence of possible inconsistencies, recently reported\cite{vanLoon2016} exclusively for the potential energy part, may have relevant implications for all many-body approximated algorithms which require the calculation of the total energy of the system (such as lattice optimization within realistic DFT+DMFT calculations). The general discussion of such implications goes obviously beyond the scope of this paper. Nonetheless, the observation of these inconsistencies in the specific case of our ladder D$\Gamma$A calculations has been already inspiring for formulating suggestions of possible improvements in the state-of-the-art D$\Gamma$A algorithms:  These suggestions exploit a more consistent enforcing of the physical sum rules at the one- {\sl and} two-particle levels. Such ideas might be also of interest  for several quantum many-body schemes built on ladder resummations beyond the weak-coupling regime, like, e.g., for all the diagrammatic extensions of DMFT.  

\subsection*{Acknowledgments}

We thank Andrey Katanin, Patrick Thunstr\"om, Thomas Sch\"afer, James LeBlanc, Angelo Valli and Karsten Held for insightful discussions. We acknowledge financial support for our research activities from the Austrian Science Fund (FWF) through the project I-610-N16 (GR, AT), the subproject I-1395-N16 as part of the DFG research unit FOR 1346 (GR) and the SFB-ViCoM F41 (AT). The numerical calculations for this project have been performed on the Vienna Scientific Cluster (VSC).

\appendix

\section{Lambda corrections for the self-energy}
\label{app:lambdacorr}

\begin{figure}[t!]
 \centering
 \includegraphics[width=0.5\textwidth]{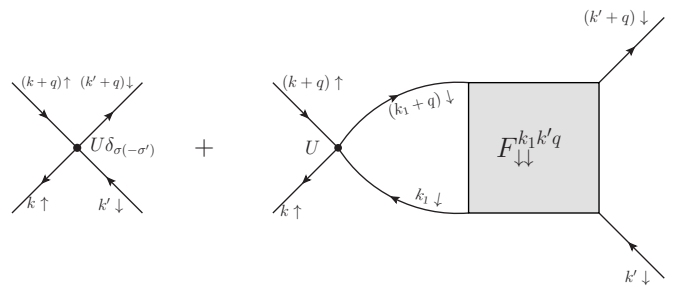}
 \caption{Diagrams for $F_{\uparrow\downarrow}^{kk'q}$ that do not depend on $\nu$ (and $\mathbf{k}$). Note that the four-vector notation $k\widehat{=}(i\nu,\mathbf{k})$ and $q\widehat{=}(i\Omega,\mathbf{q})$ has been adopted here.}
 \label{fig:equofmotionnuind}
\end{figure}
In this section we will first discuss the $\frac{1}{i\nu}$-asymptotic behavior of the self-energy $\Sigma(i\nu,\mathbf{k})$ obtained by means of the EOM (\ref{equ:EOM}) in the general case and then discuss the specific situation of ladder D$\Gamma$A. In order to keep the notation simple we will adopt here a four-vector notation for Matsubara frequencies and momenta: $k=(i\nu,\mathbf{k})$ for the fermionic case and $q=(i\Omega,\mathbf{q})$ for the bosonic case, respectively. Let us start with analyzing the single contributions to the self-energy according to Eq.~(\ref{equ:EOM}) with respect to their $\nu$ dependence: 

(i) $G(k+q)=\overset{\nu\rightarrow\infty}{\longrightarrow}\frac{1}{i\nu}+O\left[\frac{1}{(i\nu)^2}\right]$: Since $G(k+q)$ exhibits already a contribution $\frac{1}{i\nu}$, we have to single out terms which are constant with respect to $\nu$ in all the remaining parts of Eq.~(\ref{equ:EOM}) in order to get the $\frac{1}{i\nu}$ asymptotics of the entire expression. 

(ii) $F^{kk'q}_{\uparrow\downarrow}$: The $\nu$-independent contributions to the vertex function are given by the diagrams shown in Fig.~\ref{fig:equofmotionnuind} (see also Refs.~\onlinecite{Rohringer2012,Rohringer2013a}). Analytically, these terms can be written as:
 \begin{align}
 \label{equ:cap4equofmotionnuind}
  F_{\uparrow\downarrow}^{kk'q}&\sim\frac{U}{\beta}\sum_{k_1}\left[G(k_1)G(k_1+q)F_{\downarrow\downarrow}^{k_1k'q}+\beta\delta_{k_1k'}\right]\nonumber\\&=-\frac{U}{\beta}\frac{1}{G(k')G(k'+q)}\sum_{k_1}\chi_{\downarrow\downarrow}^{k_1k'q}+O\left[\frac{1}{i\nu}\right],
\end{align}
where the definition of the three-frequency and three-momentum susceptibility is analogous to the second line of Eq.~(\ref{equ:bethesalp}). Inserting this result into Eq.~(\ref{equ:EOM}) for the self-energy leads, at the order $\frac{1}{i\nu}$, to:
\begin{equation}
 \label{equ:cap4equofmotionasympt}
 \Sigma(k)=\frac{Un}{2}+\frac{1}{i\nu}\frac{U^2}{\beta^3}\sum_{k_1k'q}\chi_{\downarrow\downarrow}^{k_1k'q}+O\left[\frac{1}{(i\nu)^2}\right].
\end{equation}
For the exact $\chi_{\downarrow\downarrow}^{k_1k'q}$ of the Hubbard model (or a related AIM), the sum in Eq.~(\ref{equ:cap4equofmotionasympt}) can be evaluated analytically and yields:
\begin{equation}
 \label{equ:cap4sumchi}
 \frac{1}{\beta^3}\sum_{k_1k'q}\chi_{\downarrow\downarrow}^{k_1k'q}=\langle n_{\downarrow}n_{\downarrow}\rangle-\langle n_{\downarrow}\rangle\langle n_{\downarrow}\rangle=\frac{n}{2}\left(1-\frac{n}{2}\right).
\end{equation}
where SU(2) symmetry ($\langle\hat{n}_{\uparrow}\rangle\!=\!\langle\hat{n}_{\downarrow}\rangle\!=\!n/2$) has been used. Equations (\ref{equ:cap4equofmotionasympt}) and (\ref{equ:cap4sumchi}) yield the well-known expression for the $\frac{1}{i\nu}$-asymptotic behavior of self-energy of the Hubbard model\cite{Park2011}:
\begin{equation}
 \label{equ:cap4equofmotionasympt1}
 \Sigma(k)=\frac{Un}{2}+U^2\frac{n}{2}\left(1-\frac{n}{2}\right)\frac{1}{i\nu}+O\left[\frac{1}{(i\nu)^2}\right].
\end{equation}
In a plain-vanilla ladder version of D$\Gamma$A, we directly obtain the susceptibility $\chi_r^{kk'q}=\chi_{r,\mathbf{q}}^{\nu\nu'\omega}$ from a ladder consisting of local irreducible vertices $\Gamma_{r}^{\nu\nu'\omega}$ and DMFT Green's functions $G(i\nu,\mathbf{k})$ [see Eq.~(\ref{equ:bethesalp})], which actually corresponds\cite{Georges1996} to the definition of the momentum-dependent susceptibility of DMFT. It turns out that, in finite dimensions, where the DMFT self-consistency is guaranteed only at the one-particle level, this susceptibility violates the sum-rule~(\ref{equ:cap4sumchi}). This is reflected in a corresponding violation of the $\frac{1}{i\nu}$-asymptotic behavior of the ladder D$\Gamma$A self-energy. A solution for this problem has been successfully achieved within the ladder D$\Gamma$A approach\cite{Held2008,Katanin2009} through the introduction of a so-called Moriyasque $\lambda$ correction\cite{Moriya1985}. This works as follows: The sum rule in Eq.~(\ref{equ:cap4sumchi}) is restored by appropriately correcting $\chi_{\uparrow\uparrow}^{kk'q}$. Obviously, it is desirable to apply effective corrections to physical observable quantities rather than to intermediate-step objects such as the generalized susceptibility. To this end, one considers the {\sl physical} susceptibilities $\chi_{r,\mathbf{q}}^{\Omega}\!=\!(1/\beta^2)\sum_{kk'}\chi_{r}^{kk'q}$ rather than the generalized ones (see also Sec.~\ref{sec:formalismphasediag}). Here, $r$ can in principle refer to the spin ($r\!=\!m$), charge ($r\!=\!d$) and particle-particle ($r\!=\!pp$) channel. In its original version, which has been used for {\sl all} calculations in the present paper, the $\lambda$ correction $\chi_{r,\mathbf{q}}^{\Omega}\!\rightarrow\!\chi_{r,\mathbf{q}}^{\lambda_r,\Omega}$ is performed only for the dominating channel, which is the spin channel in the case of the half-filled Hubbard model on a bipartite lattice in $2d$ and $3d$. In Sec.~\ref{sec:lambdaimproved} however, an improved scheme of $\lambda$ corrections, taking into account also the charge channel in a self-consistent way is proposed. In a plain-vanilla ladder version of D$\Gamma$A, we directly obtain the susceptibility $\chi_r^{kk'q}=\chi_{r,\mathbf{q}}^{\nu\nu'\omega}$ from a ladder consisting of local irreducible vertices $\Gamma_{r}^{\nu\nu'\omega}$ and DMFT Green's functions $G(i\nu,\mathbf{k})$ [see Eq.~(\ref{equ:bethesalp})], which actually corresponds\cite{Georges1996} to the definition of the momentum-dependent susceptibility of DMFT. It turns out that, in finite dimensions, where the DMFT self-consistency is guaranteed only at the one-particle level, this susceptibility violates the sum rule~(\ref{equ:cap4sumchi}). This is reflected in a corresponding violation of the $\frac{1}{i\nu}$-asymptotic behavior of the ladder D$\Gamma$A self-energy. A solution for this problem has been successfully achieved within the ladder D$\Gamma$A approach\cite{Held2008,Katanin2009} through the introduction of a so-called Moriyasque $\lambda$ correction\cite{Moriya1985}. This works as follows: The sum rule in Eq.~(\ref{equ:cap4sumchi}) is restored by appropriately correcting $\chi_{\uparrow\uparrow}^{kk'q}$. Obviously, it is desirable to apply effective corrections to physical observable quantities rather than to intermediate-step objects such as the generalized susceptibility. To this end, one considers the {\sl physical} susceptibilities $\chi_{r,\mathbf{q}}^{\Omega}\!=\!(1/\beta^2)\sum_{kk'}\chi_{r}^{kk'q}$ rather than the generalized ones (see also Sec.~\ref{sec:formalismphasediag}). Here, $r$ can in principle refer to the spin ($r\!=\!m$), charge ($r\!=\!d$) and particle-particle ($r\!=\!pp$) channel. In its original version, which has been used for {\sl all} calculations in the present paper, the $\lambda$ correction $\chi_{r,\mathbf{q}}^{\Omega}\!\rightarrow\!\chi_{r,\mathbf{q}}^{\lambda_r,\Omega}$ is performed only for the dominating channel, which is the spin channel in the case of the half-filled Hubbard model on a bipartite lattice in $2d$ and $3d$. In Sec.~\ref{sec:lambdaimproved} however, an improved scheme of $\lambda$ corrections, taking into account also the charge channel in a self-consistent way is proposed.

The explicit transformation $\chi_{r,\mathbf{q}}^{\Omega}\rightarrow\chi_{r,\mathbf{q}}^{\lambda_r,\Omega}$ is given by\cite{Moriya1985,Katanin2009}:
\begin{equation}
 \label{equ:cap4lambdacorr}
 \left[\chi_{r,\mathbf{q}}^{\Omega}\right]^{-1}\rightarrow\left[\chi_{r,\mathbf{q}}^{\Omega}\right]^{-1}+\lambda_r=\left[\chi_{r,\mathbf{q}}^{\lambda_r,\Omega}\right]^{-1},
\end{equation}
where $\chi_{r,\mathbf{q}}^{\Omega}$ denotes the susceptibility obtained from the ladder calculation in Eq.~(\ref{equ:bethesalp}). Note that, in principle, the above relation can be made exact by considering a frequency- and momentum-dependent $\lambda_{r,\mathbf{q}}^{\Omega}$. In practice, as the exact expression of $\lambda_{r,\mathbf{q}}^{\Omega}$ is unknown, approximations are needed. For instance, in the dual boson approach\cite{Rubtsov2012,vanLoon2014,vanLoon2016} the propagators $\chi_{r,\mathbf{q}}^{\Omega}$ are indeed corrected by a frequency-dependent $\lambda^{\Omega}$. A static, i.e., frequency- and momentum-independent, $\lambda$ correction, on the other hand, allows for a transparent physical interpretation: Considering the Ornstein-Zernike form for the physical (in our case spin) propagator [see Eq.~(\ref{equ:ornstein})] it is obvious that the $\lambda$ correction as described in Eq.~(\ref{equ:cap4lambdacorr}) corresponds just to a renormalization of the correlation length of the system as $\xi\rightarrow\xi_{\lambda}=(\xi^{-2}+A\lambda)^{-1/2}$ (hereafter, we just consider the spin channel $r \!=\!m$ and suppress the corresponding index). In fact, for a positive value of $\lambda$ Eq.~(\ref{equ:cap4lambdacorr}) describes an effective reduction of the magnetic correlation length. This corrects appropriately the overestimation of $T_N$ by DMFT.

Moreover, the $\lambda$ correction renders the susceptibility $\chi_{\mathbf{q}}^{\Omega=0}$ positive in a temperature region where DMFT has already become thermodynamically unstable, marked by a negative value of $\chi_{\mathbf{q}}^{\Omega=0}$. In this respect, the procedure of $\lambda$ corrections makes ladder D$\Gamma$A applicable down to temperatures where a value of $\lambda$ can be found so that $\chi_{\mathbf{q}}^{\lambda,\Omega}$ is larger than $0$ and at the same time Eq.~(\ref{equ:cap4sumchi}) is fulfilled. In $2d$ this can be achieved down to $T\!=\!0$ in accordance with the Mermin-Wagner theorem\cite{Mermin1966}, while in three dimensions the Moriya corrected ladder D$\Gamma$A still finds a finite temperature phase transition albeit with a reduced transition temperature compared to DMFT.

\section{The physical meaning of the three-leg vertex}
\label{app:threeleg}

In this Appendix, we want to study more explicitly the diagrammatic and physical content of the three-legs vertex $\gamma_{r,\mathbf{q}}^{\nu\Omega}$, as defined in Eq.~(\ref{equ:defgamma}). Let us stress that this object coincides exactly with the three-legs vertex which is naturally obtained in the TRILEX approach from a functional perspective\cite{Ayral2015,Ayral2016a}. Hence, an improved understanding of the general properties of this quantity is highly interesting.

It is straightforward to demonstrate that $\gamma_{r,\mathbf{q}}^{\nu\Omega}$, as defined by Eqs.~(\ref{equ:defphi}) and (\ref{equ:defgamma}), can be equivalently expressed in the following way:
\begin{equation}
 \label{equ:gammarewrite}
 \gamma_{r,\mathbf{q}}^{\nu\Omega}=\frac{\left[\chi_{0,\mathbf{q}}^{\nu\Omega}\right]^{-1}\frac{1}{\beta}\sum_{\nu'}\chi_{r,\mathbf{q}}^{\nu\nu'\Omega}}{1-U_r\chi_{r,\mathbf{q}}^{\Omega}}=\frac{1-\frac{1}{\beta}\sum_{\nu'}F_{r,\mathbf{q}}^{\nu\nu'\Omega}\chi_{0,\mathbf{q}}^{\nu'\Omega}}{1-U_r\chi_{r,\mathbf{q}}^{\Omega}},
\end{equation}
which reproduces exactly the expression obtained in the TRILEX method\cite{Ayral2015,Ayral2016a}. In the following we will decompose $\gamma_{r,\mathbf{q}}^{\nu\Omega}$ into different classes of diagrams, in order to get a better insight into the physical content of this vertex function. To this end we will decompose the full vertex $F_{r,\mathbf{q}}^{\nu\nu'\Omega}$, which is contained in the generalized susceptibility $\chi_{r,\mathbf{q}}^{\nu\nu'\Omega}$ as indicated in the second line of Eq.~(\ref{equ:bethesalp}), into three distinct diagrammatic contributions.

\begin{figure}[t!]
 \centering
 \includegraphics[width=0.5\textwidth]{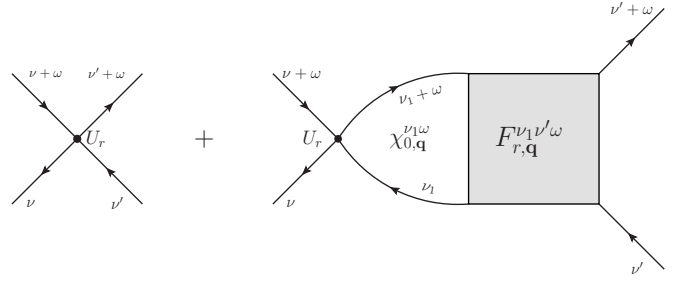}
 \caption{All Feynman diagrams  for $F_{r,\mathbf{q}}^{\nu\nu'\Omega}$ of class a. Note that this contribution to the full vertex is independent of the fermionic frequency $\nu$. }
 \label{fig:fclassa}
\end{figure}

The {\sl first class} a of diagrams for $F_{r,\mathbf{q}}^{\nu\nu'\Omega}$ is illustrated in Fig.~\ref{fig:fclassa}.
As one can see, the leftmost part of all such diagrams collapses to the bare interaction $U_r$. This renders all diagrams of class a $\nu$ independent and, hence, determines the asymptotic behavior of $\gamma_{r,\mathbf{q}}^{\nu\Omega}$ with respect to $\nu$, similar as for the self-energy obtained from $F_{r,\mathbf{q}}^{\nu\nu'\Omega}$ via the EOM [see Eq. {(\ref{equ:cap4equofmotionasympt})]. Analytically, the contribution of this diagram class, which will be denoted as $A_{r,\mathbf{q}}^{\nu\nu'\Omega}$ in the following, reads
\begin{equation}
 \label{equ:defverta}
 A_{r,\mathbf{q}}^{\nu\nu'\Omega}=U_r-\frac{U_r}{\beta}\sum_{\nu_1}\chi_{0,\mathbf{q}}^{\nu_1\Omega}F_{r,\mathbf{q}}^{\nu_1\nu'\Omega}.
\end{equation}
\begin{figure}[t!]
 \centering
 \includegraphics[width=0.5\textwidth]{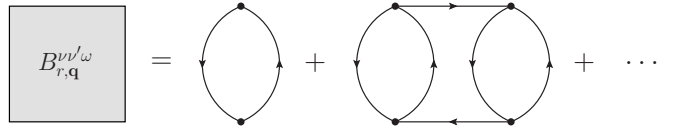}
 \caption{Examples for diagrams of class b.}
 \label{fig:fclassb}
\end{figure}
The second class of diagrams (b) is depicted in Fig.~\ref{fig:fclassb}. The defining property of this class is that the corresponding diagrams are ``irreducible in the interaction $U_r$'', i.e., they cannot be split into two parts by removing a bare interaction $U_r$ from a diagram. Furthermore, we exclude all diagrams which are already in class a, i.e., all diagrams whose leftmost part collapses to the bare interaction $U_r$. It is clear that the sum of all such diagrams (explicitly: $B_{r,\mathbf{q}}^{\nu\nu'\Omega}$) decays with the fermionic Matsubara frequency $\nu$ and, hence, does not contribute to the constant asymptotics of the three-leg vertex $\gamma$. 

\begin{figure}[t!]
 \centering
 \includegraphics[width=0.5\textwidth]{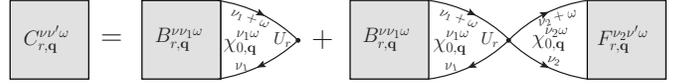}
 \caption{Bethe-Salpeter-like equation for constructing diagrams of type c.}
 \label{fig:fclassc}
\end{figure}

The third class of diagrams (c) contains all diagrams which are not yet considered in class a or b. Hence, these contributions are ``reducible in the interaction $U_r$'', and their leftmost part does not collapse to the bare interaction. Similar to the concept of reducibility in terms of Green's functions \cite{Rohringer2011} such a class of diagrams $C_{r,\mathbf{q}}^{\nu\nu'\Omega}$ can be expressed via the irreducible vertex $B_{r,\mathbf{q}}^{\nu\nu'\Omega}$ and the full vertex $F_{r,\mathbf{q}}^{\nu\nu'\Omega}$ by means of a Bethe-Salpeter-like equation. This equation is depicted diagrammatically in Fig.~\ref{fig:fclassc}. Formally this term reads
\begin{equation}
 \label{equ:defvertc}
 C_{r,\mathbf{q}}^{\nu\nu'\Omega}=\frac{U_r}{\beta^2}\sum_{\nu_1\nu_2}B_{r,\mathbf{q}}^{\nu\nu_1\Omega}\left[-\beta\chi_{0,\mathbf{q}}^{\nu_1\Omega}+\chi_{0,\mathbf{q}}^{\nu_1\Omega}\chi_{0,\mathbf{q}}^{\nu_2\nu'\Omega}F_{r,\mathbf{q}}^{\nu_2\nu'\Omega}\right].
\end{equation}
We can now proceed by inserting the three contributions a, b and c to the full vertex $F_{r,\mathbf{q}}^{\nu\nu'\Omega}$ into Eq.~(\ref{equ:gammarewrite}) for the three-leg vertex $\gamma_{r,\mathbf{q}}^{\nu\Omega}$. Considering that $\frac{1}{\beta^2}\sum_{\nu\nu'}\beta\chi_{0,\mathbf{q}}^{\nu\Omega}\delta_{\nu\nu'}-\chi_{0,\mathbf{q}}^{\nu\Omega}F_{r,\mathbf{q}}^{\nu\nu'\Omega}\chi_{0,\mathbf{q}}^{\nu'\Omega}\!=\!\frac{1}{\beta^2}\sum_{\nu_1\nu_2}\chi_{r,\mathbf{q}}^{\nu\nu'\Omega}\!\equiv\!\chi_{r,\mathbf{q}}^{\Omega}$ one observes that the denominator in Eq.~(\ref{equ:gammarewrite}) is canceled. Thus, the following expression for $\gamma_{r,\mathbf{q}}^{\nu\Omega}$ remains:
\begin{equation}
 \label{equ:gammaresult}
 \gamma_{r,\mathbf{q}}^{\nu\Omega}=\underset{\text{a}}{\underbrace{1}}-\underset{\text{b}+\text{c}}{\underbrace{\frac{1}{\beta}\sum_{\nu'}B_{r,\mathbf{q}}^{\nu\nu'\Omega}}}.
 \end{equation}
From this equation one can infer interesting interpretations. First of all, it is clear that the three-leg vertex approaches $1$ as $\nu\!\rightarrow\!\infty$ as it has been already empirically observed in the numerical results of Ref.~\onlinecite{Katanin2009}. Let us emphasize that this behavior can be found in the exact solution as well as in the $\gamma_{r,\mathbf{q}}^{\nu\Omega}$ constructed within the ladder approximation of the D$\Gamma$A. Moreover, the above considerations show that the three-legs vertex {\sl does not contain} any physical susceptibility $\chi_{r,\mathbf{q}}^{\Omega}$: This happens because $B_{r,\mathbf{q}}^{\nu\nu'\Omega}$ is constructed as the set of diagrams which do not collapse to a bare $U_r$. Instead, the physical susceptibilities are built from precisely such collapsing diagrams.

This inspires the following considerations: Since the largest nonlocal contributions to the self-energy from ladder D$\Gamma$A are due to the susceptibilities $\chi_{r,\mathbf{q}}^{\Omega}$ (in our case in particular the spin susceptibility $\chi_{m,\mathbf{q}}^{\Omega}$), it can be expected that the influence of nonlocal correlations on $\gamma_{r,\mathbf{q}}^{\nu\Omega}$ is rather moderate. For this reason, and due to the already correct asymptotics of $\gamma_{r,\mathbf{q}}^{\nu\Omega}$ we think that it is justified to perform $\lambda$ corrections for the physical susceptibilities only while the three-leg vertex remains unchanged.

Finally, the same argument can provide support of the approximation made in TRIXLEX, i.e., to entirely neglect the momentum dependence of the three-legs vertex $\gamma$.

\section{Condition for the existence of a dip in the spectral function}
\label{app:spectraldip}

\begin{figure*}
 \centering
  \includegraphics{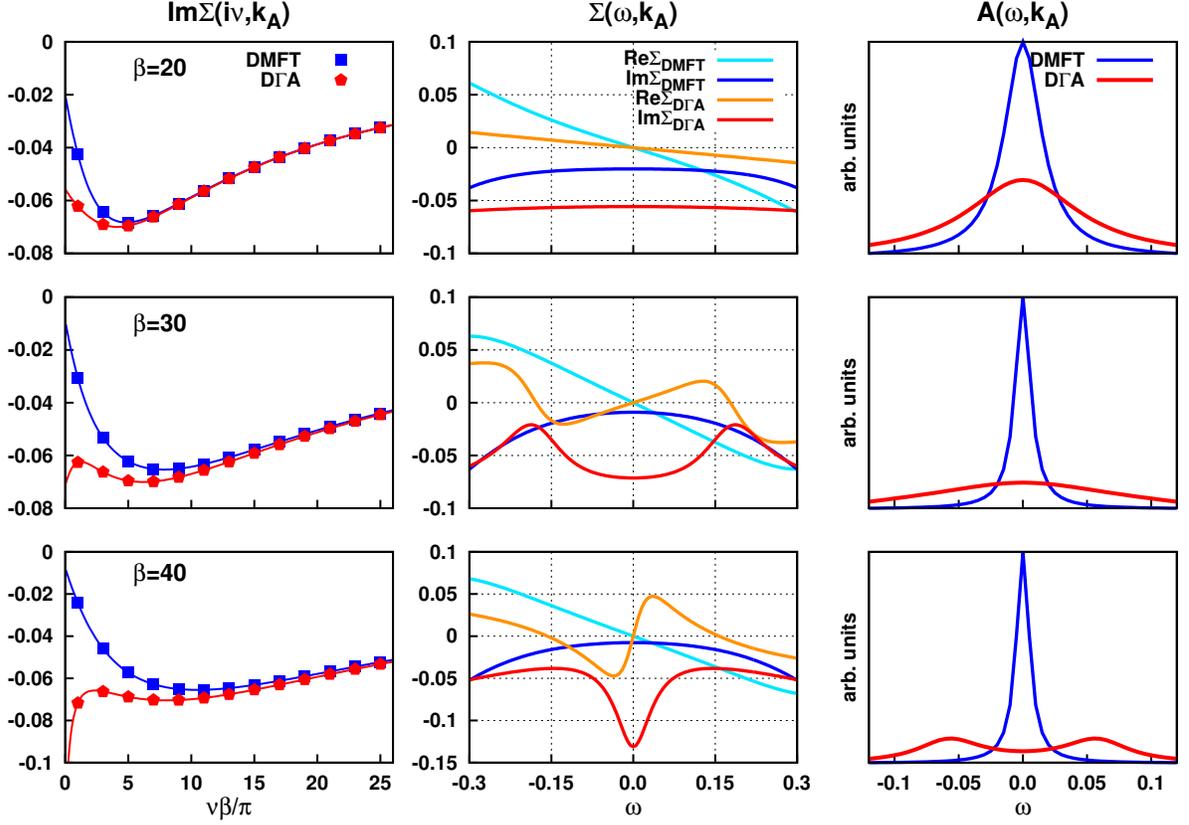}
 \caption{Self-energies and spectral functions for the antinodal point of the $2d$ Hubbard model at half filling [$\mathbf{k}_A\!=\!(\pi,0)$]. The conventions adopted are the same as those of Fig.~\ref{fig:spectra_nodal} in Sec.~\ref{subsec:sigmaspectra}.}
 \label{fig:spectra_antinodal}
\end{figure*}

In this section we provide some details about the derivation of inequality (\ref{equ:dipcondition}) of the main text. To this end we start from the explicit expression of the spectral function $A(\omega,\mathbf{k})$ in terms of the retarded self-energy $\Sigma(\omega,\mathbf{k})$ of the system:
\begin{equation}
 \label{equ:defa}
 A(\omega,\mathbf{k})=-\frac{1}{\pi}\frac{N(\omega,\mathbf{k})}{D(\omega,\mathbf{k})},
\end{equation}
where
\begin{subequations}
\label{subequ:defND}
\begin{align}
 &N(\omega,\mathbf{k})=\mbox{Im}\Sigma(\omega,\mathbf{k}), \label{equ:defN} \\
 &D(\omega,\mathbf{k})=\left[\omega+\mu-\varepsilon_{\mathbf{k}}-\mbox{Re}\Sigma(\omega,\mathbf{k})\right]^2+\left[\mbox{Im}\Sigma(\omega,\mathbf{k})\right]^2 \label{equ:defD}.
\end{align}
\end{subequations}
Expanding $\Sigma(\omega,\mathbf{k})$ around $\omega=0$ yields for the half-filled system ($\mu\!=\!U/2$) for $\mathbf{k}$ points on the Fermi surface ($\varepsilon_{\mathbf{k}}\!=\!0$):
\begin{subequations}
\label{subequ:defsigma0}
\begin{align}
 &-\mu+\mbox{Re}\Sigma(\omega=0,\mathbf{k})=0 &\mbox{Im}\Sigma(\omega=0,\mathbf{k})=-\gamma_{\mathbf{k}} \label{equ:defsigma00}\\
 &\left[\frac{\partial\mbox{Re}\Sigma}{\partial\omega}\right]_{\omega=0}=-\alpha_{\mathbf{k}} &\left[\frac{\partial\mbox{Im}\Sigma}{\partial\omega}\right]_{\omega=0}=0. \label{equ:defsigma10}
\end{align}
\end{subequations}
Differentiating expression (\ref{equ:defa}) once with respect to $\omega$ at $\omega=0$ yields $0$ when considering Eqs.~(\ref{subequ:defND}) and (\ref{subequ:defsigma0}). Hence, $A(\omega,\mathbf{k})$ has an extremum at $\omega=0$. The condition that this extremum is a minimum requires the second derivative of the spectral function being positive. A straightforward differentiation of Eq.~(\ref{equ:defa}) under consideration of Eqs.~(\ref{subequ:defND}) and Eqs. (\ref{subequ:defsigma0}) yields relation (\ref{equ:dipcondition}). 

For the spectral function at the antinodal point $\mathbf{k}_A$ shown in Fig.~\ref{fig:spectra_antinodal} this dip condition is fulfilled for $\beta\!=\!40$ while it is violated at $\beta\!=\!35$. Hence, similar as for the nodal point discussed in Sec.~\ref{subsec:sigmaspectra}, in the latter case we observe a situation where the self-energy already exhibits non-Fermi-liquid behavior while the spectrum displays still a peak at the Fermi level.

\section{Analytical approximation for D$\Gamma$A self-energy}
\label{app:dgaanalyt}

In this Appendix we outline explicitly the approximation steps which have been applied to the ladder-D$\Gamma$A self-energy in Eq. (\ref{equ:EOMrewrite}) in order to obtain the corresponding simplified expressions in Eqs. (\ref{subequ:EOMapprox}):

(i) First of all, we neglect the part of the equation describing charge fluctuations. This is justified since, at half filling, spin fluctuations dominate the physics. 

(ii) The three-leg vertex $\gamma_{m,\mathbf{q}}^{\nu\Omega}$ can be replaced by a constant, as it does not contain any spin ladders and, hence, will not be enhanced when approaching the AF phase transition upon lowering the temperature. This issue and the three-leg vertex itself have been discussed in more detail in Appendix \ref{app:threeleg}. 

(iii) We will consider $\chi_{m,\mathbf{q}}^{\Omega}$ only at its maximal value, which is assumed for $\Omega\!=\!0$ and at $\mathbf{q}=\mathbf{\Pi}$, with $\mathbf{\Pi}\!=\!(\pi,\pi)$ in $2d$ and $\mathbf{\Pi}\!=\!(\pi,\pi,\pi)$ in $3d$, respectively. This corresponds to taking into account only classical ($\Omega\!=\!0$) spin fluctuations around the predominant antiferromagnetic wave vector. This way $\chi_{m,\mathbf{q}}^{\Omega}$ can be represented analytically by a simple Ornstein-Zernike-like form:
\begin{equation}
 \label{equ:ornstein}
 \chi_{m,\mathbf{q}}^{\Omega=0}=\frac{A}{(\mathbf{q}-\mathbf{\Pi})^2+\xi^{-2}(T)},
\end{equation}
where $\xi(T)$ is the correlation length of the system and $A$ is a constant assumed to be approximately temperature independent in the following. Moreover, we will perform for convenience a shift of the integration variable $\mathbf{q}$ as $\mathbf{q}\rightarrow\mathbf{q}+\mathbf{\Pi}$ in Eq.~(\ref{equ:EOMrewrite}), i.e., in the new coordinate system the most relevant contributions due to antiferromagnetic correlations correspond to $\mathbf{q}\equiv0$.

(iv) In order to perform analytical calculations, we simplify also the DMFT Green's function $G(i\nu+i\Omega,\mathbf{k}+\mathbf{q}+\mathbf{\Pi})$ in the D$\Gamma$A equation (\ref{equ:EOMrewrite}). As for the magnetic susceptibility, we will of course restrict ourselves to the contribution $\Omega=0$. Moreover, as we are interested in the D$\Gamma$A self-energy on the real axis, we perform the analytic continuation $i\nu\rightarrow\omega+i\delta$ and expand the DMFT self-energy in real frequencies $\omega$ around  $\omega=0$:
\begin{equation}
 \label{equ:sigmaexpand}
 \Sigma(\omega)=-i\gamma-\alpha\omega+O(\omega^2),
\end{equation}
Here, we have taken into account only terms up to the first order in $\omega$, as we are interested in the D$\Gamma$A low-energy coefficients $\alpha_{\mathbf{k}}(T)$ and $\gamma_{\mathbf{k}}(T)$. Accordingly, $G(\omega,\mathbf{k}+\mathbf{q}+\mathbf{\Pi})$ can be written as
\begin{equation}
\label{equ:gapprox}
 G(\omega,\mathbf{k}+\mathbf{q}+\mathbf{\Pi})\propto\frac{1}{\omega-\widetilde{\varepsilon}_{\mathbf{k}+\mathbf{q}+\mathbf{\Pi}}+i\Gamma},
\end{equation}
where $\Gamma=\gamma/(1+\alpha)$ is the renormalized quasiparticle scattering factor (or inverse quasiparticle lifetime) of DMFT and $\widetilde{\varepsilon}_{\mathbf{k}}=\varepsilon_{\mathbf{k}}/(1+\alpha)$ is the renormalized dispersion. Let us recall that $\Gamma$ decreases with small temperatures as $T^2$, i.e., 
\begin{equation}
 \label{equ:tempdepgammadmft}
 \Gamma(T)\equiv\frac{\gamma(T)}{1+\alpha(T)}=CT^2+O(T^4),
\end{equation}
since $\gamma\sim T^2$ for low temperatures, while the renormalization factor $0<(1+\alpha)^{-1}<1$ remains finite even at $T=0$ and can be, hence, approximated as constant at low temperatures.  

(v) A final approximation is applicable to the (renormalized) dispersion $\widetilde{\varepsilon}_{\mathbf{k}+\mathbf{q}+\mathbf{\Pi}}$. As discussed above, the most relevant contribution to the $\mathbf{q}$ integral in Eq. (\ref{equ:EOMrewrite}) originates from the wave vector $\mathbf{q}=0$ (after the before mentioned shift of $\mathbf{q}$ by $\mathbf{\Pi}$). Hence, we perform a Taylor series expansion of $\widetilde{\varepsilon}_{\mathbf{k}+\mathbf{q}+\mathbf{\Pi}}$ around this point:
\begin{equation}
\label{equ:expandeps}
 \widetilde{\varepsilon}_{\mathbf{k}+\mathbf{\Pi}+\mathbf{q}}\sim\widetilde{\varepsilon}_{\mathbf{k}+\mathbf{\Pi}}+\underset{-\mathbf{v}_{\mathbf{k}}}{\underbrace{\frac{\partial\widetilde{\varepsilon}}{\partial\mathbf{k}}(\mathbf{k}+\mathbf{\Pi})}}\mathbf{q}+\frac{1}{2}\mathbf{q}\underset{-2/m_{\mathbf{k}}}{\underbrace{\frac{\partial^2\widetilde{\varepsilon}}{\partial\mathbf{k}^2}(\mathbf{k}+\mathbf{\Pi})}}\mathbf{q},
\end{equation}
where the last term is taken into account only when the second one, i.e., the Fermi velocity, vanishes (which is indeed the case for the antinodal point in $2d$). The first term on the right hand side of Eq.~(\ref{equ:expandeps}) is always $0$ for $\mathbf{k}$ vectors located on the (perfectly nested) Fermi surface. As the (simplified) magnetic susceptibility given in Eq.~(\ref{equ:ornstein}) is rotational invariant with respect to $\mathbf{q}$, we can rotate the coordinate system in the $\mathbf{q}$ integral of Eq.~(\ref{equ:EOMrewrite}) in such a way, that the Fermi velocity (if it is not $0$) points into the $q_x$ direction.  The second derivative of the renormalized dispersion in Eq. (\ref{equ:expandeps}) is given by a diagonal matrix with the entries $1$ and $-1$ and, hence, the last term in this equation can be written as $(q_x^2-q_y^2) /m_{\mathbf{k}}$ with $m_{\mathbf{k}}$ being the mass of the quasiparticle excitations renormalized by DMFT. Considering now the simplifications (i)-(v) for Eq.~(\ref{equ:EOMrewrite}) leads straightforwardly to Eqs. (\ref{subequ:EOMapprox}) in the main text of the paper.

Let us finally point out that the above approximations for the integrands of the $\mathbf{q}$ integral determining the D$\Gamma$A self-energy via Eq.~(\ref{equ:EOMrewrite}) are in principle valid only in a tiny region around $\mathbf{q}=(0,0,0)$ (corresponding to the antiferromagnetic vector $\mathbf{q}=\mathbf{\Pi}$ before the shift of integration variables). However, in the parameter regime where the most relevant contribution to these integrals stems from critical antiferromagnetic fluctuations, the leading order terms in the low-$T$ behavior of $\gamma_{\mathbf{k}}(T)$ and $\alpha_{\mathbf{k}}(T)$ will {\sl not} depend qualitatively on the limits of the integrals in Eqs.~(\ref{subequ:EOMapprox}) which can be, hence, chosen for convenience. 

\section{Integrals for calculating Fermi liquid parameters}
\label{app:qintegrals}

In this section, we provide some further details concerning the evaluation of the integrals (\ref{equ:EOMapproxnodal}) and (\ref{equ:EOMapproxantinodal}) for calculating the low energy coefficients of the self-energy $\gamma_{\mathbf{k}}(T)$ and $\alpha_{\mathbf{k}}(T)$. The actual calculations follow those of Ref.~\onlinecite{Vilk1997}. We first define explicitly the prefactor $C_{\mathbf{k}}^{\omega}$. From the D$\Gamma$A equation (\ref{equ:EOMrewrite}) it follows that
\begin{equation}
 \label{equ:defCqomega}
 C_{\mathbf{k}}^{\omega}=\frac{1}{(2\pi)^2}\frac{3}{2}U^2A\gamma_{m,\mathbf{q*}}^{\omega,\Omega=0},
\end{equation} 
where $A$ is a constant defined by the spin susceptibility in Eq.~(\ref{equ:ornstein}), and $\gamma_{m,\mathbf{q*}}^{\omega,\Omega=0}$ is the (analytically continued) three-leg vertex defined in Eq.~(\ref{equ:defgamma}), evaluated for the bosonic Matsubara frequency $\Omega\!=\!0$ and a momentum $\mathbf{q}^*$ in the Brillouin zone which is determined by applying the mean-value theorem of integral calculus to the $\mathbf{q}$ integral in Eq. (\ref{equ:EOMrewrite}). Note that we have also included the $2d$ normalization factor $1/(2\pi)^2$ in this prefactor. Hence, for $3d$ another factor $1/(2\pi)$ has to be considered explicitly in the calculation.

\subsubsection{2d, Nodal point}
In order to evaluate Eq.~(\ref{equ:EOMapproxnodal}) for $d\!=\!2$ we first perform the integral over $q_y$. As for integration extremes, we consider $[-\pi,\pi]$, i.e., we perform the integration over the Brillouin zone in the $y$ direction. As discussed at the end of Appendix~\ref{app:dgaanalyt} the choice of the extremes of the integral does not alter qualitatively the results, because the integrals are mainly controlled by the singularity of the integrand at $\mathbf{q}\!=\!(0,0)$. The result of this integration, thus, yields:
\begin{equation}
 \label{equ:intnodqy}
 \int_{-\pi}^{\pi}dq_y\frac{1}{q_y^2+q_x^2+\xi^{-2}}=\frac{2\arctan\left[\frac{\pi}{\sqrt{q_x^2+\xi^{-2}}}\right]}{\sqrt{q_x^2+\xi^{-2}}}.
\end{equation} 
This result can be then used in the $q_x$ integral of Eq.~(\ref{equ:EOMapproxnodal}) where we perform the change of variables $x\!=\!\xi q_x$. Considering the $\arctan$ in Eq.~(\ref{equ:intnodqy}) we note that $\arctan(\pi\xi/\sqrt{1+x^2})\!=\!\pi/2-\arctan[\sqrt{1+x^2}/(\pi \xi)]$. As the second term vanishes when $\xi\!\rightarrow\!\infty$, we can neglect this contribution in the previous relation and keep only the term $\pi/2$. Finally, the new limits of the $x$ integral are given by $[-\pi\xi,\pi\xi]$, which can be extended to $\infty$ as discussed before. The remaining $x$ integral hence reads
\begin{equation}
 \label{equ:intnodqx}
 \Sigma(\omega,\mathbf{k}_N)\cong \frac{\pi C_{\mathbf{k}}^{\omega}}{v_{\mathbf{k}_N}}T\xi\int_{-\infty}^{+\infty}dx\frac{1}{\sqrt{1+x^2}}\frac{1}{x+B_{\mathbf{k}_N}^{\omega}},
\end{equation}
where
\begin{equation}
 \label{equ:defB}
 B_{\mathbf{k}_N}^{\omega}=\frac{\xi}{v_{\mathbf{k}_N}}(\omega+i\Gamma),
\end{equation} 
i.e., $B_{\mathbf{k}_N}^{\omega=0}=ib_{\mathbf{k}_N}$ where the latter quantity is defined in Eq.~(\ref{equ:defak}). Integral (\ref{equ:intnodqx}) is convergent and can be performed explicitly. The final result for $\Sigma(\omega,\mathbf{k}_N)$ is then given by
\begin{align}
 \label{equ:sigmanodfinal}
 \Sigma(\omega,\mathbf{k}_N)=&\frac{2\pi C_{\mathbf{k}}^{\omega}}{v_{\mathbf{k}_N}}T\xi\frac{1}{\sqrt{(B_{\mathbf{k}_N}^{\omega})^2+1}}\nonumber\\&\times\log\left[-i\left(B_{\mathbf{k}_N}^{\omega}+\sqrt{(B_{\mathbf{k}_N}^{\omega})^2+1}\right)\right].
\end{align}
Evaluating this expression for $\omega\!=\!0$ and extracting its negative imaginary part yields the expression for $\gamma_{\mathbf{k}_N}$ as given in Eq.~(\ref{equ:gamma2dnodal}), when considering the definition $2\pi C_{\mathbf{k}}^{\omega=0}/v_{\mathbf{k}_N}\!\equiv\! C_{\mathbf{k}_N}$.  As for the calculation of $\alpha_{\mathbf{k}_N}$ we have to differentiate Eq.~(\ref{equ:sigmanodfinal}) with respect to $\omega$. In principle, we would have to consider a contribution from the $\omega$ derivative of $C_{\mathbf{k}_N}^{\omega}$. This can be neglected, since the remaining part is purely imaginary and, hence, does not contribute to $\alpha_{\mathbf{k}_N}$. Interestingly, as the remaining part of the self-energy depends on $\omega$ only via $B_{\mathbf{k}_N}^{\omega}$, $\alpha_{\mathbf{k}_N}$ can be calculated for the nodal point directly from $\gamma_{\mathbf{k}_N}$ as
\begin{equation}
 \label{equ:alphafromgamma}
 \alpha_{\mathbf{k}_N}=\frac{\partial}{\partial\Gamma}\gamma_{\mathbf{k}_N}.
\end{equation} 

\subsubsection{2d, Antinodal point}
For the antinodal point we have to evaluate the integral in Eq.~(\ref{equ:EOMapproxantinodal}) for $d\!=\!2$. This is more complicated than the corresponding calculation for the nodal case as both $q_x$ and $q_y$ appear squared in both parts of the integrand. On the other hand, this allows us to extend the domain of integration to the entire two-dimensional $\mathbf{k}$ space and perform the integration in polar coordinates. The integration over $q\!=\!\lvert\mathbf{q}\rvert$ can be thus easily done analytically and -after rearranging of the $\varphi$ integral- yields
\begin{equation}
 \label{equ:varphiintegrate}
\Sigma(\omega,\mathbf{k}_A)=2C_{\mathbf{k}}^{\omega}m_{\mathbf{k}_A}T\xi^2\int_0^{\pi}\!d\varphi\;\frac{\log\left[\frac{B_{\mathbf{k}_A}^{\omega}}{\cos\varphi}\right]}{B_{\mathbf{k}_A}^{\omega}-\cos\varphi},
\end{equation}
where $B_{\mathbf{k}_A}^{\omega}$ is defined as
\begin{equation}
 \label{equ:defBantinode}
 B_{\mathbf{k}_A}^{\omega}=m_{\mathbf{k}_A}\xi^2(\omega+i\Gamma),
\end{equation}
i.e., $B_{\mathbf{k}_A}^{\omega=0}=ib_{\mathbf{k}_A}$ as defined in Eq.~(\ref{equ:defbk}). One can clearly see that the integral in Eq.~(\ref{equ:varphiintegrate}) becomes divergent for $\Gamma\!\equiv\! 0$. This illustrates the importance of of the quasiparticle damping factor of DMFT for $\mathbf{k}$ points at the van Hove singularity. After splitting up the logarithm in Eq. (\ref{equ:varphiintegrate}) as $\log(B_{\mathbf{k}}^{\omega}/\cos\varphi)\!=\!\log(B_{\mathbf{k}}^{\omega})\!-\!\log(\cos\varphi)$ the first integral can be solved exactly. Hence, we obtain
\begin{align}
 \label{equ:defsigmaantinodefinal}
 \Sigma(\omega,&\mathbf{k}_A)=2C_{\mathbf{k}}^{\omega}\pi m_{\mathbf{k}}T\xi^2\nonumber\\&\times\left[\frac{\log (B_{\mathbf{k}}^{\omega})}{\sqrt{(B_{\mathbf{k}}^{\omega})^2-1}}-\frac{1}{\pi}\int_0^{\pi}\!d\varphi\;\frac{\log(\cos\varphi)}{B_{\mathbf{k}}^{\omega}-\cos\varphi}\right].
\end{align}
Evaluating Eq.~(\ref{equ:defsigmaantinodefinal}) for $\omega\!=\!0$ and extracting its negative imaginary part yields the expression for $\gamma_{\mathbf{k}_A}$ as given in Eq.~(\ref{equ:fl2antinodalgamma}), when considering the definition $2\pi m_{\mathbf{k}_A} C_{\mathbf{k}}^{\omega=0}\!\equiv\! C_{\mathbf{k}_A}$. Finally, $\alpha_{\mathbf{k}_A}$ can be obtained from Eq.~(\ref{equ:defsigmaantinodefinal}) by differentiating its negative real part with respect to $\omega$ at $\omega\!=\!0$. 

\subsubsection{3d}
For the three-dimensional system we have to evaluate the integral in Eq. (\ref{equ:EOMapproxnodal}) for $d\!=\!3$. First, we perform the integration of $q_y$ and $q_z$ in polar coordinates where we cut off the radial integral at $q\!=\!\lvert\mathbf{q}\rvert\!=\!\pi$ (corresponding to the border of the Brillouin zone). The result is
\begin{equation}
\label{equ:3dintqyqz}
 \int dq_ydq_z\frac{1}{\mathbf{q}^2+\xi^{-2}}=\pi\log\left[\frac{\pi^2+\xi^{-2}+q_x^2}{\xi^{-2}+q_x^2}\right].
\end{equation} 
The remaining $q_x$ integral is then extended to $q_x\!=\!\pm\infty$ by adding an infinitesimal convergence factor. This enables us to perform the integral by means of the residue theorem considering that $\log(q_x^2+C^2)\!=\!\log(q_x+iC)+\log(q_x-iC)$ for an arbitrary number $C$: We can then close the integration path in the complex half-plane where no logarithmic singularity is present. The result is
\begin{equation}
 \label{equ:defsigma3dfinal}
 \Sigma(\omega,\mathbf{k})=-i\frac{\pi C_{\mathbf{k}}^{\omega}}{v_{\mathbf{k}}}T\log\left[\frac{\sqrt{1+(\pi\xi)^2}-iB_{\mathbf{k}}^{\omega}}{1-iB_{\mathbf{k}}^{\omega}}\right],
\end{equation} 
where $B_{\mathbf{k}}^{\omega}$ is the same as in Eq.~(\ref{equ:defB}). Evaluating Eq. (\ref{equ:defsigma3dfinal}) and its first derivative w.r.t. $\omega$ at $\omega\!=\!0$ and considering $\pi C_{\mathbf{k}}^{\omega=0}/v_{\mathbf{k}}\!\equiv\! C_{\mathbf{k}}$ yields the results for $\gamma_{\mathbf{k}}$ and $\alpha_{\mathbf{k}}$, respectively, which are presented in Eqs.~(\ref{subequ:fl3d}) of the main text.

\section{Energy distributions - analytic derivations and further results}
\label{app:atomiclimit}

\subsection{Energy distribution for the AL self-energy}
\label{appsub:al}

In this section we report the analytical results for $n(\varepsilon)$ obtained by assuming the AL self-energy given in Eq.~(\ref{equ:alsigma}) in the calculation:
\begin{equation}
 \label{equ:alcalc}
 n_{\text{AL}}(\varepsilon)=\frac{2}{\beta}\sum_{\nu\mathbf{k}}\delta(\varepsilon-\varepsilon_{\mathbf{k}})\frac{1}{i\nu-\varepsilon_{\mathbf{k}}-\frac{U^2}{4i\nu}}.
\end{equation}
Note that this corresponds to an approximation where just the self-energy of the system is replaced by the corresponding atomic limit one rather than taking the atomic limit $t\rightarrow 0$ itself. As for $U=4.0$ this self-energy is, however, a good approximation for the corresponding DMFT self-energy (as we have verified for our numerical data), Eq.~(\ref{equ:alcalc}) represents indeed a reasonable approximation for $n(\varepsilon)$ at such large values of $U$.

By exploiting the definition of the DOS [$\sum_{\mathbf{k}}\delta(\varepsilon-\varepsilon_{\mathbf{k}})\!=\!D(\varepsilon)$] we can rewrite Eq.~(\ref{equ:alcalc}) as
\begin{equation}
 \label{equ:alcalc1}
  n_{\text{AL}}(\varepsilon)=2D(\varepsilon)\frac{1}{\beta}\sum_{\nu}\frac{1}{i\nu-\varepsilon-\frac{U^2}{4i\nu}}.
\end{equation}
The $\nu$ sum in Eq.~(\ref{equ:alcalc1}) can be now performed explicitly by means of standard techniques\cite{Mahan2000} yielding:
\begin{align}
 \label{equ:alresult}
 n_{\text{AL}}(\varepsilon)=&2D(\varepsilon)\frac{1}{\sqrt{\varepsilon^2+U^2}}\left[X_+(\varepsilon)f\left(X_+(\varepsilon)\right)\right.\nonumber\\&\left.-X_-(\varepsilon)f\left(X_-(\varepsilon)\right)\right],
\end{align}
where $f(x)\!=\!(1+e^{\beta x})^{-1}$ denotes the Fermi function and
\begin{equation}
 \label{equ:defX}
 X_{\pm}(\varepsilon)=\frac{1}{2}\left(\varepsilon\pm\sqrt{\varepsilon^2+U^2}\right).
\end{equation}
As $X_-(\varepsilon)\!<\!0$ and $X_+(\varepsilon)\!>\!0$, at $T\!=\!0$ we can replace the corresponding Fermi functions by $1$ and $0$, respectively, which leads exactly to the result given in Eq.~(\ref{equ:nal}). 

\subsection{Further numerical results for $n(\varepsilon)$}
\label{appsub:furtherres}

\begin{figure*}[t!]
\centering
\begin{tabular}{cc}
  $d=3$, $U=0.75$, $\beta=35$ & $d=3$, $U=4.0$, $\beta=24$ \\
 \includegraphics[width=0.45\textwidth]{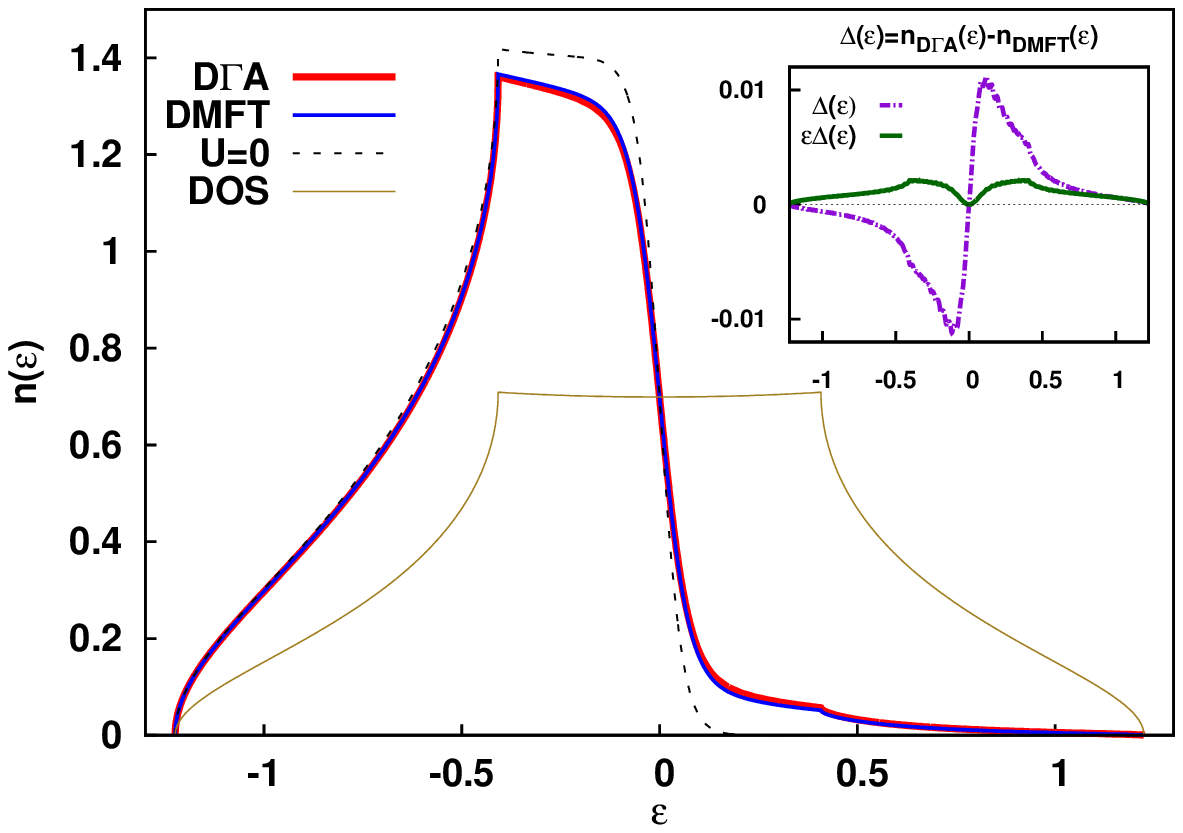}&
 \includegraphics[width=0.45\textwidth]{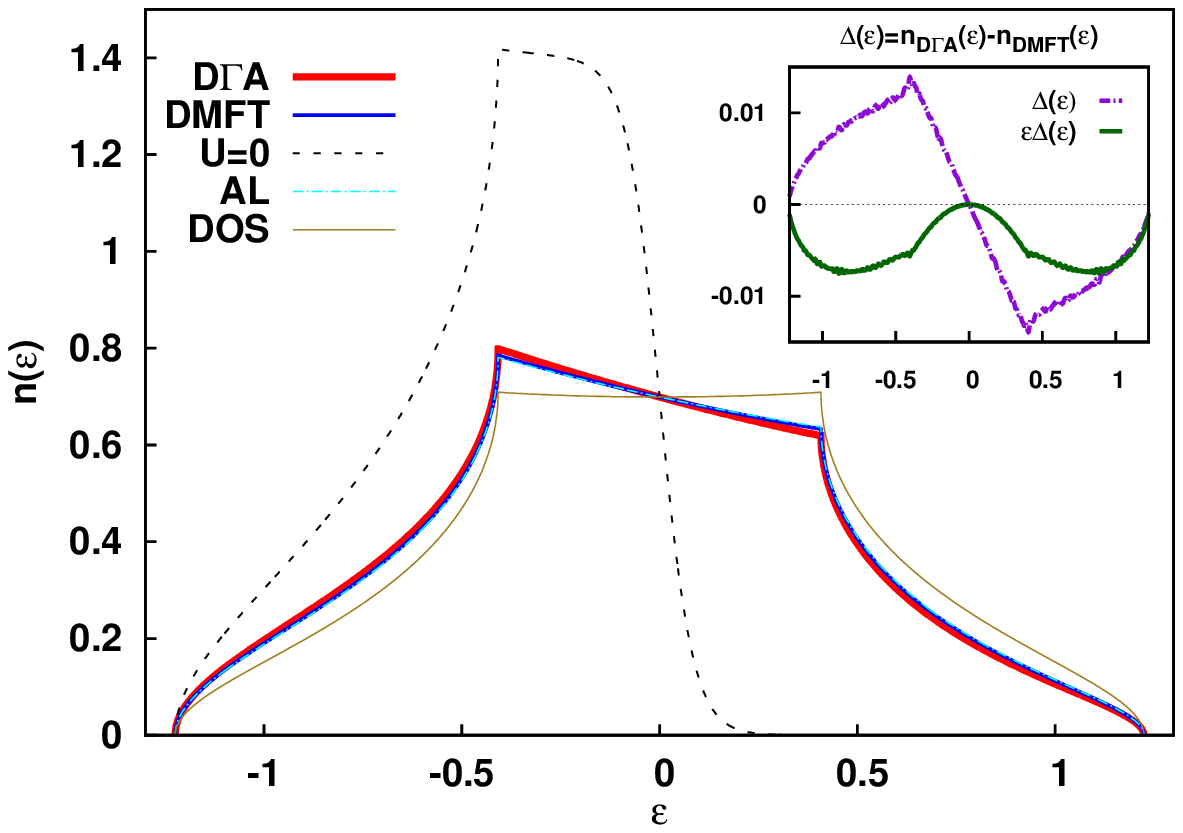}
\end{tabular}
\caption{Energy distributions $n(\varepsilon)$ for the three-dimensional Hubbard model at two different values of the interaction parameter $U$ for temperatures slightly above the D$\Gamma$A phase transition to the antiferromagnetically ordered phase. In the inset the difference $\Delta(\varepsilon)$ between the energy distributions of DMFT and D$\Gamma$A as well as the contribution $\varepsilon\Delta(\varepsilon)$ to the corresponding difference of the kinetic energies are shown.}
\label{fig:occup3d}
\end{figure*}
\begin{figure*}[th]
\centering
\begin{tabular}{cc}
  $$d=2$, U=2.375$, $\beta=80$, metallic & $d=2$, $U=2.375$, $\beta=80$, insulating \\
 \includegraphics[width=0.45\textwidth]{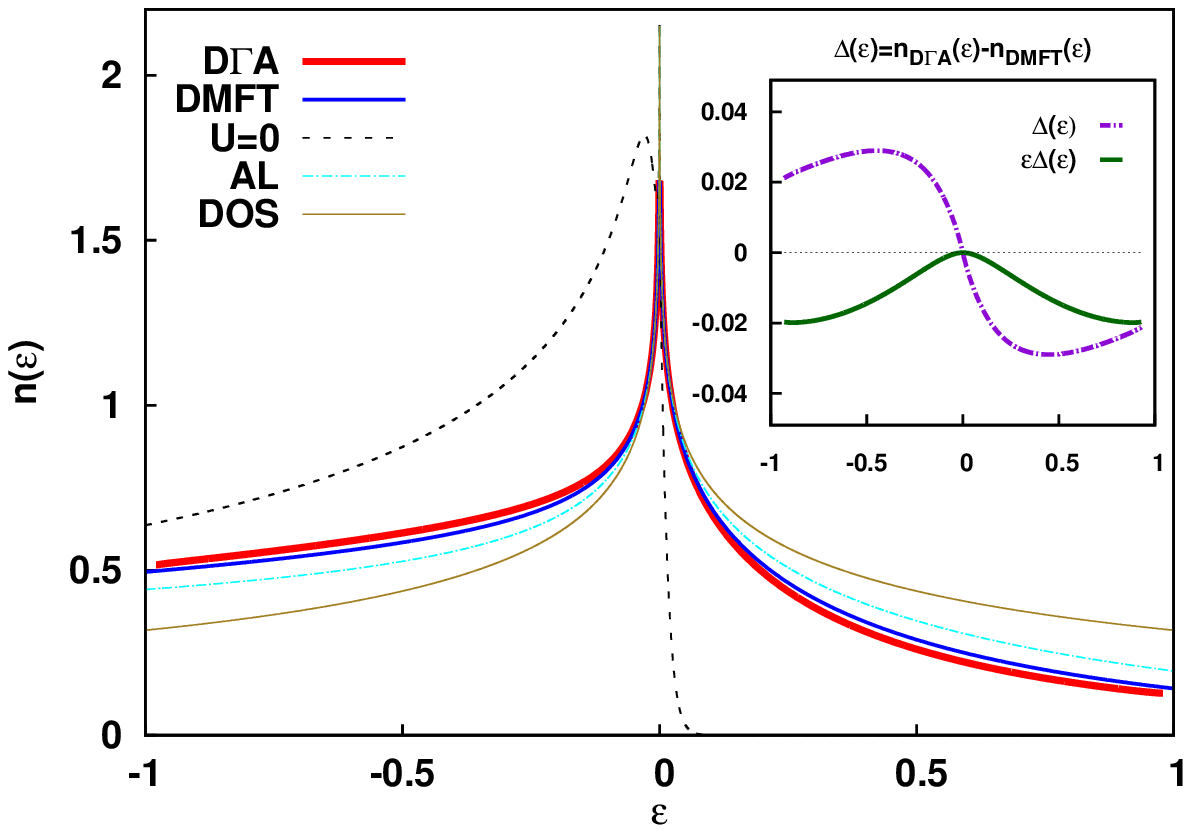}&
 \includegraphics[width=0.45\textwidth]{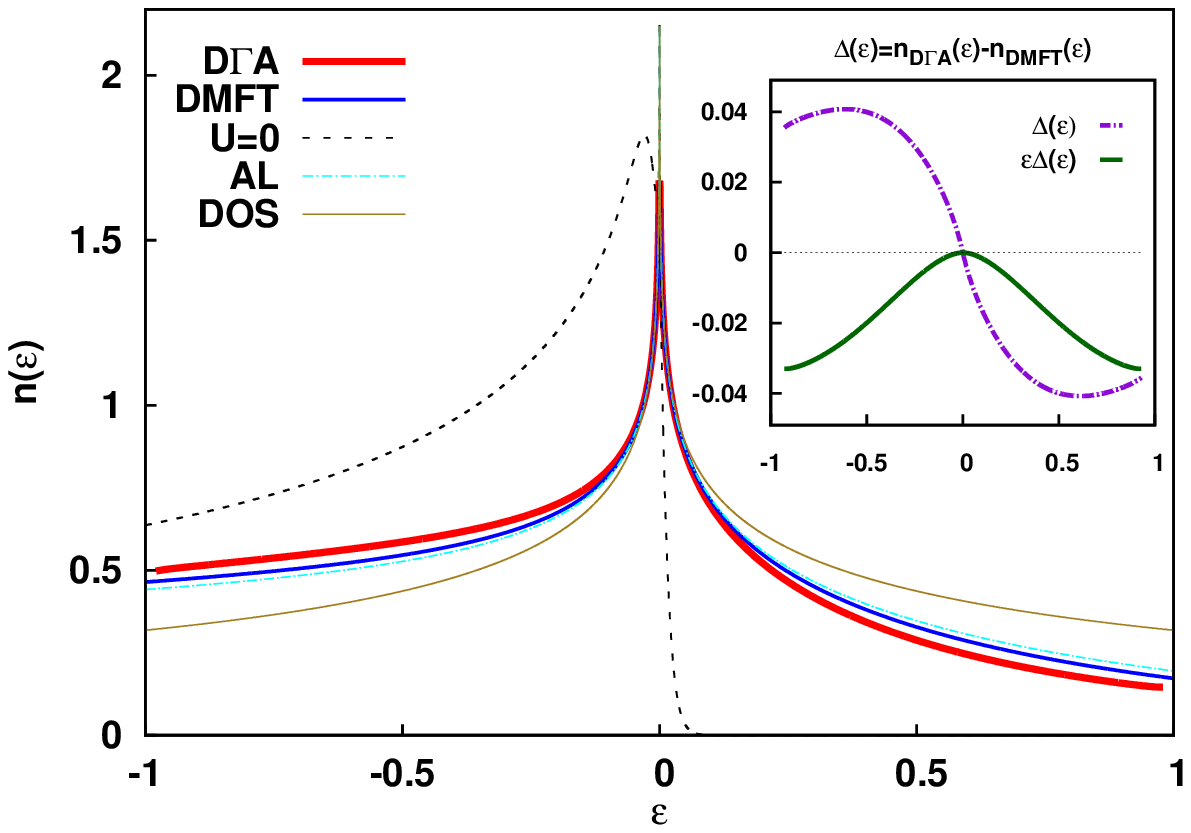}
\end{tabular}
\caption{Energy distributions $n(\varepsilon)$ for the two-dimensional Hubbard model at $U=2.375$ for $\beta=80$ corresponding to the coexistence region between an metallic and an insulating state in DMFT. In the inset the difference $\Delta(\varepsilon)$ between the energy distributions of DMFT and D$\Gamma$A as well as the contribution $\varepsilon\Delta(\varepsilon)$ to the corresponding difference of the kinetic energies are shown.}
\label{fig:occup2U2375}
\end{figure*}
In Fig.~\ref{fig:occup3d}, we present our data for $n(\varepsilon)$ in $3d$ for $U\!=\!0$, DMFT, D$\Gamma$A, and the corresponding $3d$ DOS. The situation is completely analogous to the $2d$ case in Sec. \ref{subsec:occupations}, except for an (expected) weakening of nonlocal correlation effects in $3d$. 

In Fig.~\ref{fig:occup2U2375} we show $n(\varepsilon)$ for the $2d$ Hubbard model in the coexistence region of DMFT, i.e., where a metallic and a Mott-insulating phase coexist. Here, one observes the same gain of weight at large negative energies in D$\Gamma$A w.r.t. to DMFT whereas -as expected- this strong coupling feature becomes more pronounced in the insulating case (right panel).

\section{Calculation of kinetic and potential energies}
\label{app:energies}

In this section we will give computational details about the calculation of the kinetic and potential energies.

\subsection{Kinetic energy}
\label{app:ekin}

The general expressions for the calculation of the kinetic energy of the system used in this paper is given in Eq.~(\ref{equ:ekin}). For a practical evaluation of the frequency sum one has, of course, to single out the $1/i\nu$ contribution of the summand, which in our case is given by
\begin{equation}
 \label{equ:ekin1overnu}
 2\sum_{\mathbf{k}}\varepsilon_{\mathbf{k}}.
\end{equation}
In the situation considered in this paper this term is zero due to the symmetry properties of the dispersion relation. From a numerical point of view, it is also convenient to single out the contribution $\propto 1/(i\nu)^2$ in order to achieve a better convergence of the numerical sum over the Matsubara frequencies. This term, which is proportional to the second moment of the Green's function, reads (in the general case of arbitrary filling $n$)
\begin{equation}
 \label{equ:ekin1overnu2}
 2\sum_{\mathbf{k}}\varepsilon_{\mathbf{k}}\left[\varepsilon_{\mathbf{k}}+\frac{Un}{2}-\mu\right].
\end{equation}
Hence, we can rewrite Eq.~(\ref{equ:ekin}) in the following way
\begin{align}
 \label{equ:ekinrewrite}
 E_{\text{kin}}=2\frac{1}{\beta}\sum_{\nu\mathbf{k}}&\left[\varepsilon_{\mathbf{k}}G(i\nu,\mathbf{k})-\frac{\varepsilon_{\mathbf{k}}}{i\nu}-\frac{\varepsilon_{\mathbf{k}}\left(\varepsilon_{\mathbf{k}}+\frac{Un}{2}-\mu\right)}{(i\nu)^2}\right]\nonumber\\&+2\sum_{\mathbf{k}}\varepsilon_{\mathbf{k}}-\frac{\beta}{2}\sum_{\mathbf{k}}\varepsilon_{\mathbf{k}}\left(\varepsilon_{\mathbf{k}}+\frac{Un}{2}-\mu\right),
\end{align}
where we have used that $\frac{1}{\beta}\sum_{\nu}\frac{e^{-i\nu 0^+}}{i\nu}\!=\!\frac{1}{2}$ and $\frac{1}{\beta}\sum_{\nu}\frac{1}{(i\nu)^2}\!=\!-\frac{\beta}{4}$. Since $\sum_{\nu}\frac{1}{(i\nu)^3}\equiv 0$ it immediately follows that the error due to the finite frequency summation on the r.h.s. in Eq.~(\ref{equ:ekinrewrite}) is of the order $\frac{1}{(i\bar{\nu})^3}$, where $\bar{\nu}$ denotes the frequency up to which the sum in Eq.~(\ref{equ:ekinrewrite}) is performed. 

\subsection{Potential energy}
\label{app:epot}

The expression for calculating the potential energy used in this paper is given in Eq.~(\ref{equ:epot2}). Similar as for the kinetic energy, in order to ensure/improve convergence of the frequency sum in this equation, we subtract the $1/(i\nu)$ and $1/(i\nu)^2$ contributions from the summand. Evaluating the latter analytically yields
\begin{align}
 \label{equ:epotrewrite}
 E_{\text{pot}}=&\frac{1}{\beta}\sum_{\nu\mathbf{k}}\left[G(\nu,\mathbf{k})\Sigma(\nu,\mathbf{k})-\frac{Un}{2}\frac{1}{i\nu}-\frac{Un}{2}\frac{U+\varepsilon_{\mathbf{k}}-\mu}{(i\nu)^2}\right]\nonumber\\&+\frac{Un}{4}-\frac{Un\beta}{8}\sum_{\mathbf{k}}\left[U+\varepsilon_{\mathbf{k}}-\mu\right].
\end{align}

\subsection{Frequency extrapolation}
\label{app:extrapol}

In addition to the above discussed treatment of the asymptotic behavior of $G(\nu,\mathbf{k})$ and $\Sigma(\nu,\mathbf{k})$ in the calculation of $E_{\text{kin}}$ [see Eqs.~(\ref{equ:ekin}) and (\ref{equ:ekinrewrite})] and $E_{\text{pot}}$ [see Eqs.~(\ref{equ:epot2}) and (\ref{equ:epotrewrite})] we have performed a frequency extrapolation in order to check/improve the quality of our numerical results. This has been achieved by calculating the frequency sums for the evaluation of $E_{\text{kin}}$ and $E_{\text{pot}}$ for different cutoff-frequencies $\bar{\nu}$ which defines, in turn, a function $E(\bar{\nu})$ where $E$ denotes either $E_{\text{kin}}$ or $E_{\text{pot}}$. From a typical high-frequency expansion of $G(\nu,\mathbf{k})$ and $\Sigma(\nu,\mathbf{k})$ it follows that a similar expansion hold for $E(\bar{\nu})$:
\begin{equation}
 \label{equ:expandebar}
 E_{\text{fit}}(\bar{\nu})=\sum_{i=0}^{\infty}\frac{a_i}{(i\bar{\nu})^i}.
\end{equation} 
Confining the sum in Eq.~(\ref{equ:expandebar}) to a finite value of $i=I$ yields the asymptotic behavior of $E(\bar{\nu})$, which allows us to fit our numerical data to this function. Obviously, the corresponding fit parameter $a_0$ will then represent our extrapolated result for the energy for $\bar{\nu}\rightarrow\infty$. We have, hence, performed a least-square fit of our numerical data for $E(\bar{\nu})$ to the fit function $E_{\text{fit}}(\bar{\nu})$ in an asymptotic interval $[\bar{n}_{\text{min}},\bar{n}_{\text{max}}]$ by minimizing
\begin{equation}
 \label{equ:minimize}
 \sum_{\bar{n}=\bar{n}_{\text{min}}}^{\bar{n}_{\text{max}}}\left(E_{\bar{n}}-\sum_{i=0}^{I}\frac{a_i}{i^{\bar{n}}}\right)^2,
\end{equation}
where $E_{\bar{n}}\!\equiv\! E(\bar{\nu})$ with $\bar{n}$ being the index of the Matsubara frequency $\bar{\nu}\!=\!\pi/\beta(2\bar{n}+1)$. $\bar{n}_{\text{min}}$ and $\bar{n}_{\text{max}}$ define the (asymptotic) frequency interval in which the function $E(\bar{\nu})$ is fitted and $I$ represents the maximal fitting order according to Eq.~(\ref{equ:expandebar}). The minimization of (\ref{equ:minimize}) w.r.t. the fitting parameters $a_i$ can be performed analytically as Eq.~(\ref{equ:minimize}) depends linearly on $a_i$ and yields the following linear equation for $a_i$:
\begin{equation}
 \label{equ:fitparam}
 \sum_{i=0}^{I}\underset{M_{li}}{\underbrace{\left(\sum_{\bar{n}=\bar{n}_{\text{min}}}^{\bar{n}_{\text{max}}}\frac{1}{\bar{n}^l}\frac{1}{\bar{n}^i}\right)}}a_i=\sum_{\bar{n}=\bar{n}_{\text{min}}}^{\bar{n}_{\text{max}}}\frac{E_{\bar{n}}}{\bar{n}^l}.
\end{equation}  
In order to solve this system for the fitting parameters $a_i$ we have to invert $M_{li}$ which might be challenging as this $I\times I$ matrix is rather ill conditioned. However, since $M_{li}$ does not depend on our numerical data $E_{\bar{n}}$ we can perform the inversion analytically for $I~\lesssim 10$, which yields
\begin{equation}
 \label{equ:solvefitparam}
 a_i=\sum_{\bar{n}=\bar{n}_{\text{min}}}^{\bar{n}_{\text{max}}}\underset{w_{i\bar{n}}}{\underbrace{\left(\sum_{l=0}^{I}M^{-1}_{il}\frac{1}{\bar{n}^l}\right)}}E_{\bar{n}},
\end{equation}
where we have exchanged the $\bar{n}$ and $l$ sum. Note that $w_{i\bar{n}}$ does not depend on the numerical data $E_n$ and can be pre-computed analytically avoiding any numerical instabilities. Hence, the extrapolated value for the energy $E$ is eventually given by
\begin{equation}
 \label{equ:a0extrapol}
 E(\bar{\nu}\rightarrow\infty)=a_0=\sum_{\bar{n}=\bar{n}_{\text{min}}}^{\bar{n}_{\text{max}}}w_{0\bar{n}}E_{\bar{n}}.
\end{equation}
Throughout this paper, we have extrapolated the results for the kinetic and potential energies by means of Eq.~(\ref{equ:a0extrapol}) subtracting only the $1/i\nu$ contribution in the summand and verifying that this indeed compares well with a corresponding calculation where also the $1/(i\nu)^2$ term has been subtracted in the summand. An extrapolation of the latter results yields -as expected- only very small corrections for $E$ w.r.t. to $E(\bar{\nu}_{\text{max}})$.

\section{Technical details about Pad\'e fits}
\label{app:pade}

In this section we provide some technical details about the Pad\'e approximation we have adopted for the analytical continuation of our Matsubara frequency data of the DMFT and D$\Gamma$A self-energies presented in Figs.~\ref{fig:spectra_nodal} and \ref{fig:spectra_antinodal}. The continuation has been performed by means of a ``plain'' Pad\'e fit, as it is described, e.g., in Refs. \onlinecite{Beach2000,Schoett2016}. Concretely, we have fitted a rational function $f(z)=\frac{p(z)}{q(z)}$, $z\in\mathds{C}$, with $N$ fit parameters against $N$ self-energy data points [$(z_n=i\nu_n,\Sigma(z_n))$, $n=1\ldots N$] on the Matsubara axis such that $f(z_n)=\Sigma(z_n)$. For even $N$, the functions $p(z)$ and $q(z)$ are (complex) polynomials of degree $N/2-1$ and $N/2$, respectively, whereupon the coefficient of $z^{N/2}$ in the denominator polynomial $q(z)$ can be set to $1$ w.l.o.g., as the denominator and the numerator can be always divided by this factor. In order to verify the reliability of the continuation, for sets of data presented in the paper, we have (i) carefully checked the stability of our fits by varying the set of Matsubara frequencies used for the fit. Moreover, we have (ii) explicitly verified that our Pad\'e approximants satisfy the correct $1/i\nu$ asymptotic behavior for the self-energy on the Matsubara axis. Finally, we have (iii) analyzed the pole structure of our Pad\'e fit in order to make sure that no spurious poles appear in the upper complex half-plane. 

Note that by adopting ED as impurity solver our Matsubara data are not afflicted with any statistical error. Hence, numerical inaccuracies are due to the finite precision of the double precision floating point numbers used for the evaluation of the Pad\'e fit by means of our Fortran code. However, this problem might become relevant only in situations where the continued self-energies exhibit poles on the real frequency axis\cite{Beach2000}. As this is not the case for the finite coupling/finite temperature regime considered in the paper (where the one-particle self-energy always displays a sizable imaginary part), our algorithmic treatment should not require further extensions, in agreement with the discussions of Ref. \onlinecite{Beach2000}.

One comment is in order regarding the analytic continuation of DMFT(ED) data. In principle, ED provides the exact solution of the (finite) system and, hence, the corresponding Green's function and the self-energy of the associated impurity problem could be evaluated directly on the real axis avoiding any analytical continuation procedure. Such an approach is, however, problematic for two reasons: (i) The corresponding spectral function consists of a finite collection of ($\delta$-like) peaks which have typically no physical meaning (at least in the paramagnetic phase). In fact, in most of the cases where peak-shaped excitations are not present in the local spectral function, the DMFT(ED) self-energy on the Matsubara axis must be considered just as an approximation of the ``exact'' DMFT result, which can be obtained, e.g., by means of a high statistic QMC calculation (for exceptions, see Refs. \onlinecite{Sangiovanni2006,Taranto2012}). (ii) For the D$\Gamma$A exact results on the real frequency axis are -in any case- not available. Thus, for an unbiased comparison between the two methods, the corresponding analytic continuations should be calculated on the same footing, which explains the necessity of performing a Pad\'e analytic continuation also for the DMFT data.

\bibliography{Impact_arXiv}

\end{document}